\documentclass[longbibliography,reprint,superscriptaddress,aps,prx,twocolumn]{revtex4-1}
\usepackage{float}
\usepackage{ulem}
\usepackage{caption}
\usepackage{tabularx}
\usepackage{graphicx}
\usepackage[svgnames]{xcolor}
\usepackage{color}
\usepackage{soul}
\usepackage[colorlinks=True,linkcolor=DarkRed,citecolor=ForestGreen,urlcolor=MediumBlue,pdfstartview=FitH,bookmarks=False,pdfpagemode=UseNone]{hyperref}
\hypersetup{colorlinks=true,linkcolor=blue,filecolor=blue,urlcolor=blue}
\begin{document}
\title{Extended Comment on {\it Nature} {\bf 586}, 373 (2020) by E. Snider {\it et al.}}
\author{D. van der Marel} 
\affiliation{Department of Quantum Matter Physics, University of Geneva, 24 Quai Ernest-Ansermet, 1211 Geneva, Switzerland}
\author{J. E. Hirsch}
\affiliation{Department of Physics, University of California, San Diego, La Jolla, CA 92093-0319, USA}
\begin{abstract} 
Recently the discovery of room-temperature superconductivity was announced for a carbonaceous sulfur hydride (CSH) under high pressure~\cite{snider2020}. 
The evidence for superconductivity was based on resistance and magnetic susceptibility measurements. 
In the figures showing the susceptibility it was stated that ``the background signal, determined from a non-superconducting CSH sample at 108~GPa, has been subtracted from the data".
From a thorough data analysis we show that the data are incompatible with the notion that the susceptibility data are obtained from the ``measured voltage" using a background correction. 
On the other hand the data {\it are} compatible with the reverse procedure, namely the ``measured voltage" is obtained by adding a ``background signal" containing noise to what was reported as the background-corrected susceptibility. 
For all 6 of the reported pressures our analysis leads to the conclusion that:
(i) the reported background-corrected susceptibility data are pathological, 
(ii) they were not obtained by the method described in this paper nor by any one of the alternative 3 methods that were subsequently provided by the authors and 
(iii) the ``measured voltage" data are not raw data.
\end{abstract}
\maketitle 
\section{Introduction}\label{section:Introduction}
\begin{figure}[]
\centering
\includegraphics[width=\columnwidth]{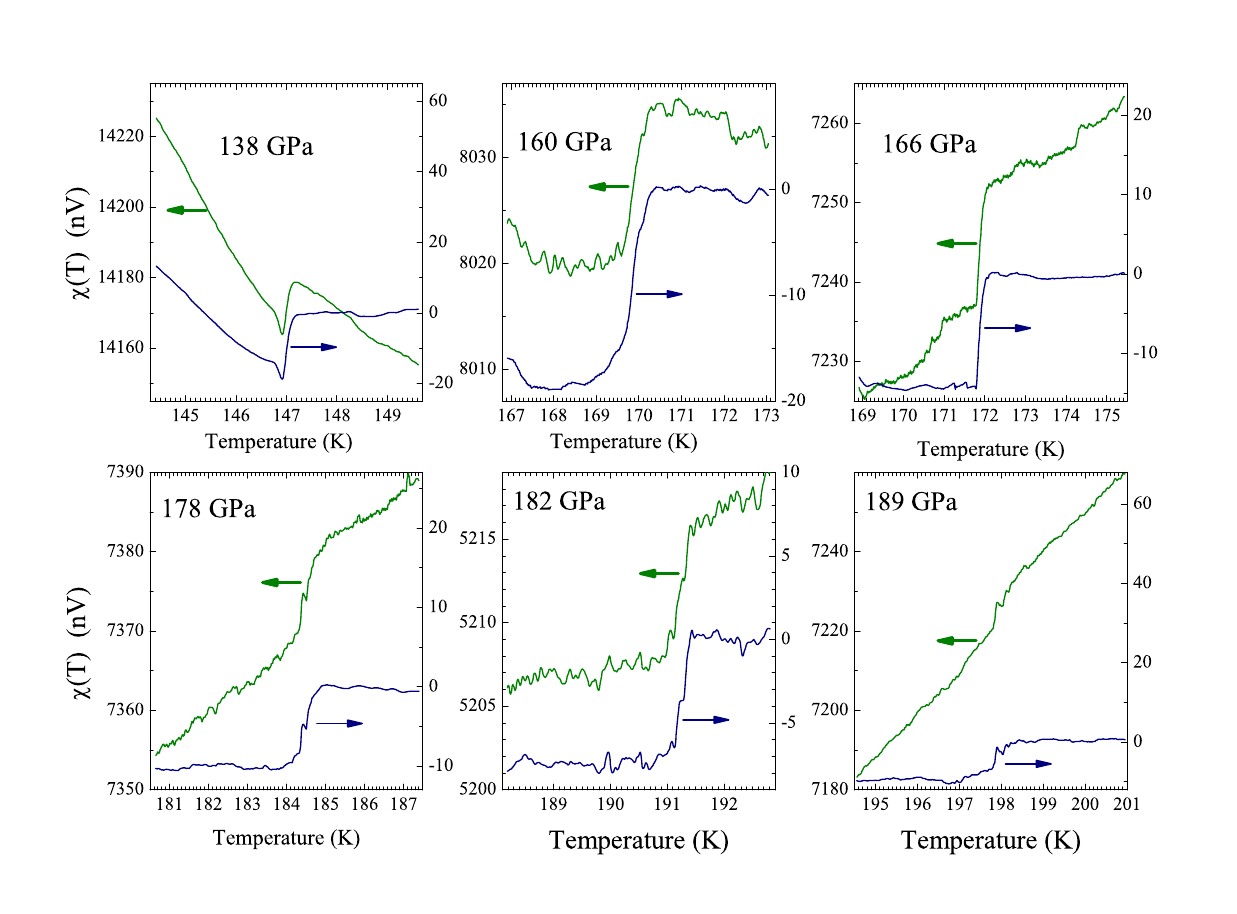}
\caption{``Measured voltage" (green) and ``superconducting signal" (dark blue) reproduced from the tables in Refs.~\cite{dias2021,opendata} for the six different pressures reported in Ref.~\onlinecite{snider2020}.}
\label{figure:susceptibility_all}
\end{figure}

In Ref.~\onlinecite{snider2020} it is reported that a material termed carbonaceous sulfur hydride (hereafter called CSH) is a room temperature superconductor. 
Data for resistance versus temperature and ac susceptibility versus temperature at six different pressures show drops suggesting superconducting transitions. 
Recently two of the authors of Ref.~\onlinecite{snider2020} have posted the numerical values of the data for the ac susceptibility curves presented in Ref.~\onlinecite{snider2020}, as well as the underlying raw data, on arXiv~\cite{dias2021}.  
The raw data and the data presented in Ref.~\onlinecite{snider2020} are  called ``measured voltage'' and ``superconducting signal'' respectively in Ref.~\onlinecite{dias2021}, for which we will use the notation $\chi^{\prime}_{mv}(T)$ and $\chi^{\prime}_{sc}(T)$ . 
In Fig.~\ref{figure:susceptibility_all}  $\chi^{\prime}_{mv}(T)$ and $\chi^{\prime}_{sc}(T)$ are displayed for all 6 pressures. In general $\chi^{\prime}_{mv}(T)$ exhibits a prominent step-like feature that is attributed in Ref.~\onlinecite{snider2020,dias2021} to the superconducting phase transition.

Initial analysis of these data was reported in Refs.~\onlinecite{hirsch2022} and~\onlinecite{mre2022}, where we pointed out that the fact that the data show much less noise than the raw data and the background signal is incompatible with the statement in Ref.~\onlinecite{snider2020} that the background signal was obtained in an independent measurement. Furthermore, we called attention to the fact that the data for pressure 160~GPa show highly anomalous regularities, and showed that they result from the superposition of a quantized component and a smooth curve~\cite{mre2022}.
	
In this paper we give an analysis  of the ac susceptibility data  for all pressures. We begin with $\chi^{\prime}_{sc}(T)$ and $\chi^{\prime}_{mv}(T)$ at $p=160$~GPa. Next the $\chi^{\prime}_{sc}(T)$ and  $\chi^{\prime}_{mv}(T)$  for all other pressures are analyzed. Our analysis leads to the conclusion  that the statements in Refs~\onlinecite{snider2020},~\onlinecite{dias2021} and~\onlinecite{dias2022} that the reported ``Measured voltage'' was a measured voltage is not compatible with the facts.

\begin{figure*} [!!ht!!]
\includegraphics[width=1.5\columnwidth]{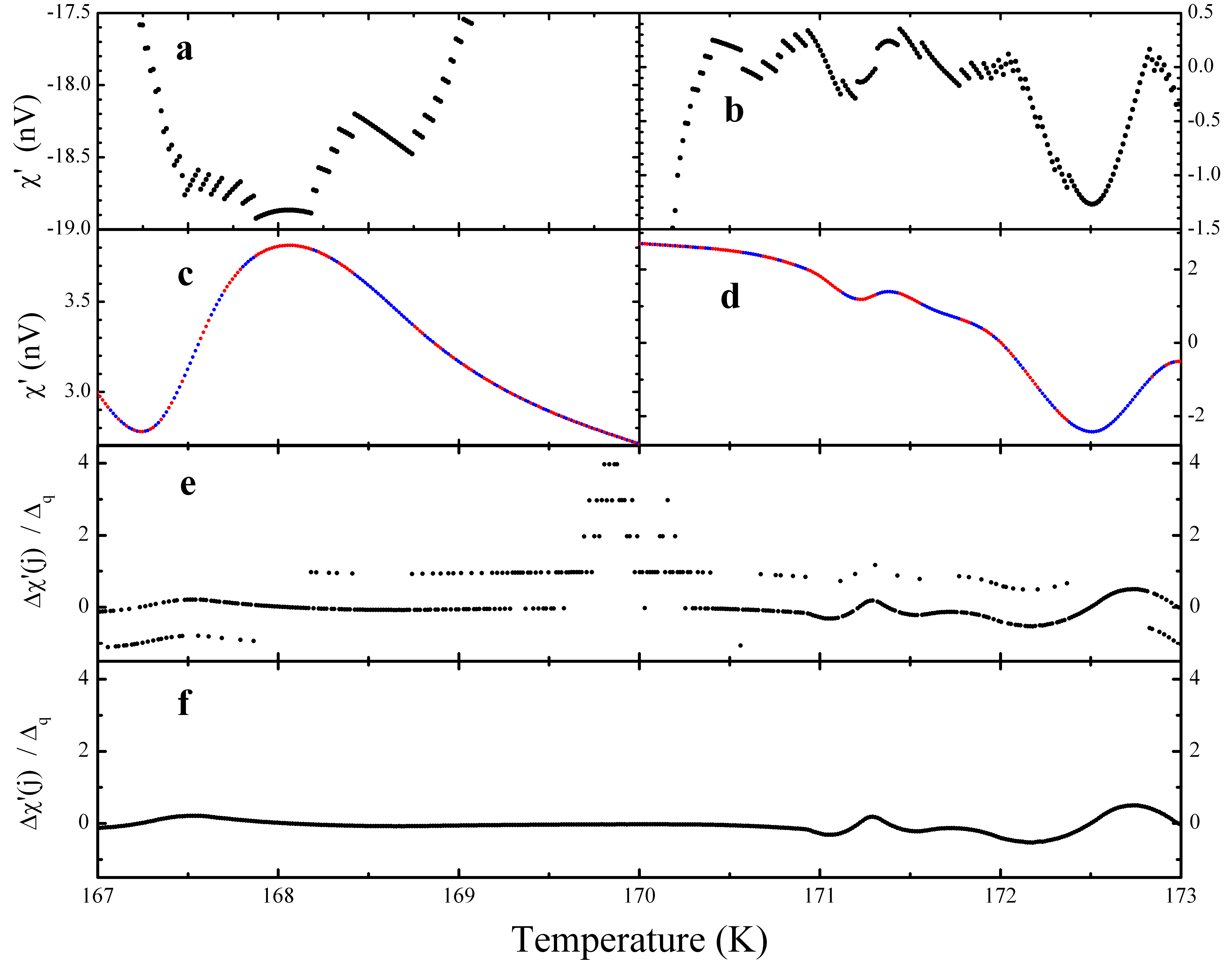}
\caption {{\bf a} and {\bf b}. Enlarged scale of the susceptibility data (``superconducting signal'') for CSH at pressure 160~GPa, from the numerical data of Table~5 of Ref.~\onlinecite{dias2021}. 
{\bf c} and {\bf d}, The data of panels {\bf a} and {\bf b} after unwrapping with integer multiples of $\Delta_q=0.16555$~nV.
The different colors  refer to disconnected segments of panels  {\bf a} and {\bf b}.
{\bf e}. The difference between neighboring points of panel {\bf a} and  {\bf b} divided by $\Delta_q=0.16555$~nV. 
{\bf f}. Same as panel {\bf e} but now using the unwrapped data of panels {\bf c} and {\bf d}. 
The Excel file with the quantities shown in this figure can be downloaded from Ref.~\onlinecite{opendata}.}
\label{figure:unwrapping}
\end{figure*}

\begin{figure} []
\centering
\includegraphics[width=\columnwidth]{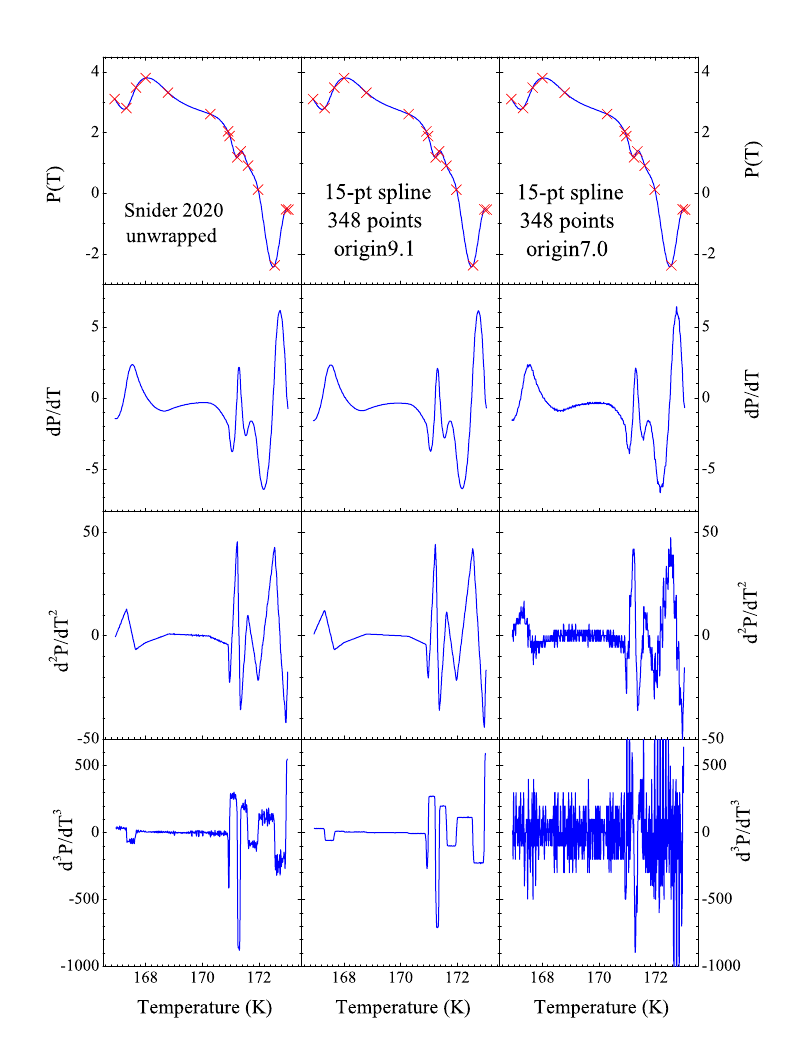}
\includegraphics[width=0.8\columnwidth]{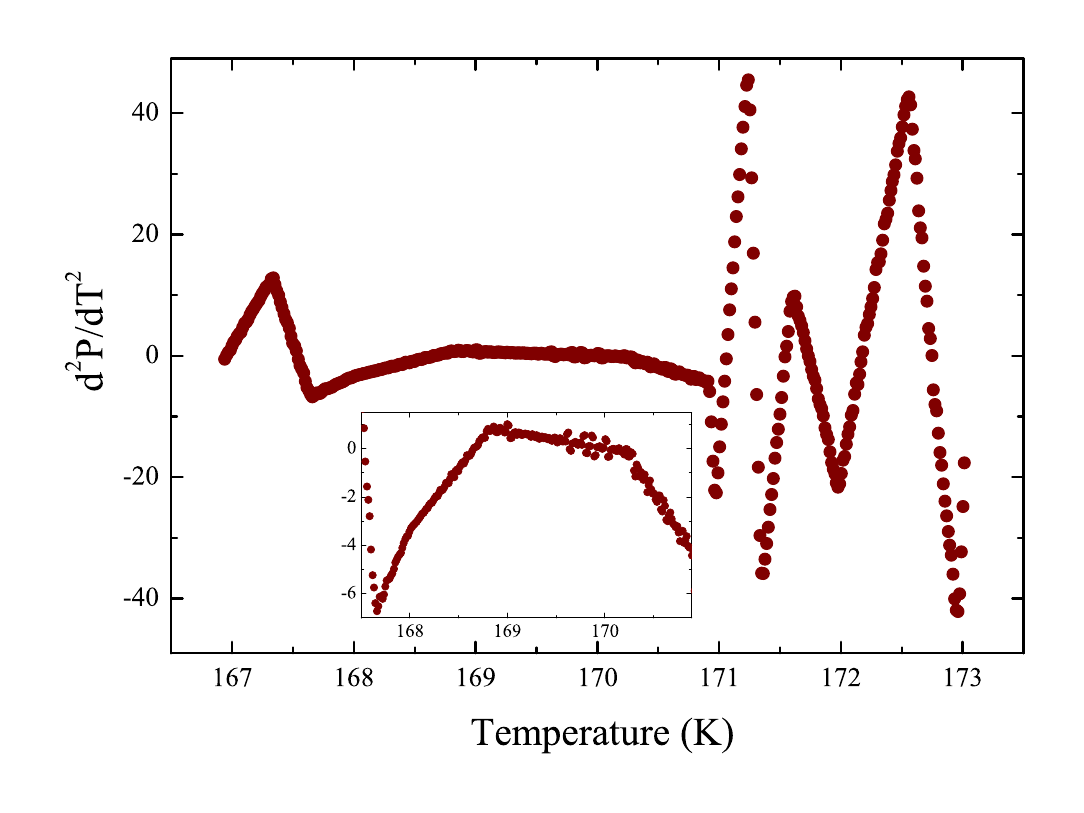} 
\caption{{\bf Top panels.} Left from top to bottom: Unwrapped component of the susceptibility and it's first, second, and third derivatives. 
Right (middle) from top to bottom: 15-node cubic spline with natural boundary conditions generated with an older (newer) version of commercial plotting software and it's first, second, and third derivatives. The nodes are indicated as red crosses in the top panels.
{\bf Bottom panel.} Enlarged view of $d^2P/dT^2$, showing the fourteen linear segments for the second derivative. The inset shows the center part enlarged further.
The Excel file with the quantities shown in this figure can be downloaded from Ref.~\onlinecite{opendata}.
\label{figure:spline_compare}}
\end{figure}

\begin{figure*} []
\includegraphics[width=1.5\columnwidth]{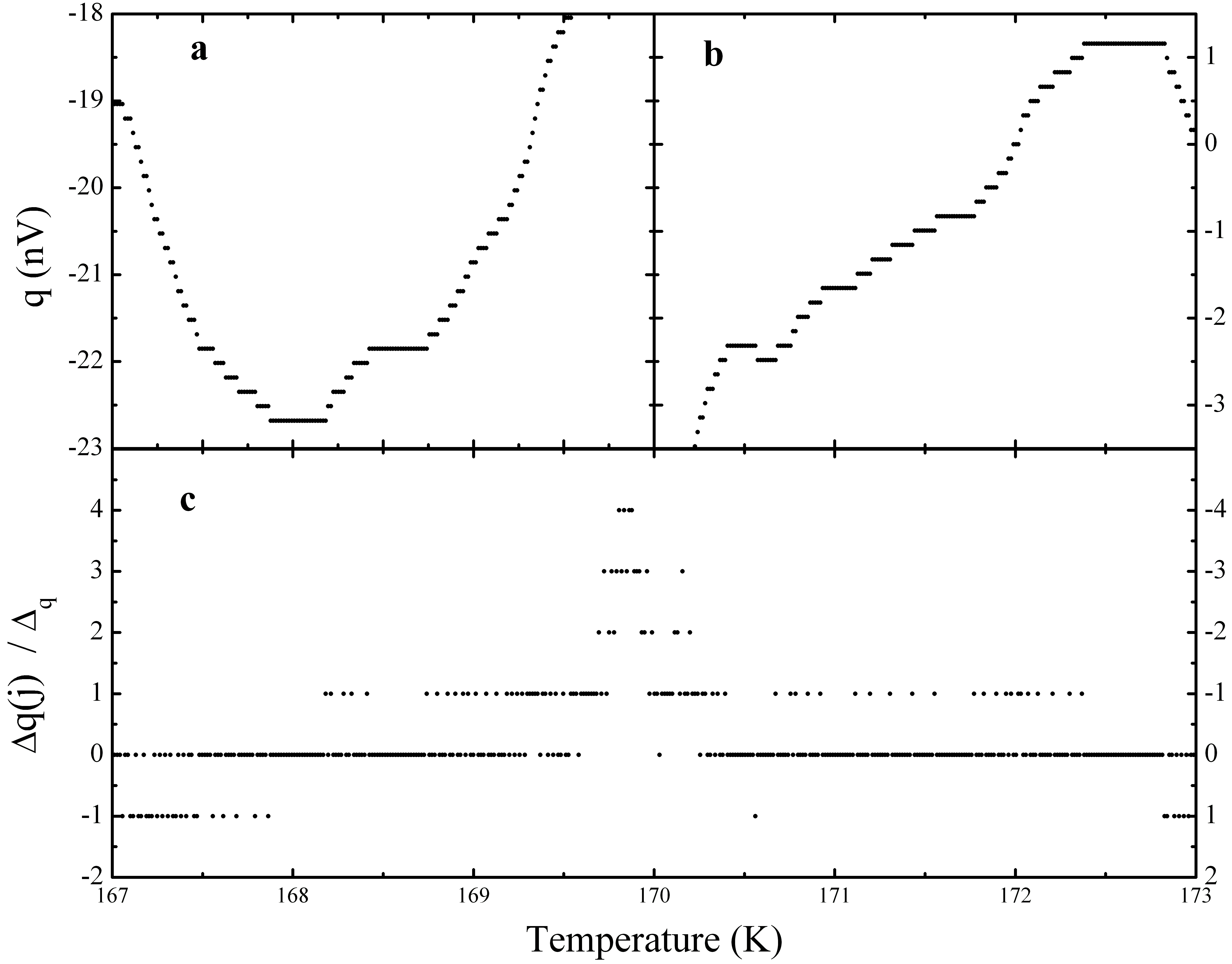} 
\caption {{\bf a} and {\bf b}, Quantized component of susceptibility data (``superconducting signal'') for CSH at pressure 160~GPa.
{\bf c}, The difference between neighboring points of panel {\bf a,b} divided by 0.16555 nV.}
\label{figure:quantized}
\end{figure*}

\begin{figure} [!!b!!]
\includegraphics[width=0.8\columnwidth]{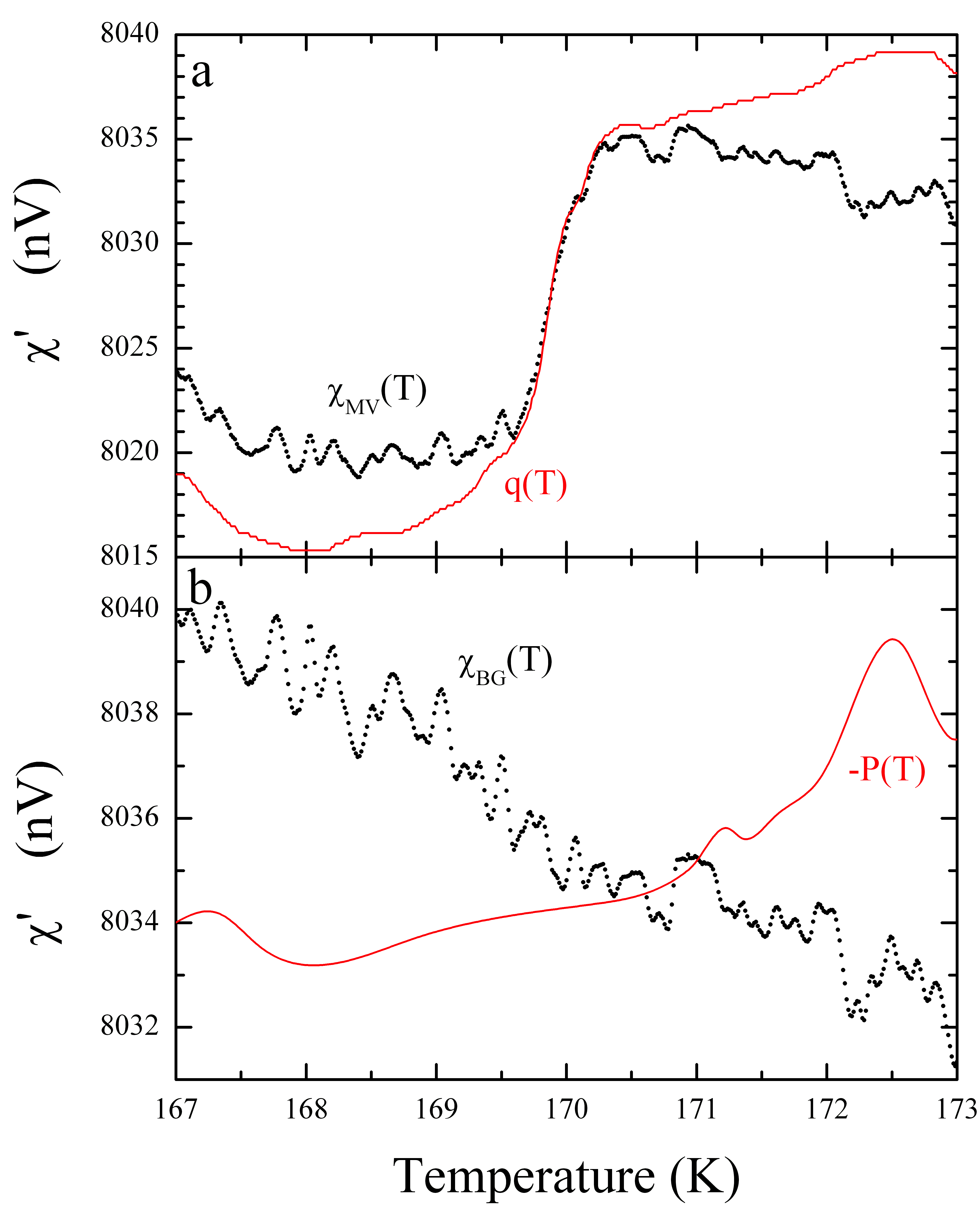} 
\caption {{\bf a}, ``measured voltage" reported in Ref.~\onlinecite{dias2021} for 160~GPa (black points), compared with quantized component $q(T)$ of $\chi^{\prime}_{sc}$ (red points).
{\bf b}, Background signal inferred from $\chi^{\prime}_{bg}=\chi^{\prime}_{mv}-\chi^{\prime}_{sc}$ in Ref.~\onlinecite{dias2021} (black points), compared with background signal $-P(T)$ inferred from unwrapping of the susceptibility data (red points). $-P(T)$ and $q(T)$ have been shifted vertically for ease of comparison.}
\label{figure:compare}
\end{figure}

\section{Analysis of the 160~GPa data}\label{section:Analysis}
Fig.~\ref{figure:unwrapping}{\bf a,b} shows the susceptibility corresponding to pressure 160~GPa shown in Extended Data Figure~7{\bf d} of Ref.~\onlinecite{snider2020}. The numerical values are given in the second column of Table~5 of Ref.~\onlinecite{dias2021} (labeled ``superconducting signal''). 
A superconducting transition appears to take place around $T=170~K$. Because of the steep rise at 170~K the regions above and below 170~K need to be displayed in separate panels. A similar zoom of the 160~GPa curve was previously shown in Fig.~9 of Ref.~\onlinecite{hirsch2022}. 
One of the striking features is a series of discontinuous steps. These steps are directly visible to the eye in the temperature ranges where $\chi^{\prime}_{sc}(T)$  has a weak temperature dependence. However,  they are also present in the range where $\chi^{\prime}_{sc}(T)$  rises steeply as a function of temperature, as can be seen by calculating the difference between neighboring points
\begin{equation}
\Delta\chi^{\prime}_{sc}(j)=\chi^{\prime}_{sc}(T_j)-\chi^{\prime}_{sc}(T_{j-1}) .
\label{equation:Delta_chi}
\end{equation}
This quantity, shown in Fig.~\ref{figure:unwrapping}{\bf e}, exhibits an intriguing ``aliasing'' effect in the ``shadow curves'' displaced vertically by integer multiples of $0.1655$ (below this will be refined to $0.16555\pm 0.00005$ by examining the noise of the 3d derivative). To make this crisp, the vertical axis of  Fig.~\ref{figure:unwrapping}{\bf e} corresponds to $\Delta\chi^{\prime}_{sc}(j)/0.16555$. 
Clearly this is a set of curves vertically offset by an integer $n=-1, 0,1,2,3$ and $4$. The most systematic offsets in sign and size occur between 169.6~K and 170.1~K.

By shifting continuous segments of the curves by an amount $0.16555n$, with $n$ integers that can be read off from Fig.~\ref{figure:unwrapping}{\bf e}, it is a simple and straightforward task to apply a commonly used procedure for removing mod-$2\pi$ discontinuities of the phase of a periodic signal, termed ``unwrapping''  of the vertical offsets. 
From here on we will reserve the notation $P(T)$ for the resulting ``unwrapped curve''.  The result for the two separate ranges above and below 170~K is displayed in Fig.~\ref{figure:unwrapping}{\bf c,d}. Comparing panels {\bf c,d} to {\bf a,b} it is possible to verify that the resulting curves are extremely smooth and completely free of discontinuities. The steep rise at 170~K is absent from panels {\bf c,d}. As a consistency check $\Delta P(j)$ of panels {\bf c,d} was finally calculated and is shown in panel {\bf f}. Comparing the result shown in panel {\bf f} with that in panel {\bf e} (shown with the same vertical scale to facilitate comparison) it can be seen that there are no shadow curves in panel {\bf f}, demonstrating that not only the temperature dependence of panels {\bf c,d} is smooth, the differential shown in panel {\bf f} is, surprisingly for an experimental quantity, also completely smooth. 
The remarkable smoothness of $P(T)$ is further illustrated by Fig.~\ref{figure:spline_compare} where this curve is displayed together with the first, second, and third temperature derivate. The noise level is the lowest for the offset in the range $0.16555\pm 0.00005$~nV.

In the second derivative graph, shown in more detail in the lower panel of Fig.~\ref{figure:spline_compare}, all segments are straight lines. 
This clearly demonstrates that $P(T)$  is a chain of 14 polynomials of order 3:
\begin{equation}
\chi(T)=a_n + b_n(T - T_n)+ c_n(T_- T_n)^2+d_n(T- T_n)^3.
\label{eq:cubicspline}
\end{equation}
The corresponding coefficients for each segment are provided in table \ref{table:coefficients}.
The  curve, and its 1st, 2nd and 3rd derivatives fits the profile of a cubic spline~\cite{denk2017} with 15 nodes.
The nodes are shown as crosses in the top panels of Fig.~\ref{figure:spline_compare}.
The second derivatives extrapolate to zero at the extremal nodes, which corresponds to the so-called ``natural" boundary conditions of the cubic spline~\cite{denk2017}~\footnote{A cubic spline smoothly connects $n$ nodes with coordinates $(x_j,y_j)$ ($1 \le j \le n$) by applying the condition of continuity of first and second derivatives at the $n-2$ internal nodes. Two additional boundary condition are needed to fully define the spline function. Ref.~\onlinecite{denk2017} specifies the following types of additional boundary conditions: 
(i) Natural ($d^2y/dx^2=0$ at the two ends), 
(ii) Not-a-knot ($d^3y/dx^3$ are continuous for $j=2$ and $j=N-1$)
(iii) Periodic  ($dy/dx$ and $d^2y/dx^2$ are equal for $j=1$ and $j=N$)
(iv) Quadratic  (the first and the last segment are quadratic).}. This is the standard spline option of commercial plotting software. Applying the cubic spline with natural boundary conditions of different versions of commercial plotting software to the 15 nodes, and exporting on the temperature range defined by the two extremal nodes, gives the result in the middle and righthand panels of Fig.~\ref{figure:spline_compare}. 
The smaller noise of the middle column as compared to the right hand column (especially the third derivative) signals an improvement over time of the spline accuracy in subsequent versions of this software.

The behavior of the data shown in Fig.~\ref{figure:unwrapping}{\bf c,d}, together with the fact that the segments can be joined by vertical shifts that are all of the same form $0.16555 n$, indicates that the disconnected segments are portions of a continuous curve that has been broken up by quantized steps. The sequence of steps form together a quantized component $q(T)$ which is entirely responsible for the steep rise of $\chi^{\prime}_{sc}(T)$ at 170~K in 
Fig.~\ref{figure:unwrapping}{\bf a,b}. 
The ``superconducting signal" $\chi^{\prime}_{sc}(T)$ shown in Fig.~\ref{figure:unwrapping}{\bf a,b} can be expressed as:
\begin{equation}
\chi^{\prime}_{sc}(T) = q(T) + P(T)
\label{equation:decomposition}
\end{equation}
where  the unwrapped curve $P(T)$ is given in Fig.~\ref{figure:unwrapping}{\bf c,d}. Fig.~\ref{figure:quantized}  shows 
the same information as Fig.~\ref{figure:unwrapping} for the quantized component $q(T)$. The connected segments in Fig.~\ref{figure:unwrapping}{\bf a,b} are now horizontal, and the $y$-values in panel {\bf c} are integers.

According to Ref.~\onlinecite{snider2020}, a background signal $\chi^{\prime}_{bg}(T)$ measured at 108~GPa was subtracted from $\chi^{\prime}_{mv}(T)$  in obtaining $\chi^{\prime}_{sc}(T)$ that was published  in Ref.~\onlinecite{snider2020}.
In other words,
\begin{equation}
\chi^{\prime}_{sc}(T) = \chi^{\prime}_{mv}(T)-\chi^{\prime}_{bg}(T)
\label{equation:calibration} 
\end{equation}
Comparison of Eq. (4) and Eq. (3) would suggest that $\chi^{\prime}_{mv}(T)$ corresponds to $q(T)$ and  $\chi^{\prime}_{bg}(T)$ to  $-P(T)$ respectively. 
Assuming that's the case,  we have to  understand   (a)  why the measured voltage is a series of flat steps separated by jumps of a fixed magnitude $0.16555 \:  nV$, and (b)  why the background signal is a smooth curve without experimental noise.

(a) A digital lock-in amplifier will yield discrete values for the measured voltages, where the size of the step between neighboring values of measured voltages is given by the instrumental resolution.
If $q(T)$ is the raw data  (``measured voltage"), this would indicate that the resolution of the instrument in this measurement was of order 0.2 nV. 
Such a low resolution could result from setting the digitizer range of the lock-in amplifier to a large value, approximately $100 \mu$ V~\cite{brad}. 

(b) The smooth behavior of the background signal $-P(T)$  ($-1\times $ panels {\bf c,d} of Fig.~\ref{figure:unwrapping})  could be explained if, rather than  measured values of the background signal, a polynomial fit to the measured values was subtracted from the raw data (``measured voltage"). However, Ref.~\onlinecite{snider2020} does not mention  such a procedure. 

We have thus arrived at a possible way to understand the very unusual nature of the  susceptibility data for 160~GPa reported in Ref.~\onlinecite{snider2020}, namely the conjecture that measured raw data correspond to the quantized component of the superconducting signal shown in panels {\bf a,b} of Fig.~\ref{figure:quantized}, and the background signal to the negative of the unwrapped curve displayed in Fig.~\ref{figure:unwrapping}{\bf c,d}.

On the other hand, the superconducting signal as well as the measured raw data were reported in Ref.~\onlinecite{dias2021} Table~5, and we can infer from them the background signal simply by subtraction.
Therefore we compare, in Figs.~\ref{figure:compare}{\bf a} and {\bf b}, the reported raw data and the background signal inferred from the reported raw data (``measured voltage") and the reported data (``superconducting signal")\cite{dias2021} with our hypothesized raw data and background signal $q(T)$ and $-P(T)$.
It can be seen in Fig.~\ref{figure:compare}  that there is a complete disconnect between the raw data (``measured voltage") and the background signal inferred from the numbers reported in Ref.~\onlinecite{dias2021}, and the raw data and background signal inferred from the analysis of
the superconducting signal \cite{snider2020} (numerical values given in Ref.~\onlinecite{dias2021}). In particular, there is certainly no way that a polynomial fit of the black points in Fig.~\ref{figure:compare}{\bf b} would have any resemblance to the red curve shown in Fig.~\ref{figure:compare}{\bf b},
and there is a significant difference between the black and red curves in Fig.~\ref{figure:compare}{\bf a}. 

\begin{center}
\begin{table*}[!!t!!]
\begin{tabular}{|c|c|c|c|c|c|c|}
\hline
$n$&$T_n$&$T_{n+1}$&$a_n$&$b_n$&$c_n$&$d_n$\\
\hline					
&K&K&nV&nVK$^{-1}$&nVK$^{-2}$&nVK$^{-3}$\\
\hline
$1$&$173.0403$&$172.9601$&$-0.5341667$&$-0.831967885$&$-1.49856942$&$+91.64778904$\\
$2$&$172.9601$&$172.5577$&$-0.5232667$&$+1.109787341$&$-23.37545976$&$-36.48274797$\\
$3$&$172.5577$&$171.9764$&$-2.3770167$&$+2.067066472$&$+22.01346341$&$+19.04094347$\\
$4$&$171.9764$&$171.6215$&$+0.1218833$&$-4.22534782$&$-10.91711562$&$-15.11550441$\\
$5$&$171.6215$&$171.3609$&$+0.9189167$&$-2.05715895$&$+7.615487151$&$+32.88542602$\\
$6$&$171.3609$&$171.2379$&$+1.3920167$&$+0.525694384$&$-26.80460669$&$-144.7971468$\\
$7$&$171.2379$&$170.9758$&$+1.1875666$&$+0.594273457$&$+24.45470318$&$+45.53297675$\\
$8$&$170.9758$&$170.9066$&$+1.8941667$&$-2.851417507$&$-14.19295181$&$-68.69964351$\\
$9$&$170.9066$&$170.2829$&$+2.0458833$&$-1.872857591$&$-2.192399589$&$-1.056277196$\\
$10$&$170.2829$&$168.7992$&$+2.6164$&$-0.355780736$&$-0.099419915$&$-0.122458732$\\
$11$&$168.7992$&$168.0085$&$+3.32481$&$-0.856801015$&$+0.365108139$&$+0.853754437$\\
$12$&$168.0085$&$167.6593$&$+3.808773$&$+0.16991766$&$-1.59661518$&$+1.657661377$\\
$13$&$167.6593$&$167.3225$&$+3.484567$&$+1.86328885$&$-3.976717692$&$-10.72445083$\\
$14$&$167.3225$&$166.9137$&$+2.814148$&$+0.911730129$&$+6.406661165$&$+5.764676053$\\
\hline
\end{tabular}
\caption{
Coefficients of Eq.~\ref{eq:cubicspline} for the unwrapped component of the susceptibility, $P(T)$, at 160~GPa. The ordering of the temperatures in Table 5 of Ref.~\onlinecite{dias2021} (from high to low) has been maintained.  The first column gives the segment number, second and third columns give the upper and lower temperature of the two nodes defining the segment, the fourth column gives $a_n=P(T_n)$. The coefficients in columns 5, 6 and 7 were calculated from the temperature derivatives shown in Fig.~\ref{figure:spline_compare} using, in this order, $d_n=\langle d^3P/dT^3\rangle_n^{n+1}/6$, $c_n=\langle d^2P/dT^2 - 6d_n(T-T_n)\rangle_n^{n+1}/2$, $b_n=\langle dP/dT - 2c_n(T-T_n)-3d_n(T-T_n)^2\rangle_n^{n+1}$, where $\langle..\rangle_n^{n+1}$ is the average in the temperature interval $\{T_n,...,T_{n+1}\}$. The quantities inside the brackets are, by construction and apart from noise, constant within each interval. The Excel file with this calculation can be downloaded from Ref.~\onlinecite{opendata}.
\label{table:coefficients}}  
\end{table*}
\end{center}
Although we will demonstrate below using correlation maps that the raw data in Ref.~\onlinecite{dias2021} for 138 and 160GPa contain a quantized component, and using correlation functions that this is in fact the case for all reported pressures, this component is superimposed on a noisy background. Consequently the raw data are not themselves integer multiples of a fixed step-size.
The reported values of the raw data are given in Table 5 of Ref.~\onlinecite{dias2021} with 11 significant digits. This is {\it not} necessarily the experimental resolution. The experimental resolution is set by the complete analogue and digital chain of which the analogue-to-digital converter is the last element. The smallest step between neighboring temperatures in Table 5  of Ref.~\onlinecite{dias2021} is of order 0.0001 nV. Hence the  resolution of the experimental setup is 0.0001 nV or a smaller number. This is about three orders of magnitude higher resolution than the resolution of the measuring device that would yield the quantized component (red curve in Fig.~\ref{figure:compare}{\bf a}) as measured raw data.

It can also be seen in Fig.~\ref{figure:compare} that there is much larger noise in the raw data and background signal reported in Ref.~\onlinecite{dias2021} than there is in the red curves that were deduced from the reported superconducting signal in section~\ref{section:Analysis}. The fact that the reported superconducting signal is significantly less noisy than the reported raw data  was already noted  in Ref.~\onlinecite{hirsch2022}, not only for pressure 160~GPa but for all other pressures as well. This is independent of the unwrapping analysis discussed above. 
\section{Protocols 1 and 2}\label{section:DA1}
Ref.~\onlinecite{snider2020} described the following protocol (hereafter Protocol 1) for background correction: ``The background signal, determined from a non-superconducting C-S-H sample at 108 GPa, has been subtracted from the data". Later in Ref.~\onlinecite{dias2022} the authors stated ``We note here that we did not use the measured voltage values of 108 GPa as the background" and provided a different protocol (hereafter Protocol 2):  ``We use the temperature dependence of the measured voltage above and below the $T_c$ of each pressure measurement and scale to determine a user defined background." The method illustrated in Fig.~2a of Ref.~\onlinecite{dias2022} does {\it not} lead to the ``superconducting signal" reported for 160~GPa, {\it i.e.} the superposition of a 15-node cubic spline and a quantized component. To introduce those two features in the ``superconducting signal" requires that they are already present in the raw data, in the ``user defined background", or in both.

\begin{figure}
\centering
\begin{minipage}{1\columnwidth}
\includegraphics[width=\columnwidth]{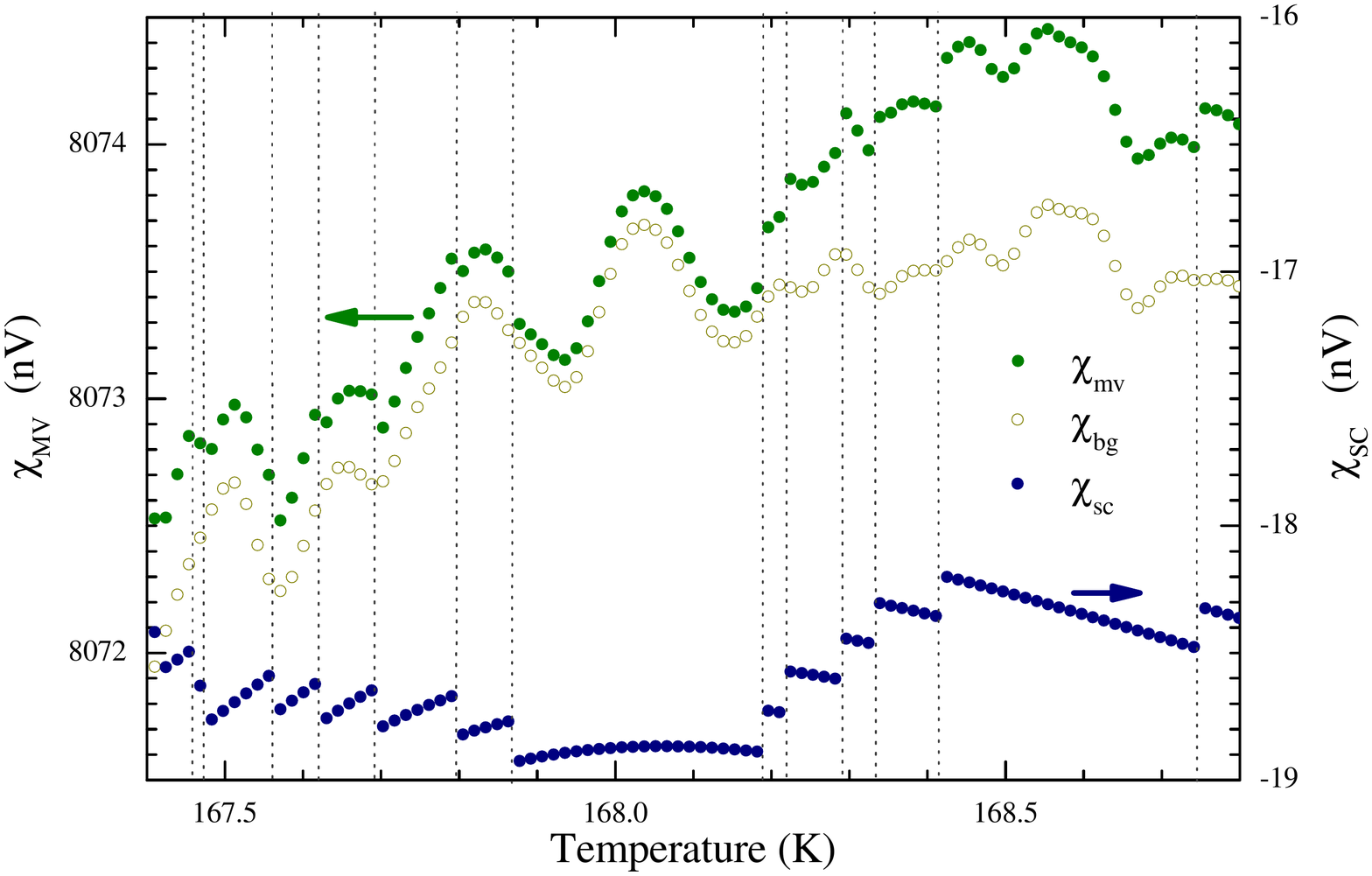}
\caption{Simulation of Eq.~\ref{equation:calibration2}.   ``Superconducting signal" ($\chi^{\prime}_{sc}$, dark blue symbols, obtained from the numerical values reported in Table~5 of Ref.~\onlinecite{dias2021}), ``background signal" (dark yellow open symbols, simulated data) and  ``Measured voltage" ($\chi^{\prime}_{mv}=\chi^{\prime}_{sc}+\chi^{\prime}_{mv}$, green symbols, shifted vertically to avoid clutter). To facilitate comparison vertical dotted lines have been drawn for each of the steps of $\chi^{\prime}_{sc}$.}
\label{figure:simulation}
\end{minipage}
\begin{minipage}{1\columnwidth}
\includegraphics[width=\columnwidth]{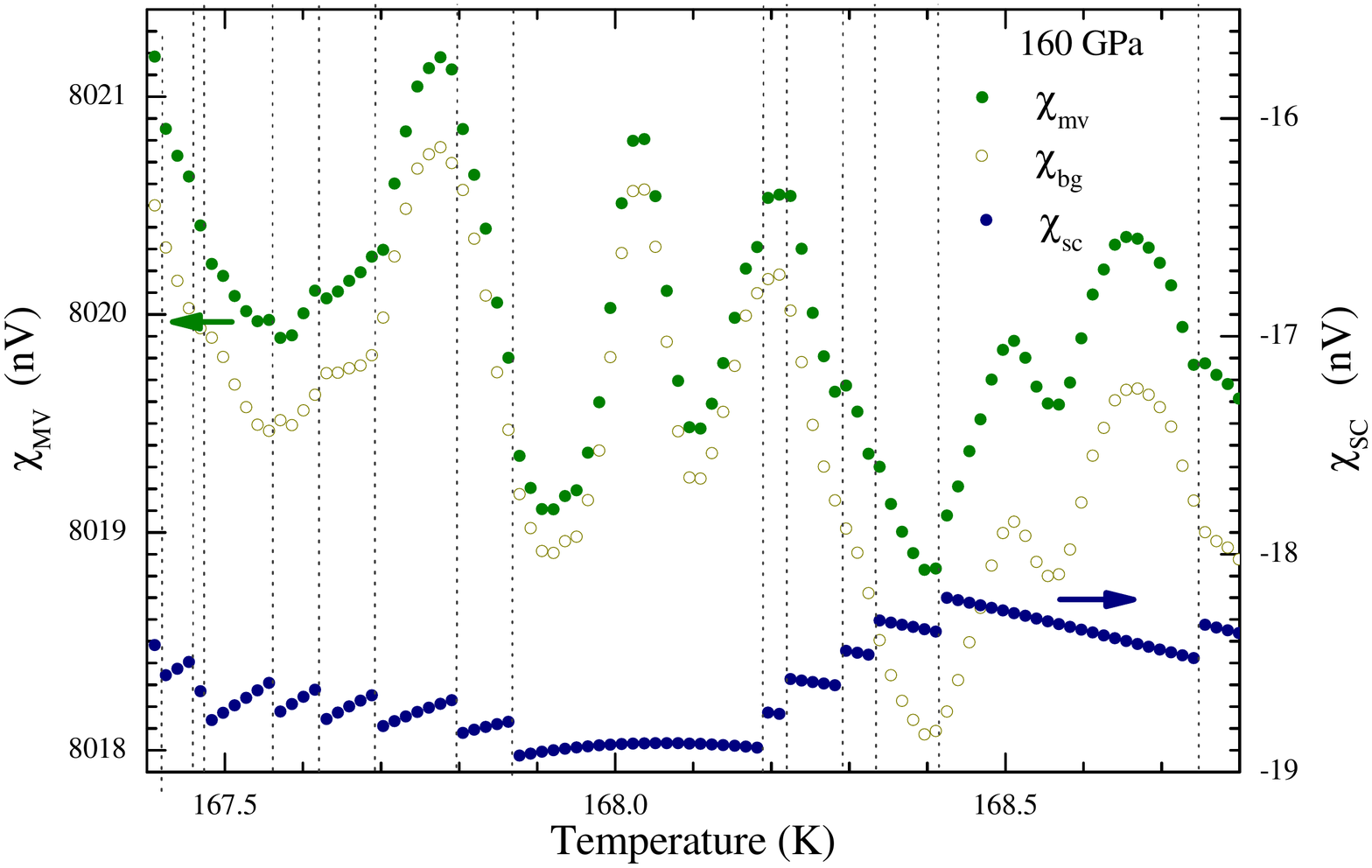}
 \caption{``Measured voltage" ($\chi^{\prime}_{mv}$, green symbols),  ``superconducting signal" ($\chi^{\prime}_{sc}$, dark blue symbols) obtained from the numerical values reported in Table~5 of Ref.~\onlinecite{dias2021}, and ``background signal" ($\chi^{\prime}_{bg}=\chi^{\prime}_{mv}-\chi^{\prime}_{sc}$, dark yellow open symbols, shifted vertically to avoid clutter). To facilitate comparison vertical dotted lines have been drawn for each of the steps of $\chi^{\prime}_{sc}$.}
\label{figure:MV_SC_160}
\end{minipage}
\end{figure}

\begin{figure} []
\includegraphics[width=1\columnwidth]{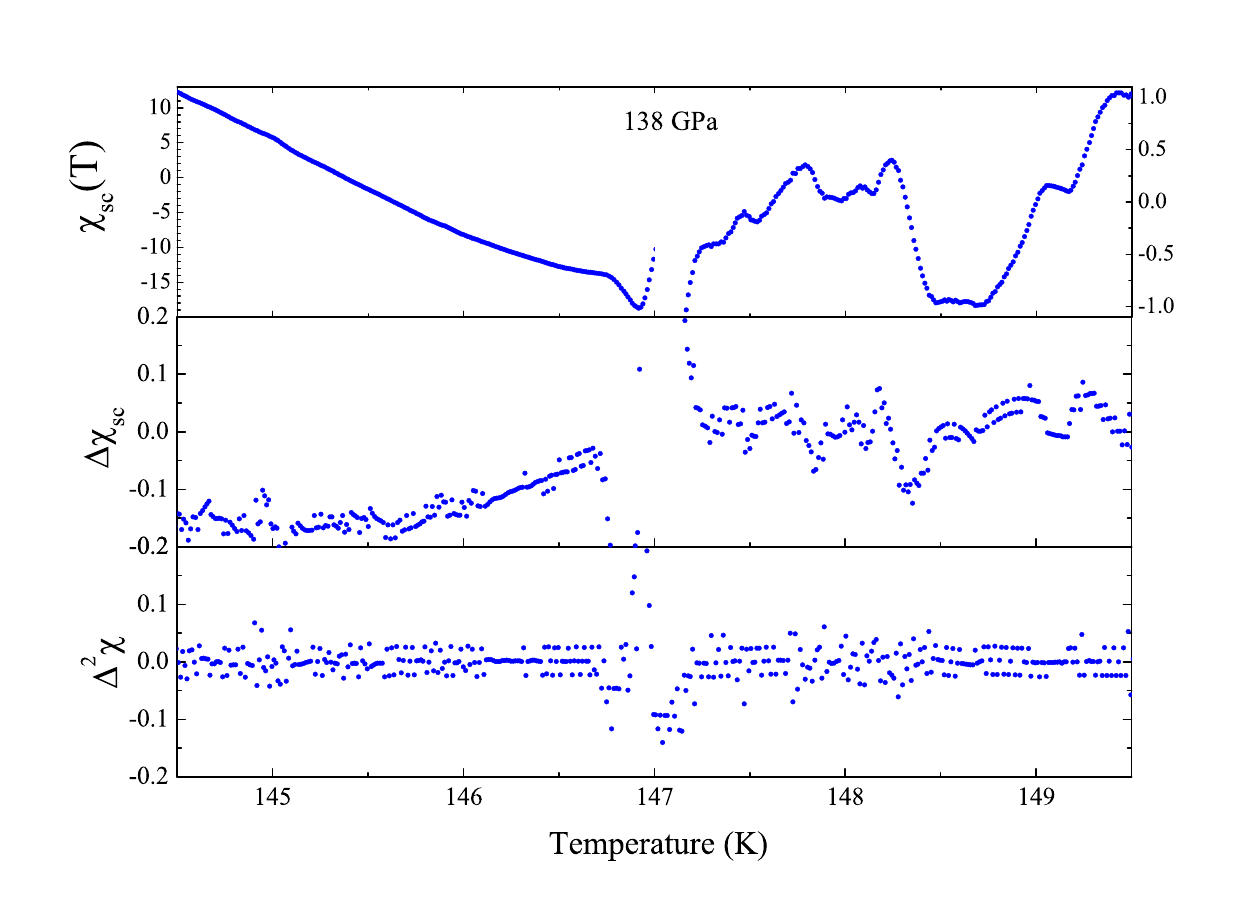}
\includegraphics[width=1\columnwidth]{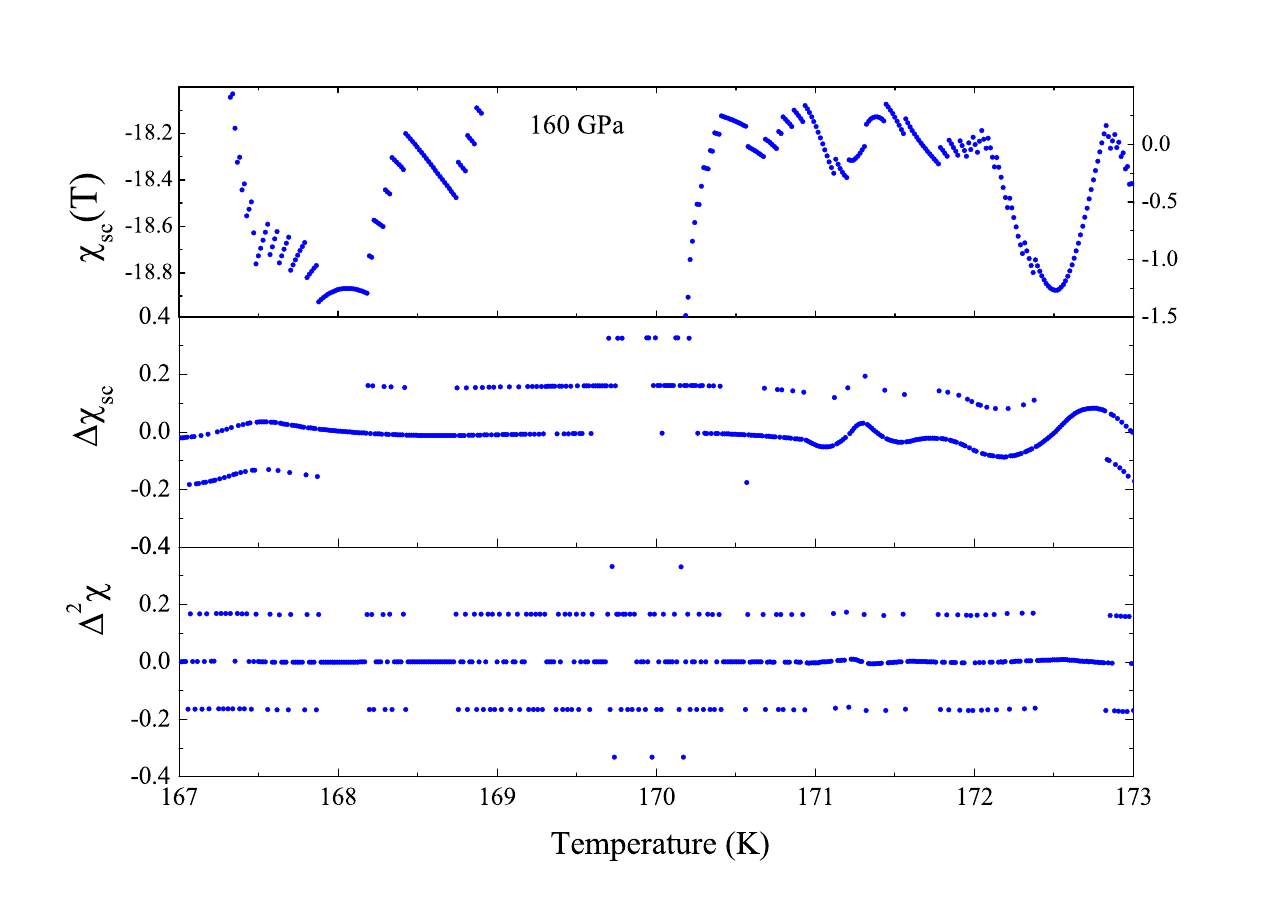}
\caption {Superconducting signal $\chi^{\prime}_{sc}(j)$ for 138~GPa (top) and 160~GPa (bottom), the first discrete differential $\Delta\chi^{\prime}_{sc}(j)=\chi^{\prime}_{sc}(j+1)-\chi^{\prime}_{sc}(j)$ and second discrete differential $\Delta^2\chi^{\prime}_{sc}(j)=\Delta\chi^{\prime}_{sc}(j+1)-\Delta\chi^{\prime}_{sc}(j)$. }
\label{figure:quantization138-160}
\end{figure}

\begin{figure} []
\includegraphics[width=1\columnwidth]{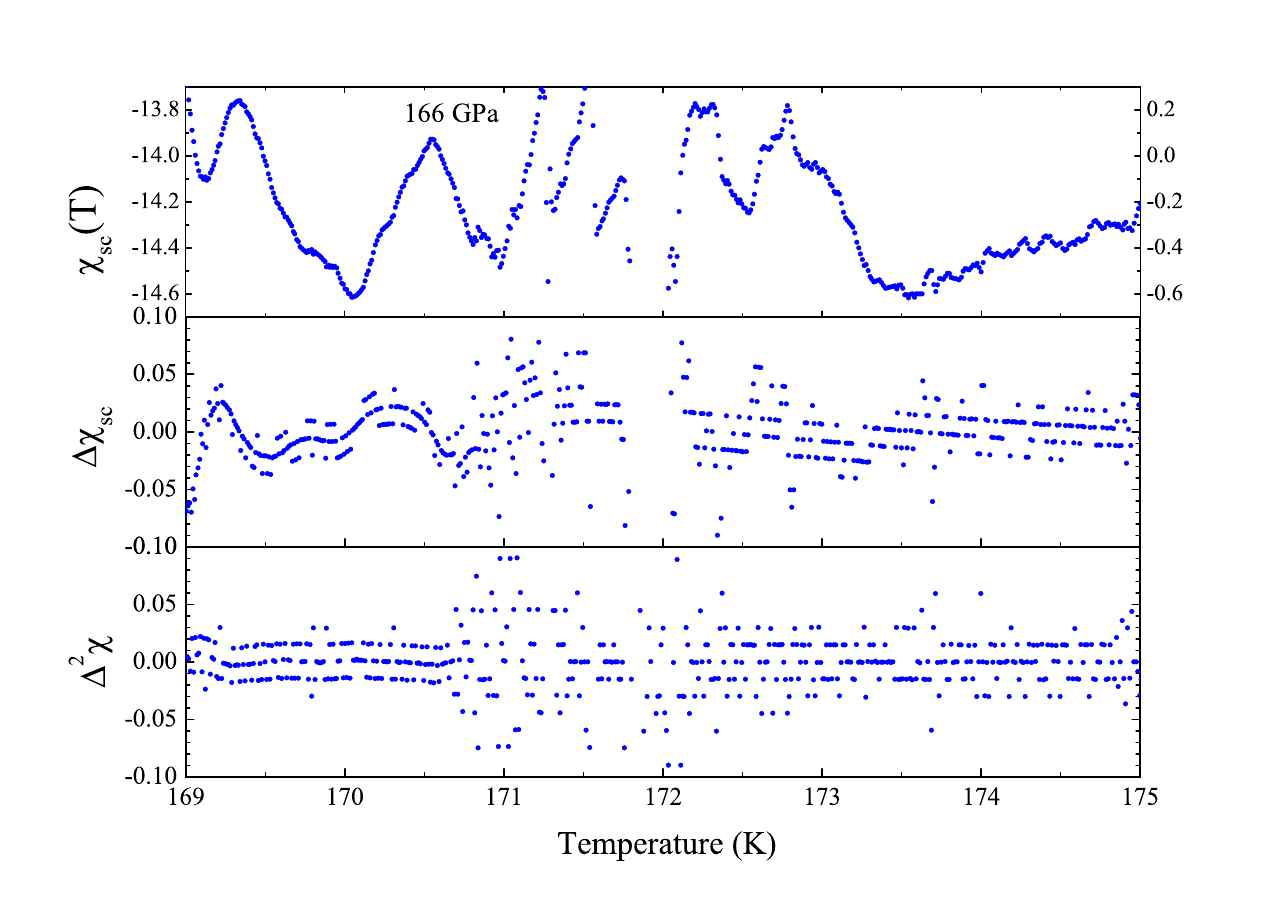}
\includegraphics[width=1\columnwidth]{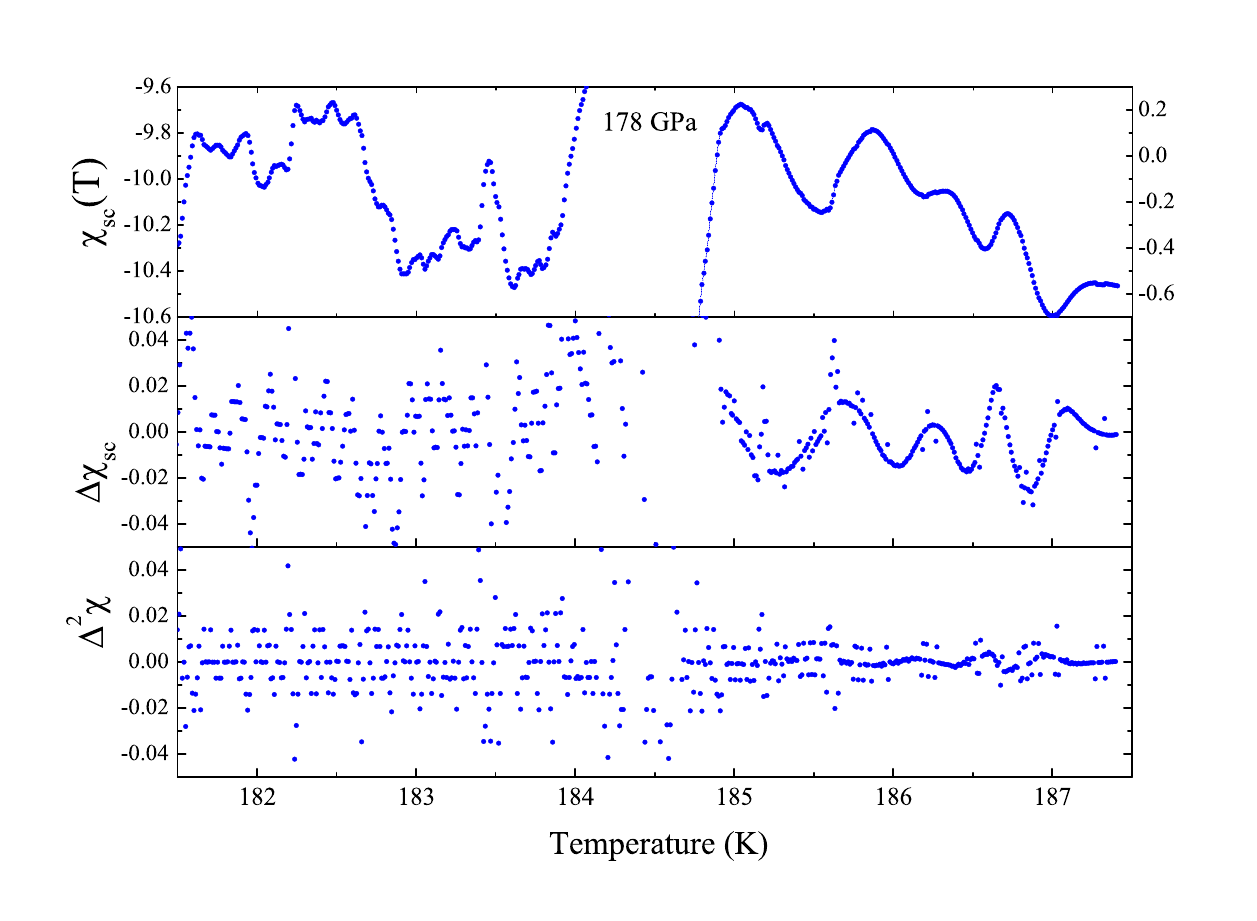}
\caption {Superconducting signal $\chi^{\prime}_{sc}(j)$ for 166~GPa (top) and 178~GPa (bottom), the first discrete differential $\Delta\chi^{\prime}_{sc}(j)=\chi^{\prime}_{sc}(j+1)-\chi^{\prime}_{sc}(j)$ and second discrete differential $\Delta^2\chi^{\prime}_{sc}(j)=\Delta\chi^{\prime}_{sc}(j+1)-\Delta\chi^{\prime}_{sc}(j)$. }
\label{figure:quantization166-178}
\end{figure}

\begin{figure} []
\includegraphics[width=1\columnwidth]{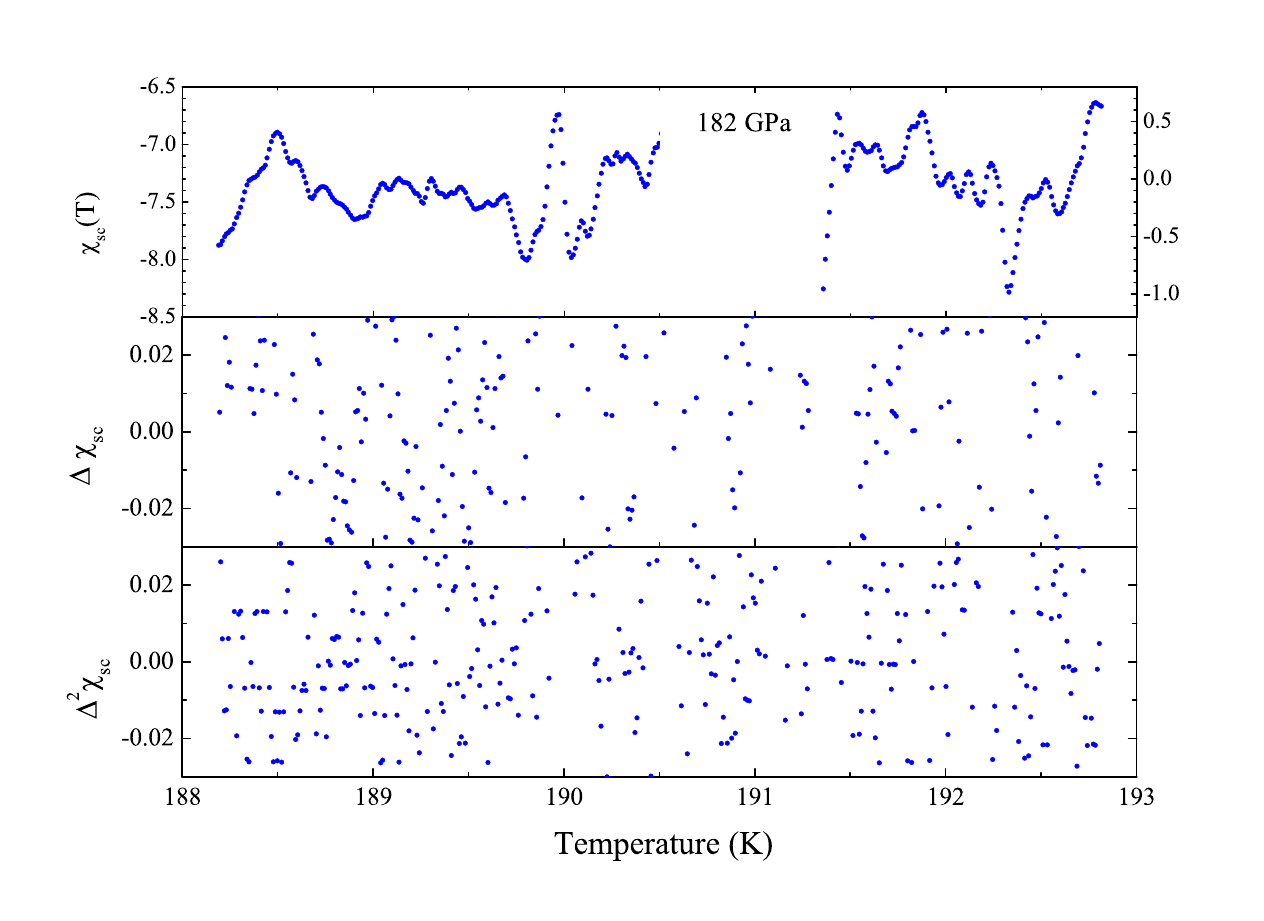}
\includegraphics[width=1\columnwidth]{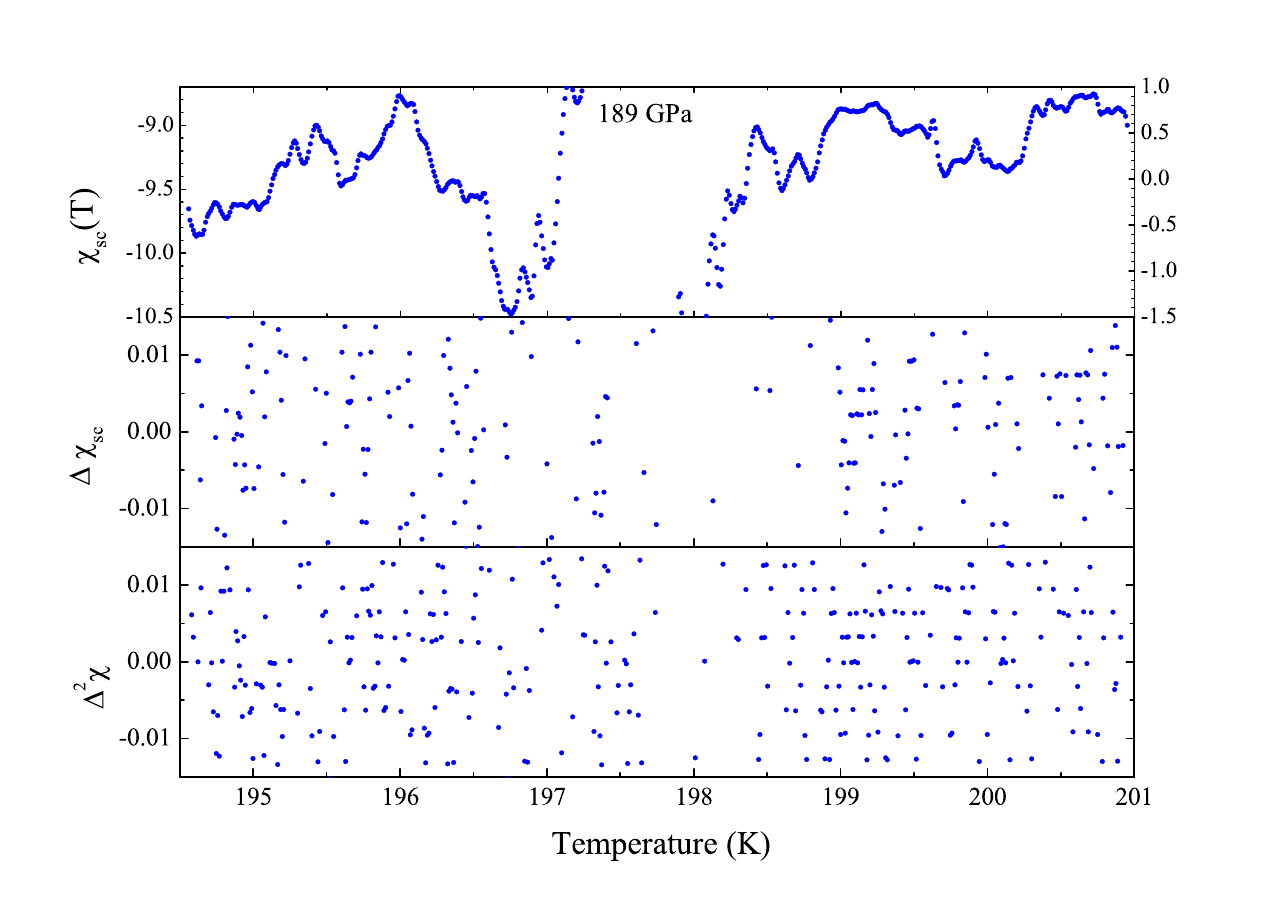}
\caption {Superconducting signal $\chi^{\prime}_{sc}(j)$ for 182~GPa (top) and 189~GPa (bottom), the first discrete differential $\Delta\chi^{\prime}_{sc}(j)=\chi^{\prime}_{sc}(j+1)-\chi^{\prime}_{sc}(j)$ and second discrete differential $\Delta^2\chi^{\prime}_{sc}(j)=\Delta\chi^{\prime}_{sc}(j+1)-\Delta\chi^{\prime}_{sc}(j)$. }
\label{figure:quantization182-189}
\end{figure}

\begin{figure} []
 \includegraphics[width=\columnwidth]{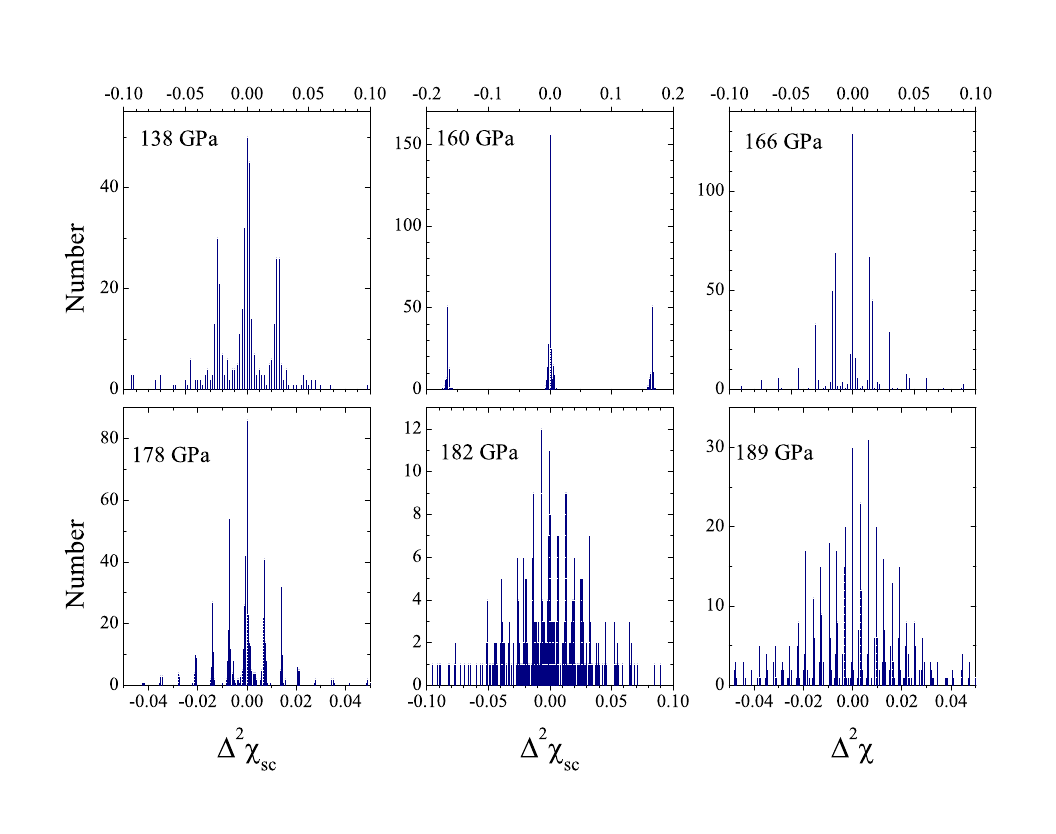}
 \caption{Histograms of $\Delta^2\chi^{\prime}_{sc}$ of the ``superconducting signal" at all pressures. This representation of the data was introduced by Dukwon~\cite{dukwon}.}
 \label{figure:histogram-sc}
\end{figure}

\begin{figure} []
  \includegraphics[width=0.8\columnwidth]{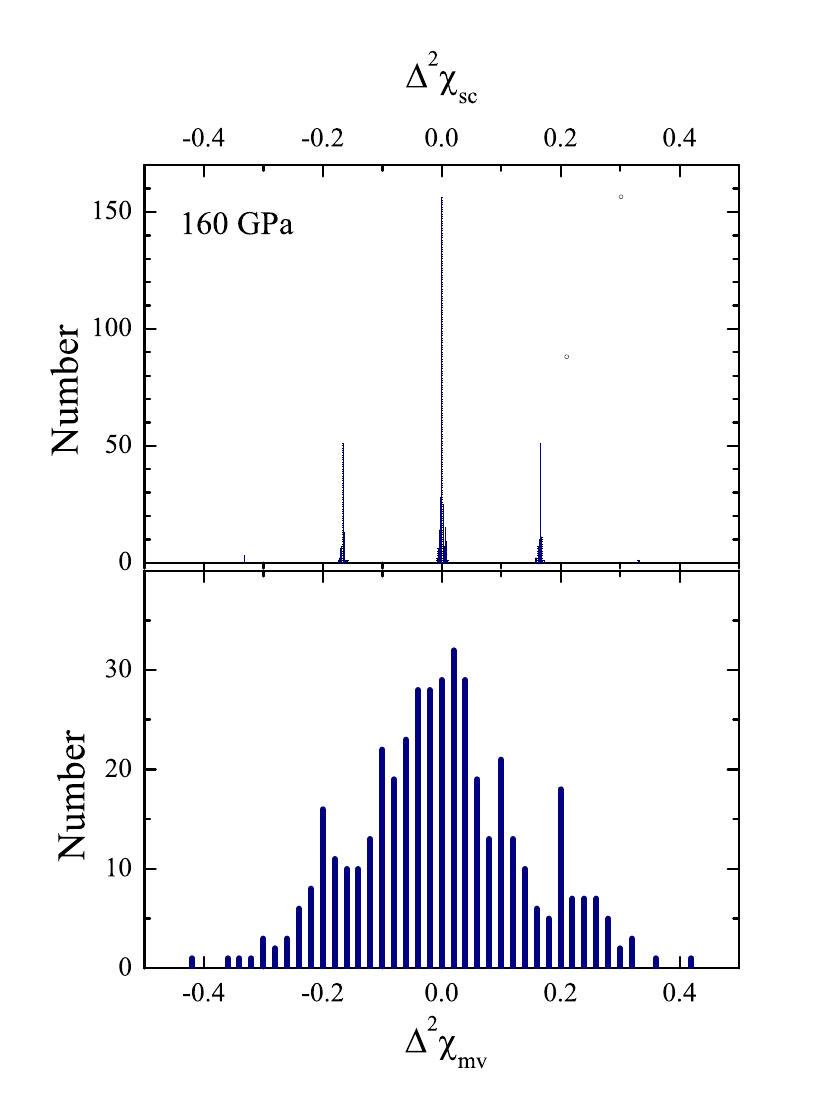}
 \caption{Histograms of $\Delta^2\chi^{\prime}_{sc}$ and $\Delta^2\chi^{\prime}_{mv}$ of the ``measured voltage" at 160~GPa. }
 \label{figure:histogram-mv160}
\end{figure}

\begin{figure} []
 \includegraphics[width=0.8\columnwidth]{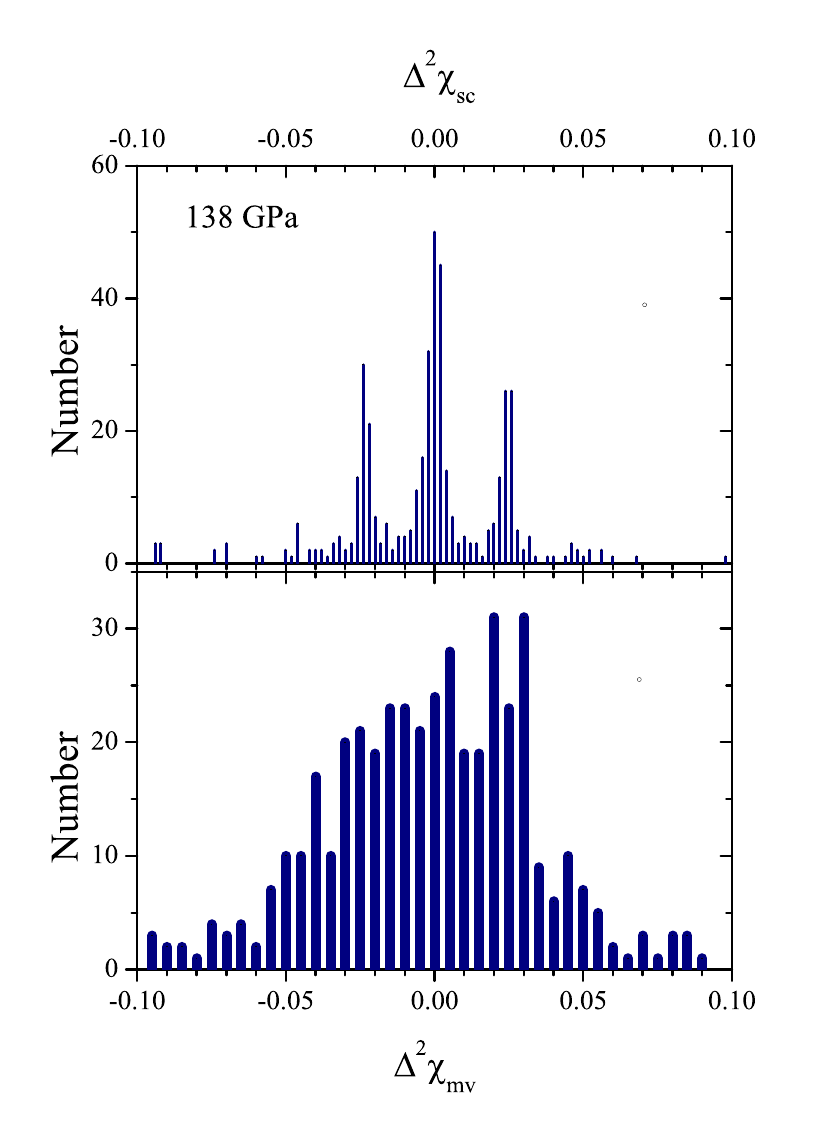}
 \caption{Histograms of $\Delta^2\chi^{\prime}_{sc}$ and $\Delta^2\chi^{\prime}_{mv}$ of the ``measured voltage" at 138~GPa. }
 \label{figure:histogram-mv138}
\end{figure}

\begin{figure}
\centering
\includegraphics[width=0.8\columnwidth]{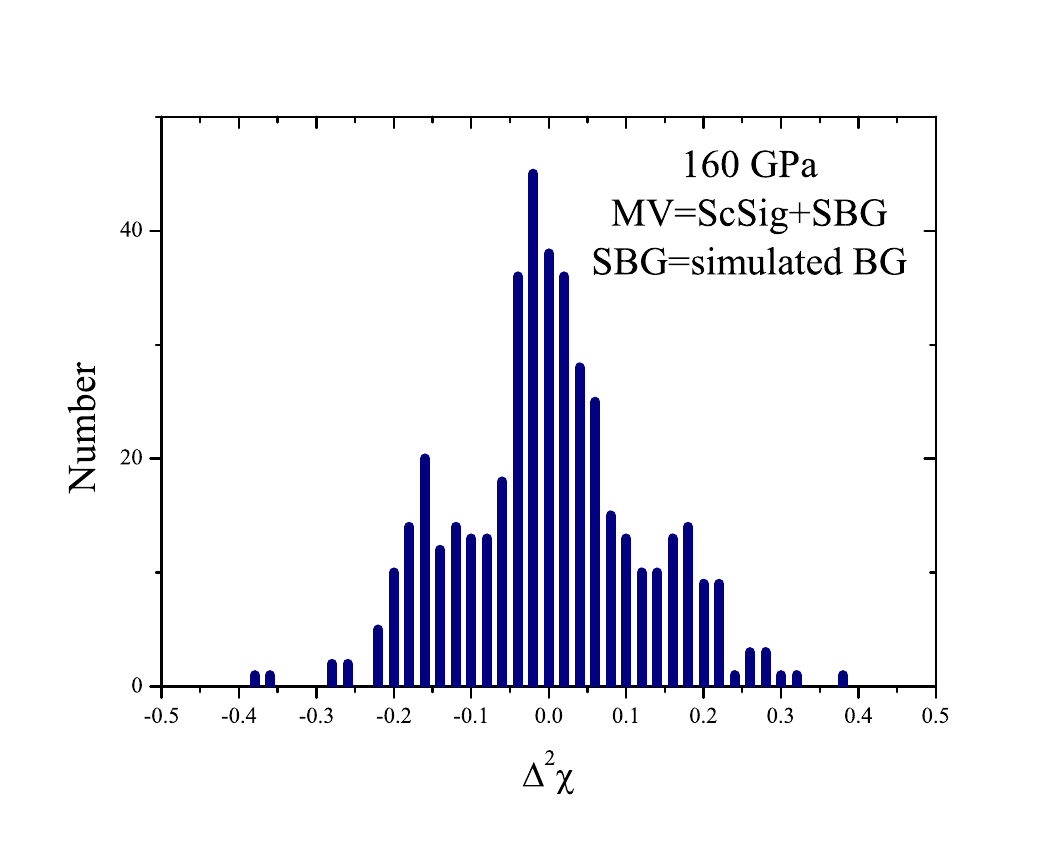}
\caption{Histogram of $\Delta^2\chi^{\prime}_{mv}$, corresponding to Fig.\ref{figure:simulation}}
\label{figure:simulationhistogram}
\end{figure}

\begin{figure*}[]
\includegraphics[width=1.6\columnwidth]{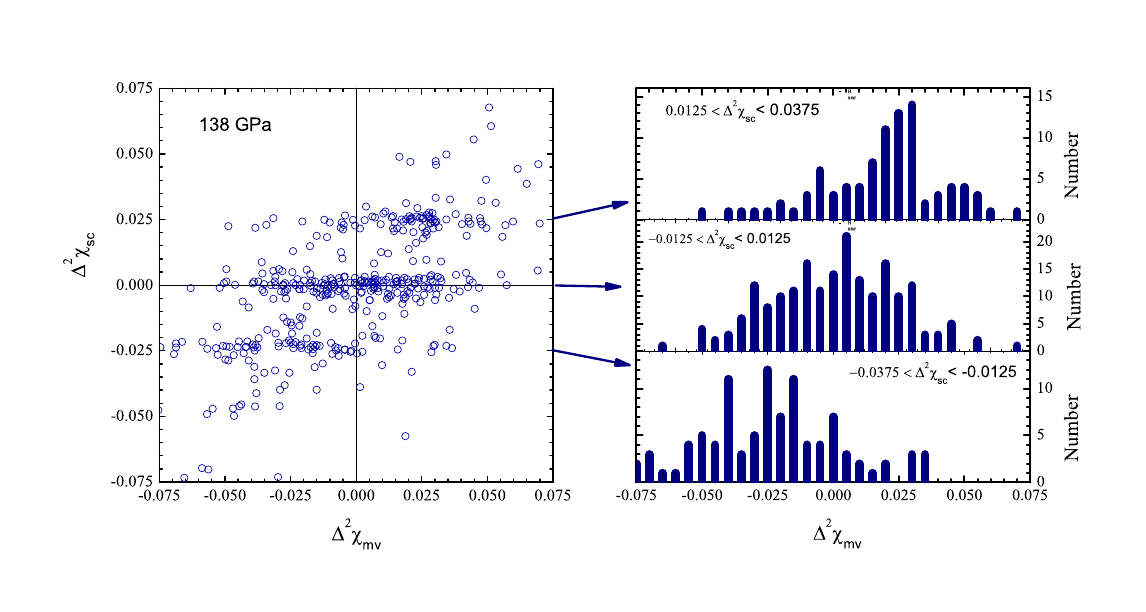}
\caption {Left: $\Delta^2\chi^{\prime}_{sc}$ versus $\Delta^2\chi^{\prime}_{mv}$ at 138~GPa.
Right: Histograms of $\Delta^2\chi^{\prime}_{mv}$ in the slots corresponding to the horizontal stripes.}
\label{figure:correlation_138}
\end{figure*}

\begin{figure*}[]
\includegraphics[width=1.6\columnwidth]{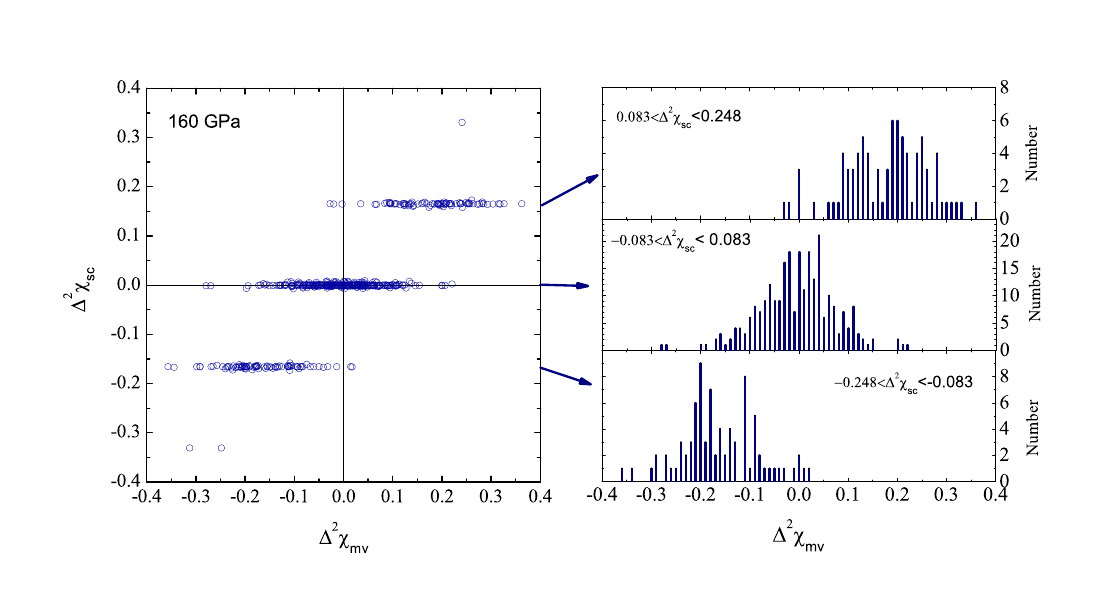}
\caption {Left: $\Delta^2\chi^{\prime}_{sc}$ versus $\Delta^2\chi^{\prime}_{mv}$ at 160~GPa.
Right: Histograms of $\Delta^2\chi^{\prime}_{mv}$ in the slots corresponding to the horizontal stripes.}
\label{figure:correlation_160} 
\end{figure*}
\section{Protocol 3}\label{section:DA2}
The inconsistencies pointed out in section~\ref{section:DA1} do not occur if, instead of assuming that $\chi^{\prime}_{mv}(T)$ and $\chi^{\prime}_{bg}(T)$ are independent signals, we assume that $\chi^{\prime}_{sc}(T)$ and $\chi^{\prime}_{bg}(T)$ are independent and rearrange Eq.~\ref{equation:calibration} as follows~\footnote{This equation was previously inferred in Ref.~\onlinecite{preprint1} from the fact that the noise in the data is significantly smaller than in the raw data}:
\begin{equation}
\chi^{\prime}_{mv}(T) = \chi^{\prime}_{sc}(T)+\chi^{\prime}_{bg}(T)
\label{equation:calibration2} 
\end{equation}
This also removes the incongruences of the noise: The noise of $\chi^{\prime}_{mv}(T)$ is now the sum of the noises of $\chi^{\prime}_{sc}(T)$ and $\chi^{\prime}_{bg}(T)$ and therefore bigger than the noise of each of the latter two.
A simulation is shown in Fig.~\ref{figure:simulation}, where for the background an arbitrary curve was taken having a noise structure similar to that shown in Fig.~2a of Ref.~\onlinecite{dias2022}. 
In Fig.~\ref{figure:MV_SC_160} we show the ``measured voltage" data points for 160 GPa corresponding to connected segments of the ``superconducting signal" points. It is apparent that several of the jumps in the data points are reflected in the raw data. Globally the features of  $\chi^{\prime}_{mv}(T)$ are similar to those in Fig.~\ref{figure:simulation}. Despite differences in  the noise structure, points in common of Fig.~\ref{figure:MV_SC_160} and Fig.~\ref{figure:simulation} are that (i) the  $\chi^{\prime}_{mv}(T)$ is noisier than  $\chi^{\prime}_{sc}(T)$ and  $\chi^{\prime}_{bg}(T)$ and (ii) several (but not all) of the steps present in the  $\chi^{\prime}_{sc}(T)$ also show up  in  $\chi^{\prime}_{mv}(T)$. 
In section~\ref{section:MV} we return to this point using a correlation method, and find that the quantized steps of $\chi^{\prime}_{sc}$ and those of $\chi^{\prime}_{mv}$ are correlated in statistically significant manner.  
This indicates that whatever the explanation is for the existence of these jumps in the data points has to also account for  their existence in the raw data.

\section{``superconducting signal" at all pressures}\label{section:SCsig_allpressures}
The data at 160 GPa stood out because of the striking sawtooth shape of the ``superconducting signal", which is in fact the superposition of a smooth curve composed of third order polynomials (15-node cubic spline with natural boundary conditions) and a quantized component in increments of 0.16555~nV.
For the other pressures the ``superconducting signal" does not have the conspicuous sawtooth appearance. 
However, just like for 160~GPa the reported data~\cite{dias2021} show much larger noise in the  $\chi^{\prime}_{mv}(T)$ and  $\chi^{\prime}_{bg}(T)$ than in  $\chi^{\prime}_{sc}(T)$ obtained by subtraction~\cite{preprint1,hirsch2022}. 
From Figs.~\ref{figure:quantization138-160},~\ref{figure:quantization166-178},~\ref{figure:quantization182-189}  showing the quantities obtained from~\cite{snider2020,dias2021,opendata}
\begin{eqnarray}
\Delta\chi^{\prime}_{sc}(j) &=& \chi^{\prime}_{sc}(j) -\chi^{\prime}_{sc}(j-1) \nonumber\\
\Delta^2\chi^{\prime}_{sc}(j)&=& \Delta\chi^{\prime}_{sc}(j) - \Delta\chi^{\prime}_{sc}(j-1) 
\end{eqnarray}
and Fig.~\ref{figure:histogram-sc} showing the distribution of $\Delta^2\chi^{\prime}_{sc}$ in the form of histograms (this method was suggested by Dukwon~\cite{dukwon}), we see that for all pressures there are underlying steps present in the temperature traces. 
The size of the steps differs from one pressure to another.
The quantization steps (Figs.~\ref{figure:quantization138-160},~\ref{figure:quantization166-178},~\ref{figure:quantization182-189} and~\ref{figure:histogram-sc},) are the smallest for 189~GPa.  
The corresponding quantization values are summarized in Table~\ref{table:properties}\\
\begin{table}[!!h!!]
\begin{center}
\begin{tabular}{| c | c | c | c | c | c |}
\hline
Pressure  & 
 $q(T)$ in ScSig ?  & 
$\Delta_q$ & 
$\delta\chi^{\prime}_{mv}$ & 
$\delta\chi^{\prime}_{mv}/\Delta_q$  
& $q(T)$ in MV ?\\
\hline
GPa &     & nV    & nV   &  &  \\
\hline
138 &yes &0.025 &0.03 & 1.2 &yes\\
160 &yes &0.17 &0.14 & 0.8 &yes\\
166 &yes &0.015 &0.2 & 13 &yes\\
178 &yes &0.007 &0.1 & 14 &yes\\
182 &yes &0.006 &0.08 &13 &yes\\
189 &yes &0.003 &0.12 & 40 &yes/no\\
\hline
\end{tabular}
\caption{Properties of the ``superconducting signal" and the ``measured voltage" at all pressures. The second (last) column is the result of the present section (section~\ref{section:correlation_functions}). In the fourth column $\delta\chi^{\prime}_{mv}$ is the variance of the distribution of $\Delta^2\chi^{\prime}_{mv}$.}
\label{table:properties}
\end{center}
\end{table}
\noindent
This suggests that for all pressures the ``superconducting signal" is the sum of a quantized component $q(T)$ and a smooth function $P(T)$.
In all cases the distribution of $\Delta^2\chi^{\prime}_{sc}$ has some broadening, and this effect is most important when the step size is smallest.
That there is no -or neglible- random noise in the superconducting signals for all pressures is born out by Figs.~\ref{figure:quantization138-160},~\ref{figure:quantization166-178},~\ref{figure:quantization182-189}, where numerous smooth sections show up prominently in $\Delta \chi^{\prime}_{sc}$ and $\Delta^2\chi^{\prime}_{sc}$. It is further illustrated by the simulation of Fig.~\ref{figure:simulation_Deltazp03} in Appendix~\ref{appendix:ED}, where the ``superconducting signal" has steps $\Delta_q=$0.03~nV, and the effect of random noise $\in\{-0.03;0.03\}$ nV is to completely obliterate the aliasing structure in  $\Delta\chi^{\prime}_{sc}$  and  $\Delta^2\chi^{\prime}_{sc}$.

In all cases the aliasing effect is most clearly seen in $\Delta^2\chi^{\prime}_{sc}$, demonstrating that for all pressures  $\chi^{\prime}_{sc}(T)$ is a superposition of a quantized component $q(T)$ and a smooth function $P(T)$, {\it i.e.} random noise -if present- has smaller amplitude than the step-size. 

Summarizing this section (i) for all pressures the ``superconducting signal" is the superposition of a quantized component and a smooth component, (ii) for the 160~GPa data the smooth component is a 15 point cubic spline with natural boundary conditions. 
\section{measured voltage at all pressures}\label{section:MV}
\subsection{Correlation maps}\label{section:correlation_maps}
The temperature traces of the ``measured voltage" at 160~GPa shown in Fig.~\ref{figure:MV_SC_160} and the corresponding histogram of $\Delta^2\chi^{\prime}_{mv}$ displayed in Fig.~\ref{figure:histogram-mv160}, show clear signatures of the quantized steps in the ``measured voltage"  reported in~\cite{snider2020,dias2021} at $\pm 0.2$~nV. 
We have verified that all steps contributing to the maxima at $\pm 0.2$~nV coincide with steps of the ``superconducting signal"~\footnote{The fact that the measured voltage peaks at $\pm 0.2$~nV instead of $\pm 0.165$~nV of the superconducting signal is a statistical fluctuation. To verify this we computed $\sqrt{\left<(\Delta^2\chi^{\prime}_{mv})^2\right>_{{i.s.}}}$ for the set of temperatures corresponding to isolated steps in $\chi^{\prime}_{sc}(T)$. The total number of isolated steps is 57. The result is $\sqrt{\left<(\Delta^2\chi^{\prime}_{mv})^2\right>_{{i.s.}}}  = 0.185 \pm 0.075~nV$. We see that the maximum at $\pm 0.2$~nV falls comfortably within this range of values.}. For 138~GPa the histogram of $\Delta^2\chi^{\prime}_{mv}$ shown in Fig.~\ref{figure:histogram-mv138}, does not provide a clear indication of quantization, which is expected given the smaller step-size. Below we will show using correlation maps that in fact the ``measured voltage" at 138~GPa has clear signatures of the same quantized steps as the ``superconducting signal" at that pressure, and using correlation functions that this is the case for all 6 reported pressures.
Despite their smaller size, these steps are nonetheless manifested by the aliasing effect in the temperature traces and histograms of $\Delta^2\chi^{\prime}_{sc}$. 
In the ``measured voltage" $\chi^{\prime}_{mv}$ the steps are obscured by the larger amplitude of the noise present in the background signal $\chi^{\prime}_{bg}$ that, as we will argue below, has been added to  $\chi^{\prime}_{sc}$. 
The exception is 160~GPa, where the steps {\it do} show up in the ``measured voltage" due to the fact that their size (0.17~nV) is much larger than for the other pressures ($< 0.025$~nV). 

A simulation is shown in Fig.~\ref{figure:simulation}, where for the background signal an arbitrary curve was taken having a noise structure similar to that shown in Fig.~2a of Ref.~\onlinecite{dias2022}. 
(i) The ``measured voltage" is noisier than ``superconducting signal" and ``background signal" and (ii) several (but not all) of the steps present in the ``superconducting signal" also show up  in ``measured voltage". 
In the corresponding histogram of  $\Delta^2\chi^{\prime}_{mv}$ (Fig.~\ref{figure:simulationhistogram}) the peaks due to the quantized steps around $\pm 0.17$~nV remain visible despite the broadening due to the noise of the added ``background" signal,
just as in Fig.~\ref{figure:histogram-mv160}. 

\begin{figure}[!!t!!]
\centering
\includegraphics[width=\columnwidth]{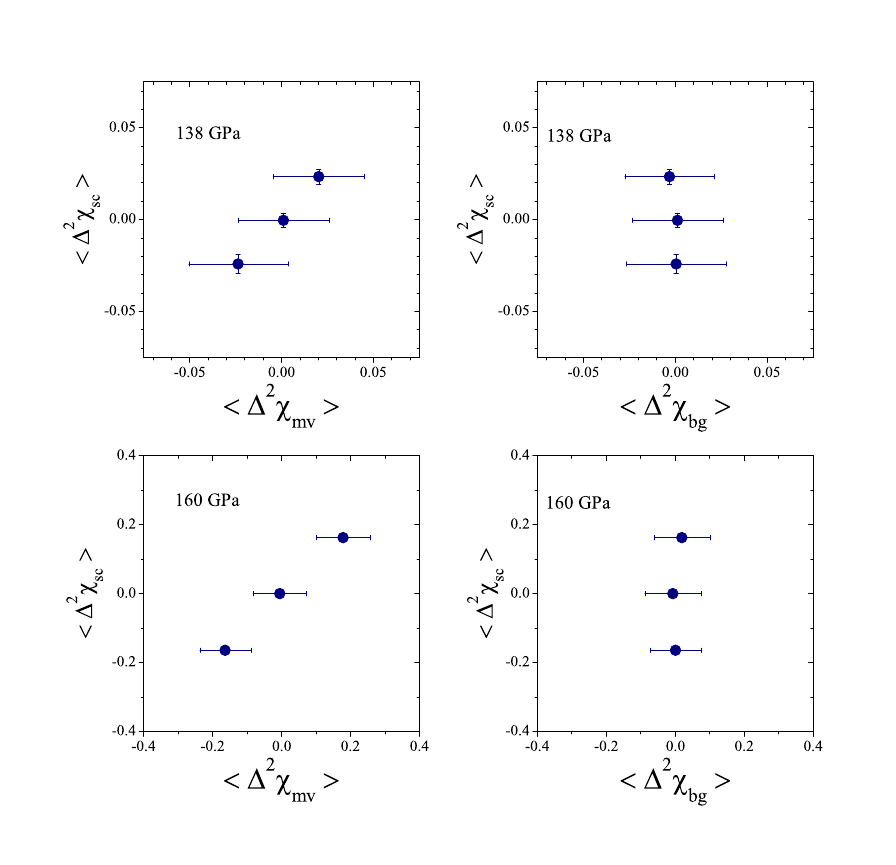}
\caption{Average values of ($\Delta^2\chi^{\prime}_{mv},\Delta^2\chi^{\prime}_{sc})$ (left) 
and ($\Delta^2\chi^{\prime}_{bg}, \Delta^2\chi^{\prime}_{sc})$ (right) with the corresponding variances obtained by averaging over the horizontal stripes in Figs.~\ref{figure:correlation_138},~\ref{figure:correlation_160},~\ref{figure:correlation_BG_SC_138},~\ref{figure:correlation_BG_SC_160}.}
\label{figure:bg_mv_sc_av}
\end{figure}

Additional insight is provided from correlations between $\Delta^2\chi^{\prime}_{mv}$ and $\Delta^2\chi^{\prime}_{sc}$. 
In Figs.~\ref{figure:correlation_138} and~\ref{figure:correlation_160} correlation maps in the $(\Delta^2\chi^{\prime}_{mv},\Delta^2\chi^{\prime}_{sc})$ plane are displayed for 138 and 160~GPa.  
These maps are parametric plots of the ($\Delta^2\chi^{\prime}_{mv}(T_j)$,$\Delta^2\chi^{\prime}_{sc}(T_j)$) coordinate for all values of $T_j$. 
The scale of the maps has been adjusted according to the quantization step size $\Delta_q$ deduced from $\Delta^2\chi^{\prime}_{sc}$ (Figs.~\ref{figure:quantization138-160},~\ref{figure:histogram-sc}, Table~\ref{table:properties}). 
For 138 and 160~GPa the noise in the ``measured voltage"  is sufficiently small to reveal that $\Delta^2\chi^{\prime}_{mv}$ has the same step structure as $\Delta^2\chi^{\prime}_{sc}$. 
This shows up in the map for 138~GPa as a pile-up of points around $(-0.025,-0.025)$ and $(0.025,0.025)$ and in the map for 160~GPa around $(-0.17,-0.17)$ and $(0.17,0.17)$. 
In the right hand panels this is shown in the form of histograms of  $\Delta^2\chi^{\prime}_{mv}$ corresponding to the horizontal stripes in the corresponding correlations maps. 
Applying the same analysis to ($\Delta^2\chi^{\prime}_{mv}(T_j)$,$\Delta^2\chi^{\prime}_{sc}(T_j)$) we see (Figs.~\ref{figure:correlation_BG_SC_138} and~\ref{figure:correlation_BG_SC_160} of Appendix~\ref{appendix:ED}) that the degree of correlation between ``background signal" and ``superconducting signal" is zero. 
Using these distributions we obtain the average values and variances summarized in Fig.~\ref{figure:bg_mv_sc_av}. This figure demonstrates unequivocally that the $q(T)$ is 100 percent present in the ``measured voltage" and totally absent from the ``background signal".
This result for 138 and 160~GPa is puzzling for the following reasons: Since the ``superconducting signal" is claimed to be obtained from the raw data (``measured voltage") by a background-correction method~\cite{dias2022}, such a method would have to distill the pathological feature $q(T) + P(T)$ out of the noisy ``measured voltage" signal. Even if that were technically possible, there is no obvious motivation to interpret $q(T) + P(T)$ as the ``superconducting signal".

From Table~\ref{table:properties} we see that for 166, 178, and 182~GPa  the noise amplitude in $\Delta^2\chi^{\prime}_{mv}\sim 13\Delta_q$, and for 189~GPa even $40\Delta_q$, which is too large compared to the step size to display a statistically significant correlation with $\Delta^2\chi^{\prime}_{sc}$. This is illustrated by the case of 166~GPa in Fig.~\ref{figure:correlation_MV_SC_166} in Appendix~\ref{appendix:ED}. 
\subsection{Correlation functions}\label{section:correlation_functions}
\begin{figure}[]
\centering
\includegraphics[width=\columnwidth]{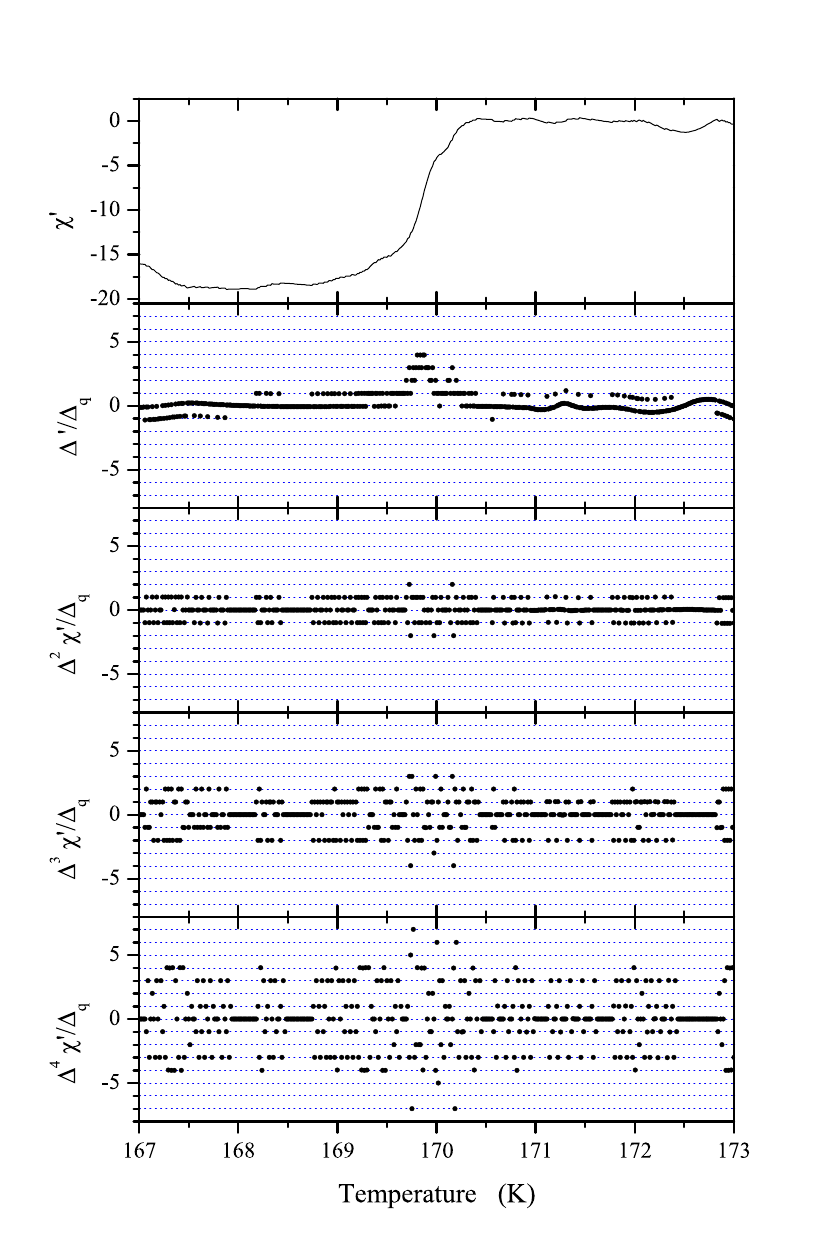}
\caption{``Superconducting signal"  at 160 GPa reproduced from the tables in Refs.~\cite{dias2021}, and the first 4 discrete derivatives normalized by $\Delta_q=0.16555$ nV.}
\label{figure:160all_D}
\end{figure} 
\begin{figure*}[]
\centering
\includegraphics[width=0.4\columnwidth]{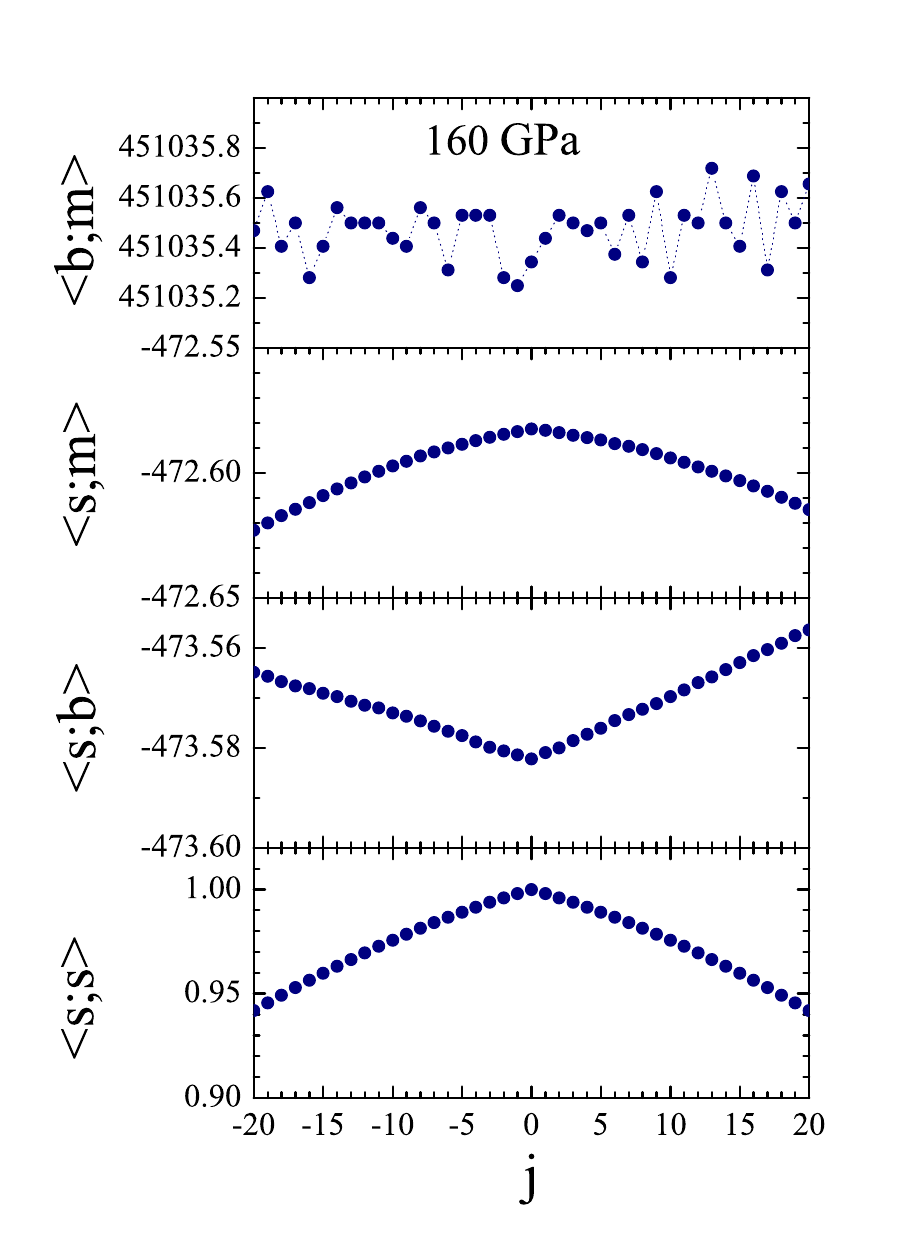}
\includegraphics[width=0.4\columnwidth]{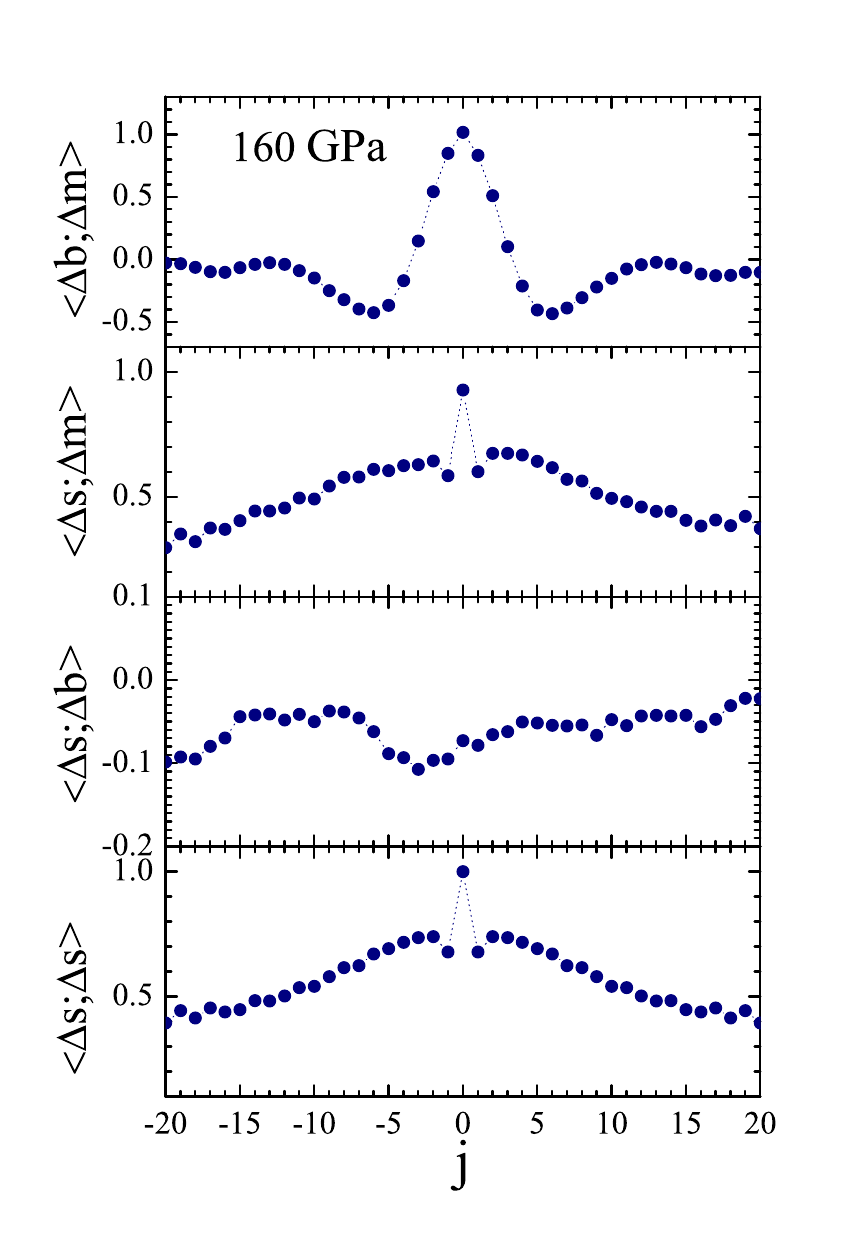}
\includegraphics[width=0.4\columnwidth]{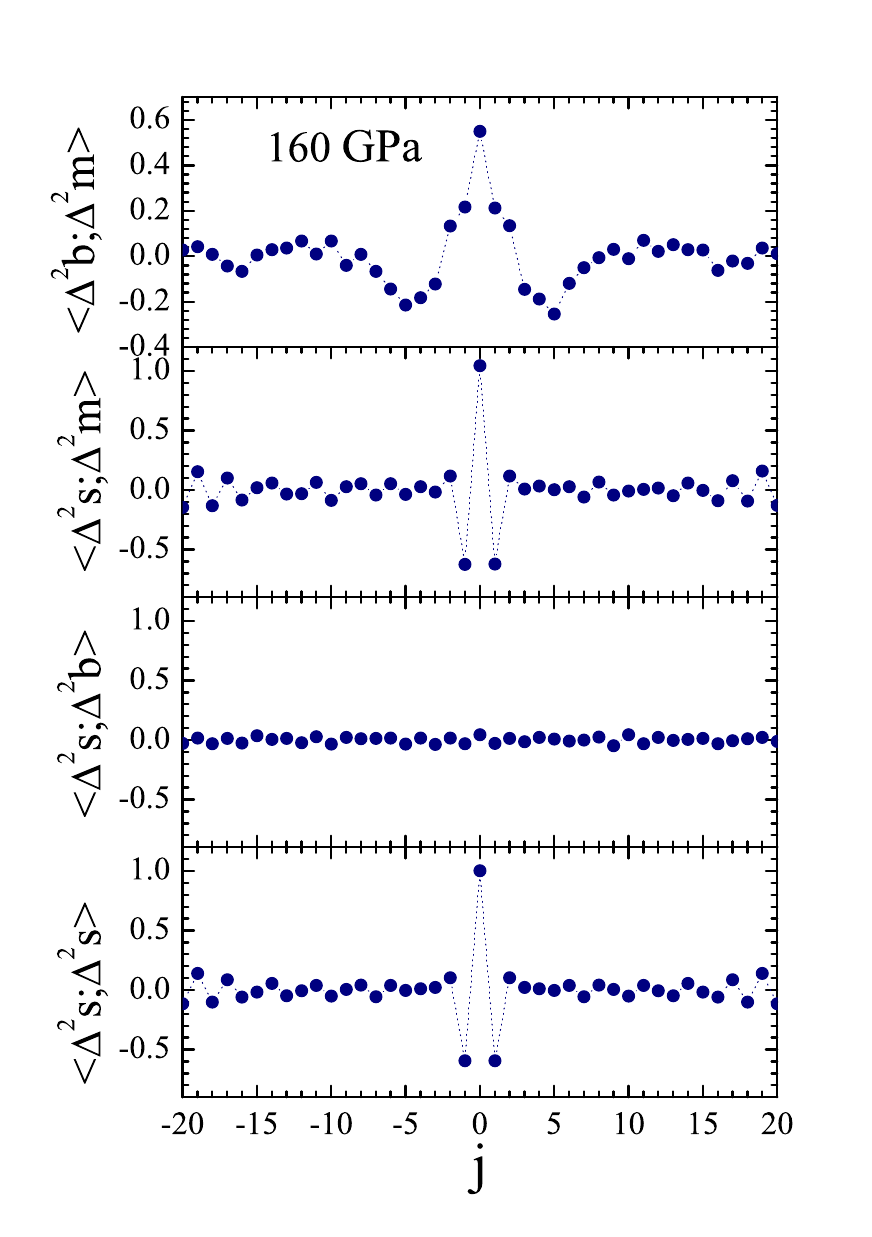}
\includegraphics[width=0.4\columnwidth]{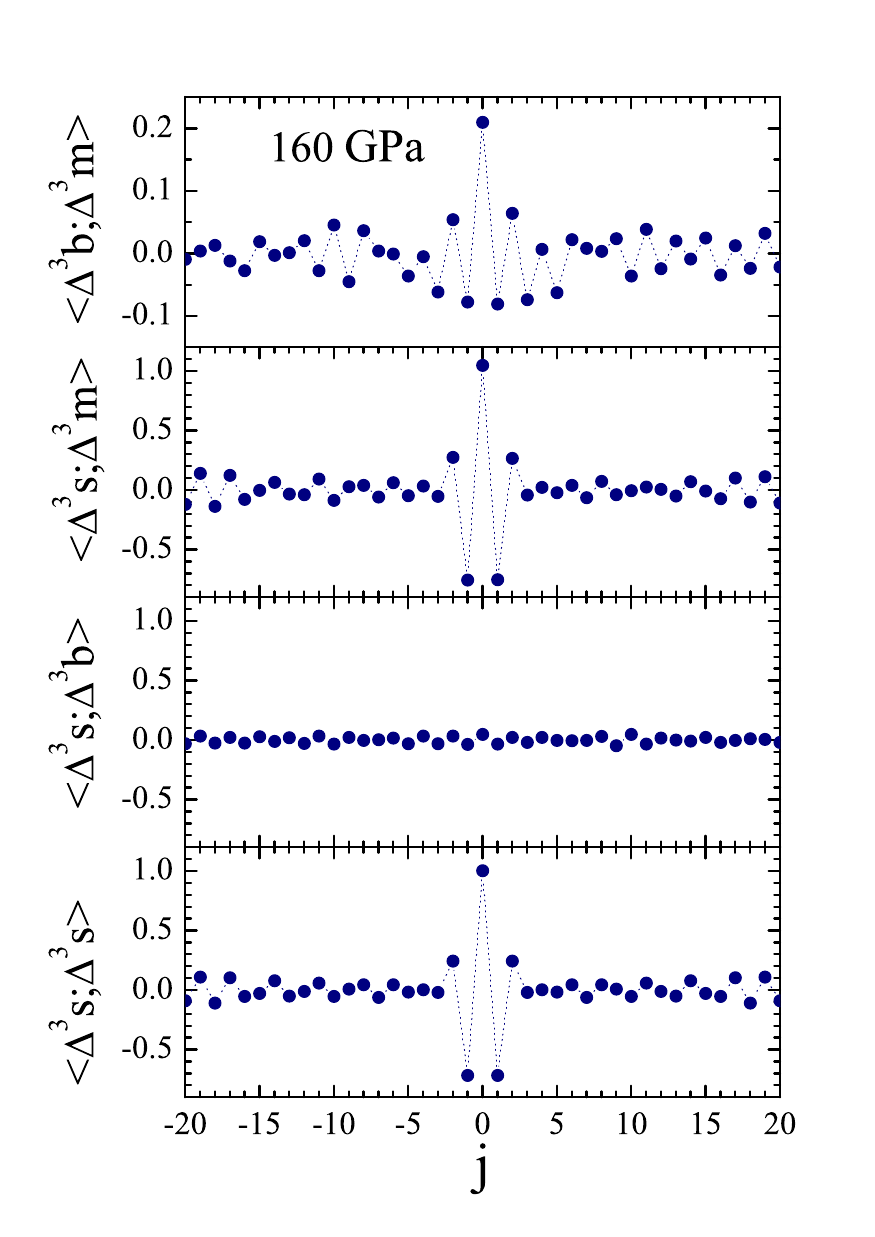}
\includegraphics[width=0.4\columnwidth]{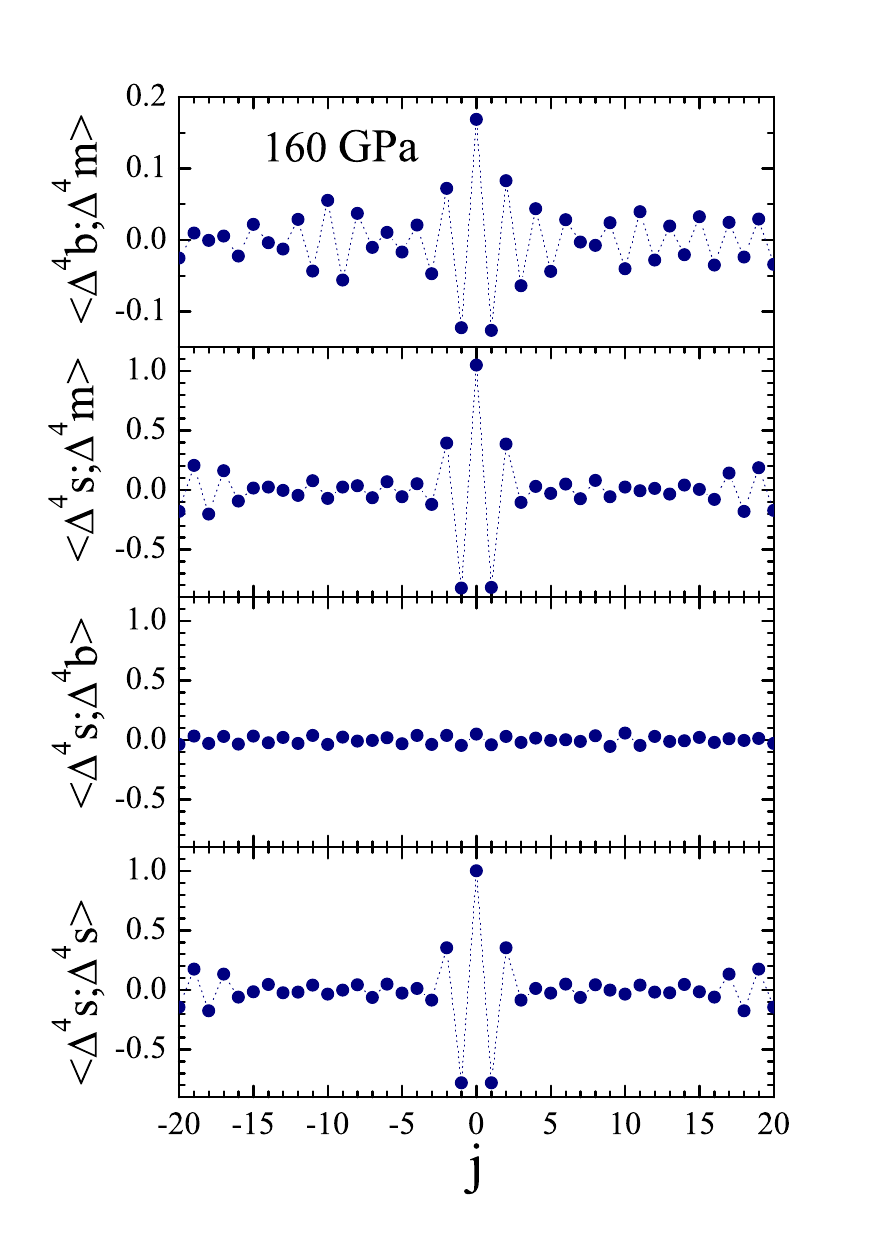}
\caption{Correlation functions $g_{\alpha;\beta}^{(n)}=<\Delta^n\chi^{\prime}_{\alpha};\Delta^n\chi^{\prime}_{\beta}>=<\Delta^n{\alpha};\Delta^n{\beta}>$ (where the latter is the shorthand notation along the ordinate) at 160~GPa. From left to right: order $n=0$,1,2,3 and 4. From top to bottom: $g_{bg;mv}^{(n)}(j)$, $g_{sc;mv}^{(n)}(j)$,  $g_{sc;bg}^{(n)}(j)$, $g_{sc;sc}^{(n)}(j)$. All values have been normalized by $g_{sc;sc}^{(n)}(0)$. 
The computer codes for obtaining $g_{\alpha;\beta}^{(n)}(j)$ can be downloaded from Ref.~\onlinecite{opendata}.
}
\label{figure:160corr_all_orders}
\end{figure*} 
\begin{figure*}[]
\centering
\includegraphics[width=0.4\columnwidth]{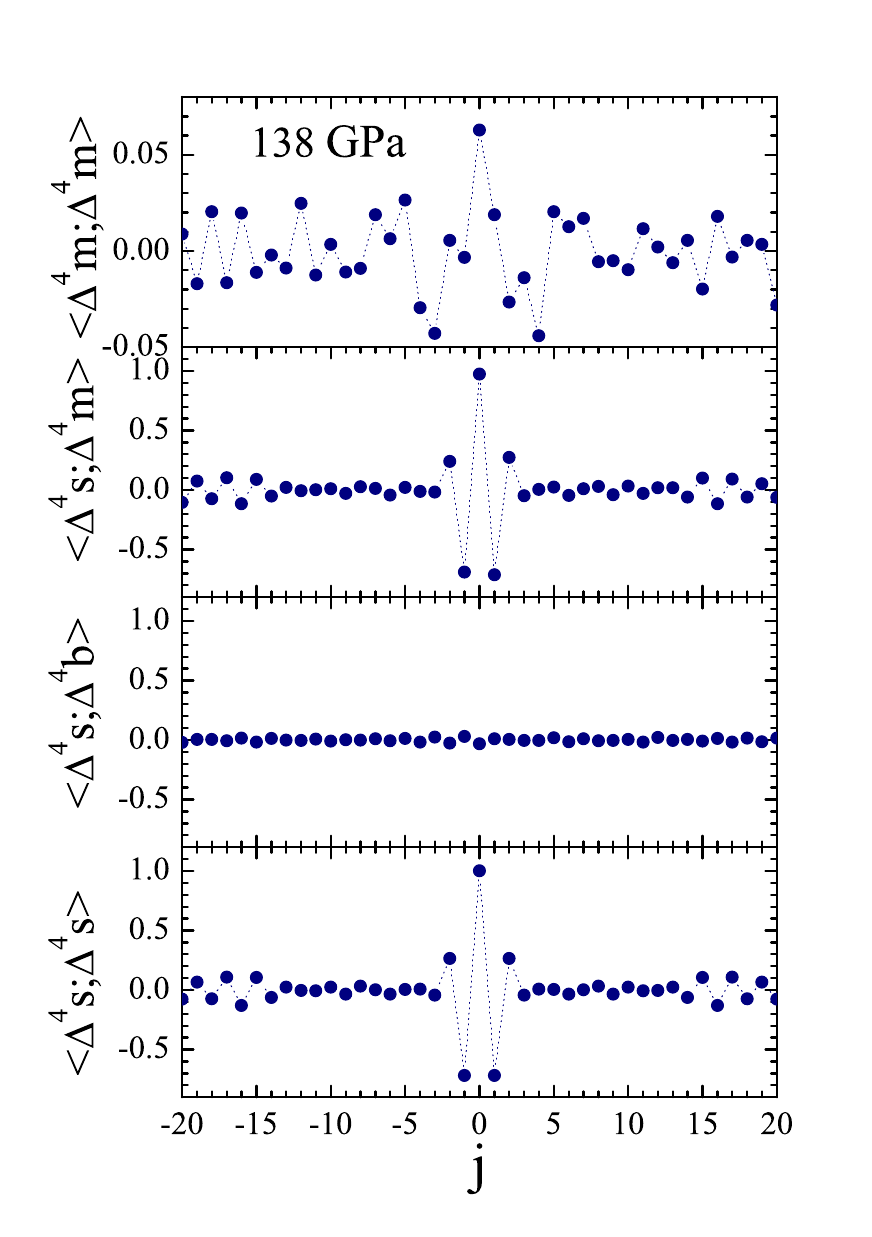}
\includegraphics[width=0.4\columnwidth]{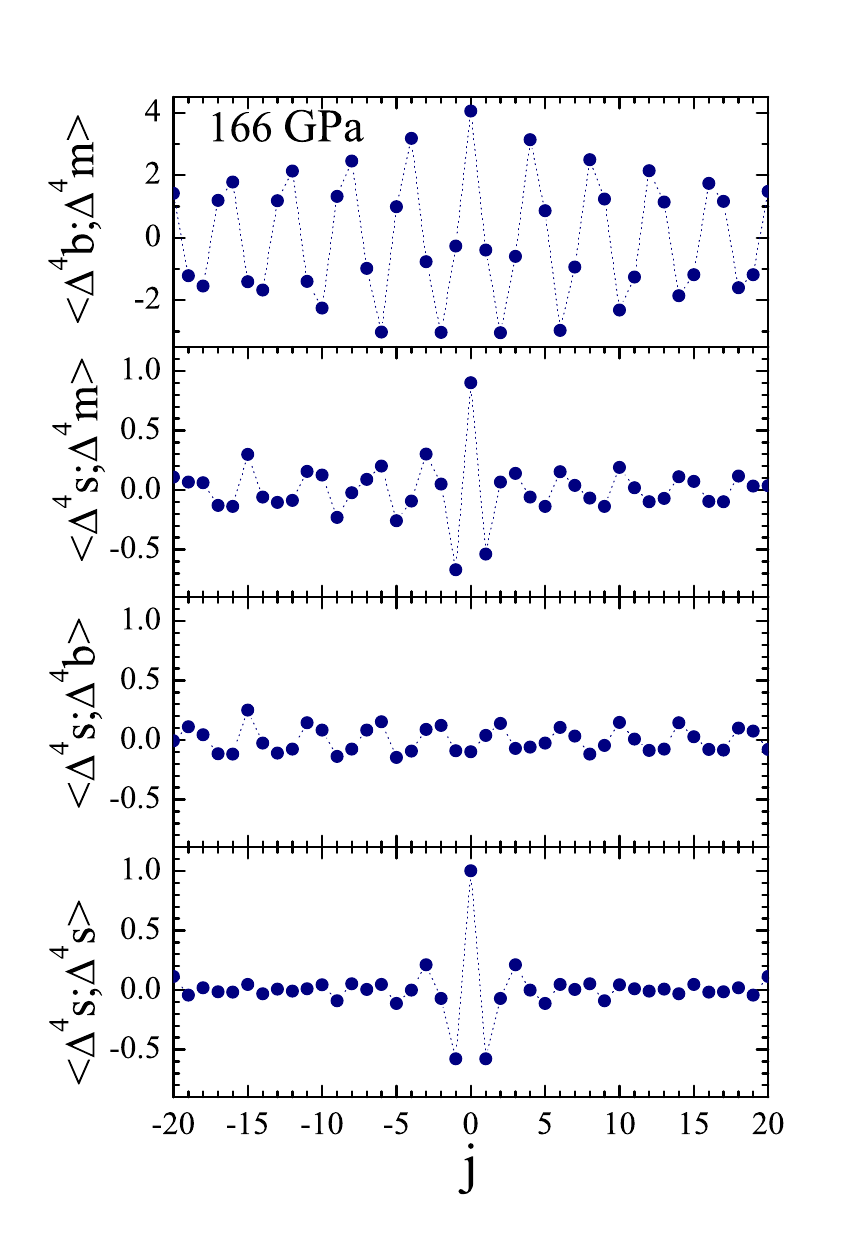}
\includegraphics[width=0.4\columnwidth]{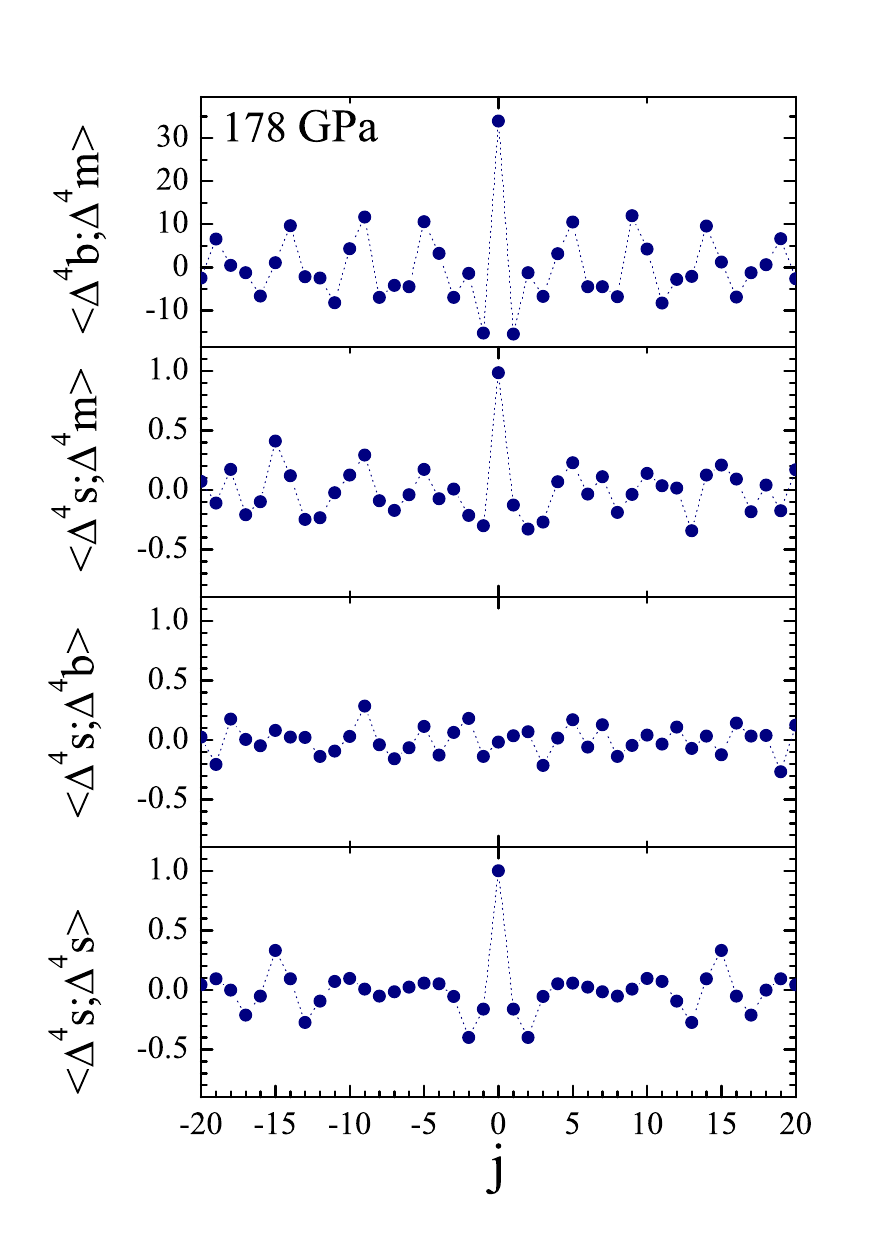}
\includegraphics[width=0.4\columnwidth]{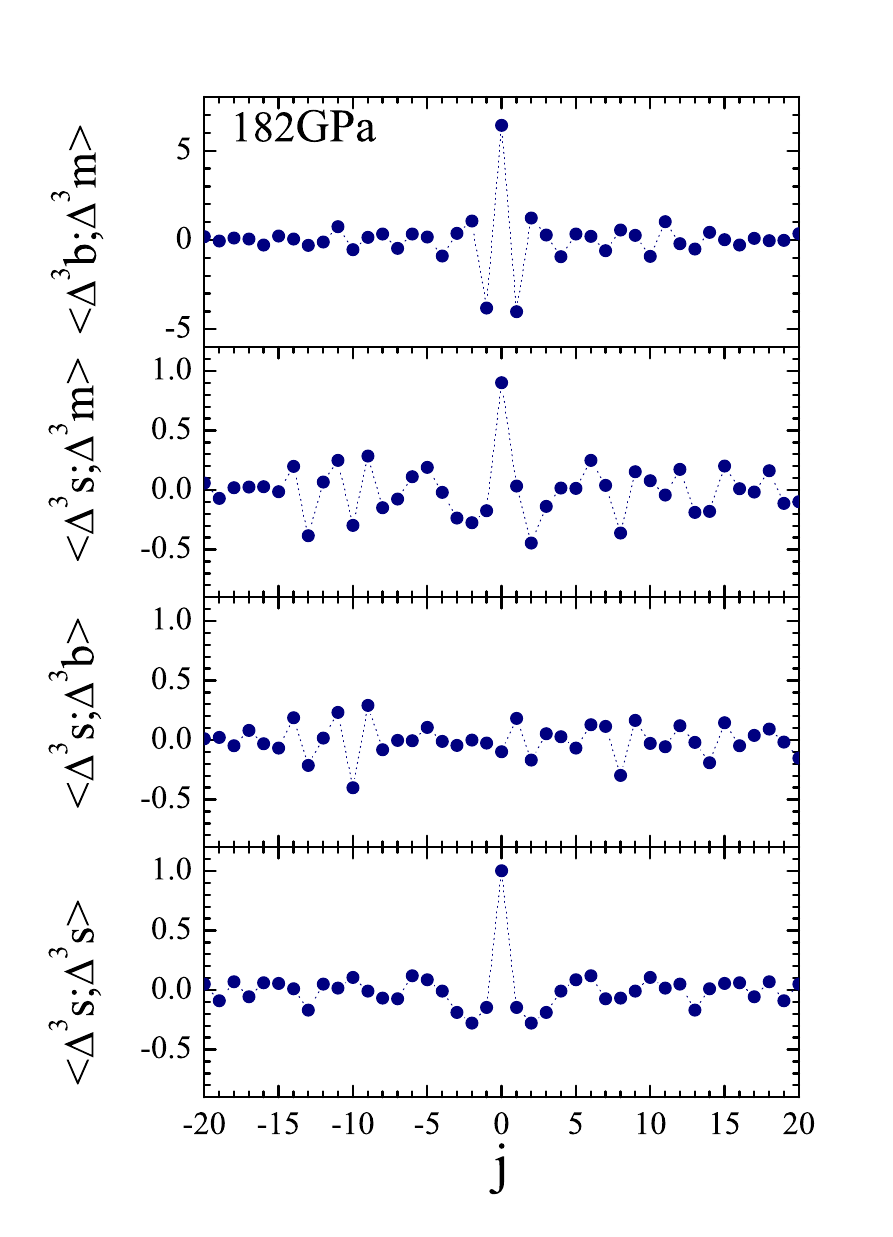}
\includegraphics[width=0.4\columnwidth]{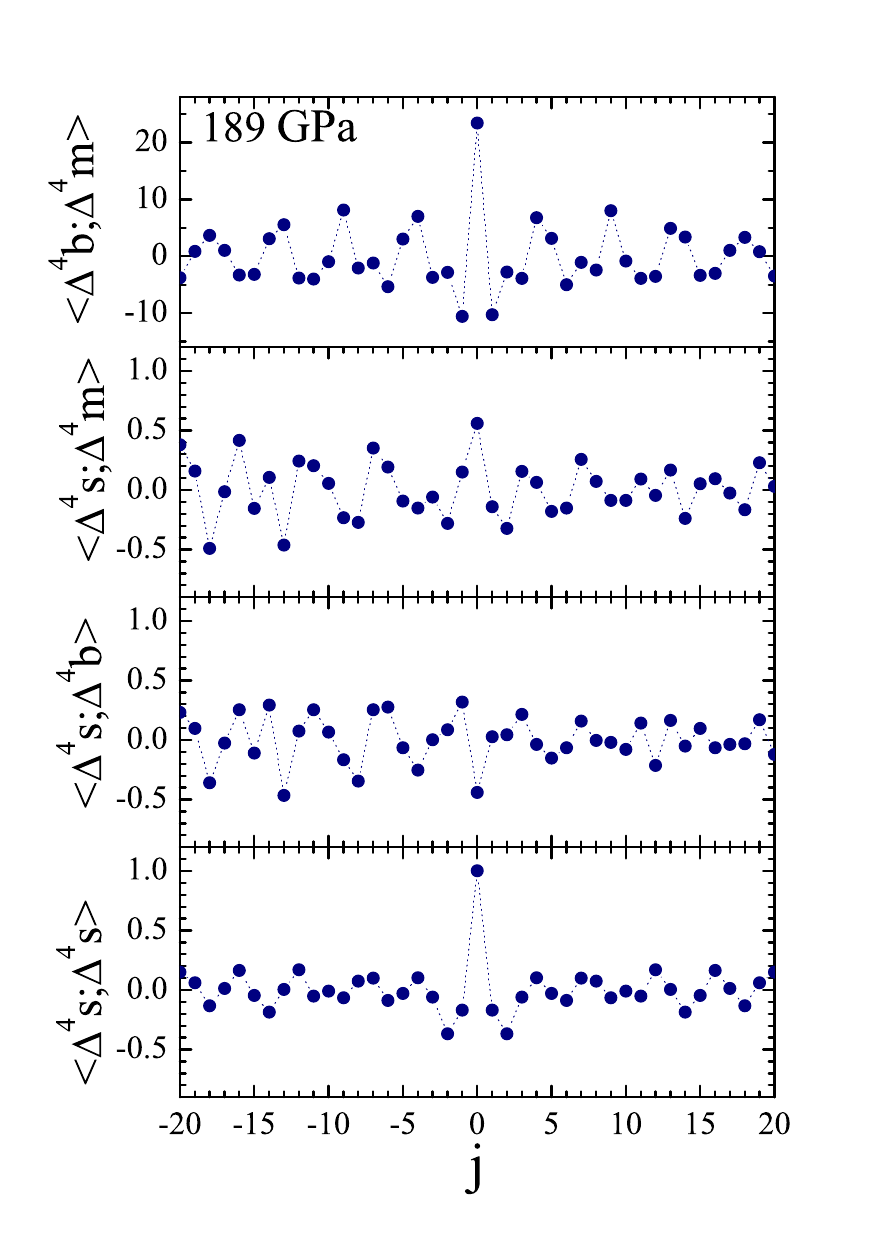}
\caption{From left to right: correlation functions $g_{\alpha;\beta}^{(4)}$ at 138, 166, 178, 182 and 189~GPa. From top to bottom: $g_{bg;mv}^{(4)}(j)$, $g_{sc;mv}^{(4)}(j)$,  $g_{sc;bg}^{(4)}(j)$, $g_{sc;sc}^{(4)}(j)$. 
}
\label{figure:6pressures_order4}
\end{figure*} 
We now turn to an analysis of the degree of correlation between $\chi^{\prime}_{mv}$, $\chi^{\prime}_{bg}$ and $\chi^{\prime}_{sc}$. To that end we consider the n$^{th}$ discrete differential
\begin{equation}
\Delta^n\chi^{\prime}_{\alpha}(j)= \Delta^{n-1}\chi^{\prime}_{\alpha}(j) - \Delta^{n-1}\chi^{\prime}_{\alpha}(j-1) 
\end{equation}
and we define the correlation functions
\begin{equation}
g_{\alpha;\beta}^{(n)}(j)= \sum_{k}^{N-1} \Delta^n\chi^{\prime}_{\alpha}(k)\Delta^n\chi^{\prime}_{\beta}(j+k)
\end{equation}
where $j$, $k$ and $j+k$ are taken modulo $N$.
Before we decide which order $n$ we will use for the correlation functions, we take a look at $\Delta^n\chi^{\prime}_{sc}(j)$ for $n=0,1,2,3,4$ for the case of 160~GPa, shown in Fig.~\ref{figure:160all_D}. For $n\ge 1$ the values are normalized by $\Delta_q=0.16555$ nV. It is immediately clear that $n=0$ is not a good choice, since it results in a large vertical offset of the correlation functions. This is offset is absent for $n\ge 1$, but for $n=1$ there is considerable warping. The warping is smaller -but still finite- for  $n = 2$. For $n = 3$ there is no warping, but since $\chi^{\prime}_{sc}(T)=P(T_j)+q(T_j)$, and $P(T_j)$ is a third order spline, the function $\Delta^3P(T_j)$ is almost~\footnote{$\Delta^3P(T_j)$ is not exactly constant because (i) the discrete derivative is not an exact derivative,  (ii) there are slight variations of the temperature steps, and (iii) $P(T)$ is non-analytical at the nodes} a constant, but unequal to zero. The fourth derivative $\Delta^4\chi^{\prime}_{sc}(j)$ contains no contribution from $P(T_j)$ other than noise, and is therefore fully representative of the quantized component $q(T_j)$. 
In Fig.~\ref{figure:160corr_all_orders} the autocorrelation functions are displayed of the ``superconducting signal" $g_{sc;sc}^{(n)}(j)$ at 160 GPa for order $n=0,1,2,3,4$. From this example we see, that for all orders $n \ge 1$ the autocorrelation function $g_{sc;sc}^{(n)}(j)$ has a narrow peak at $j=0$. Going from lower to higher order, the number of satellites left and right of the central peak increases. This peak is exactly reproduced, with the same amplitude and same satellite structure, in $g_{sc;mv}^{(n)}(j)$, and it is completely absent in $g_{sc;bg}^{(n)}(j)$. We furthermore observe (i) a peak centered at $j=0$ in $g_{bg;mv}^{(n)}(j)$, which (ii) is significantly broader than the one of $g_{sc;sc}^{(n)}(j)$,. This indicates that (i) measured voltage and background signal are interdepend, and (ii) there is some correlation between adjacent data points. Point (i) contradicts the statement of Ref.~\onlinecite{snider2020} that the background signal was obtained in a separate experiment using a different pressure. Point (ii) is a natural consequence of 
smoothing, as is detailed in Appendix~\ref{appendix:smoothing}.

From here on we use $n=4$ and apply the same analysis to the other 5 pressures. The result is shown in Fig.~\ref{figure:6pressures_order4} for $g_{sc;sc}$, $g_{sc;bg}$ $g_{sc;mv}$, and $g_{bg;mv}$. In Appendix~\ref{appendix:ED} Figs.~\ref{figure:correlations138-160-166} and~\ref{figure:correlations178-182-189} these functions and also $g_{bg;bg}$ and $g_{mv;mv}$ are shown in a 
broader range.
\begin{enumerate}
\item
For all pressures the measured voltage strongly correlates with the background signal. 
\item
For none of the pressures the background signal correlates with the superconducting signal. 
\item
For 138, 160, 166, 178 and 182 GPa the measured voltage correlates with the superconducting signal {\it and at $j=0$ the MV-SC correlation has the same amplitude as the SC-SC autocorrelation function}. 
\item
For 189 GPa the measured voltage appears to have a reduced correlation with the superconducting signal {\it i.e.} the amplitude for $j=0$ is about half that of the autocorrelation function of the superconducting signal. Most likely this is the consequence of noise: The ratio $\delta \chi^{\prime}_{mv}/\Delta_q \sim 40$ for this pressure, where $\delta \chi^{\prime}_{mv}$ is the noise of $\Delta^2\chi^{\prime}_{mv}$ and $\Delta_q$ is the quantized step size. In comparison for 166, 178 and 182 GPa $\delta \chi^{\prime}_{mv}/\Delta_q \sim 13 $, and $\delta \chi^{\prime}_{mv}/\Delta_q \sim 1$ for 138 and 160 GPa. 189 GPa being the only pressure out six where the correlation with the superconducting signal is less clear,  we will assume from here on that the weaker correlation in this case is due to noise. 
\end{enumerate}
To arrive at a crisp interpretation, we will now do two simulations calculated according to protocol 1 and 3. No simulation is attempted for protocol 2, because we do not know how to obtain the ``user defined background" (UDB-1) of Ref.~\onlinecite{dias2022}.

\vspace{0.5\baselineskip}
\noindent {\bf Simulation using Protocol 1}
\begin{itemize}
\item
Determine ``measured voltage" including random noise~\cite{lock-in}
\item
Determine ``background signal" including random noise~\cite{lock-in}
\item
Calculate ``superconducting signal" = ``measured voltage"  - ``background signal"
\end{itemize}

\begin{figure*}[!!th!!]
\centering
\begin{minipage}{2\columnwidth}
\includegraphics[width=0.4\columnwidth]{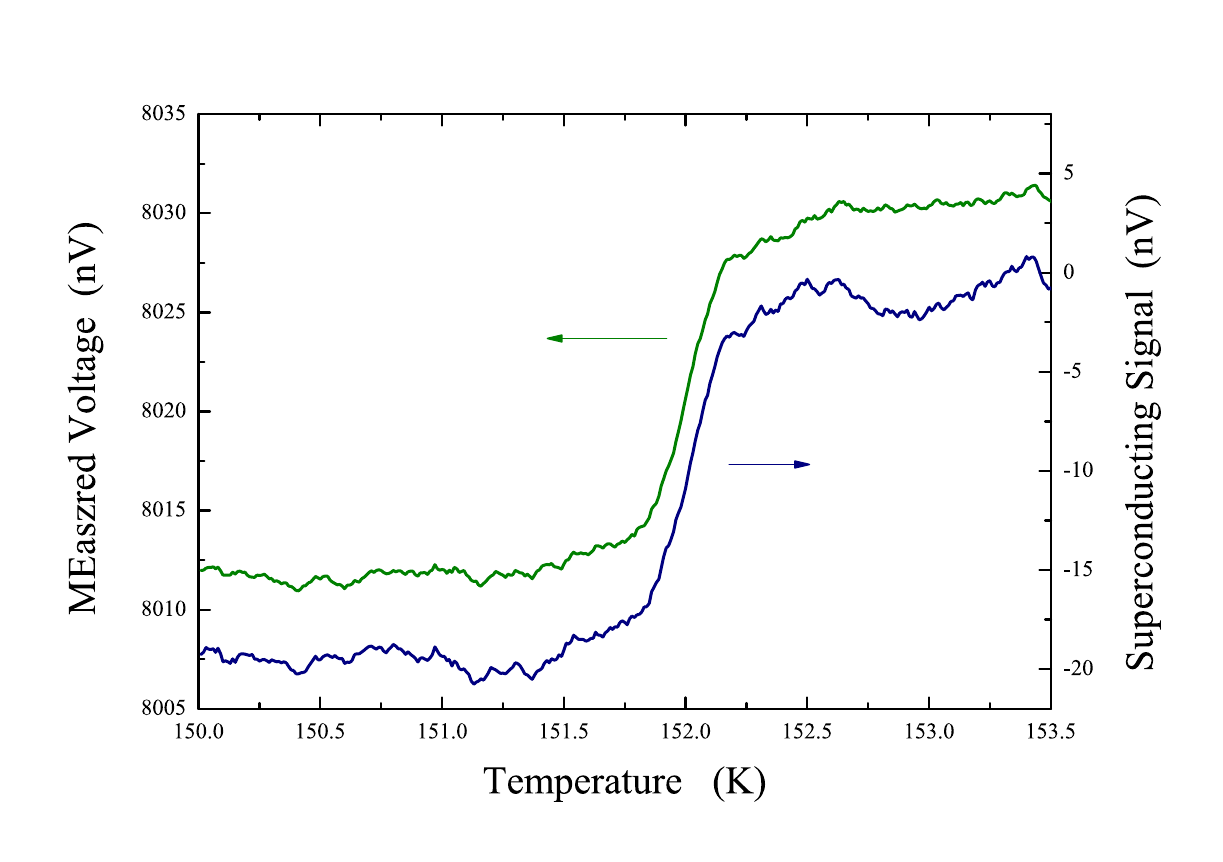}
\includegraphics[width=0.4\columnwidth]{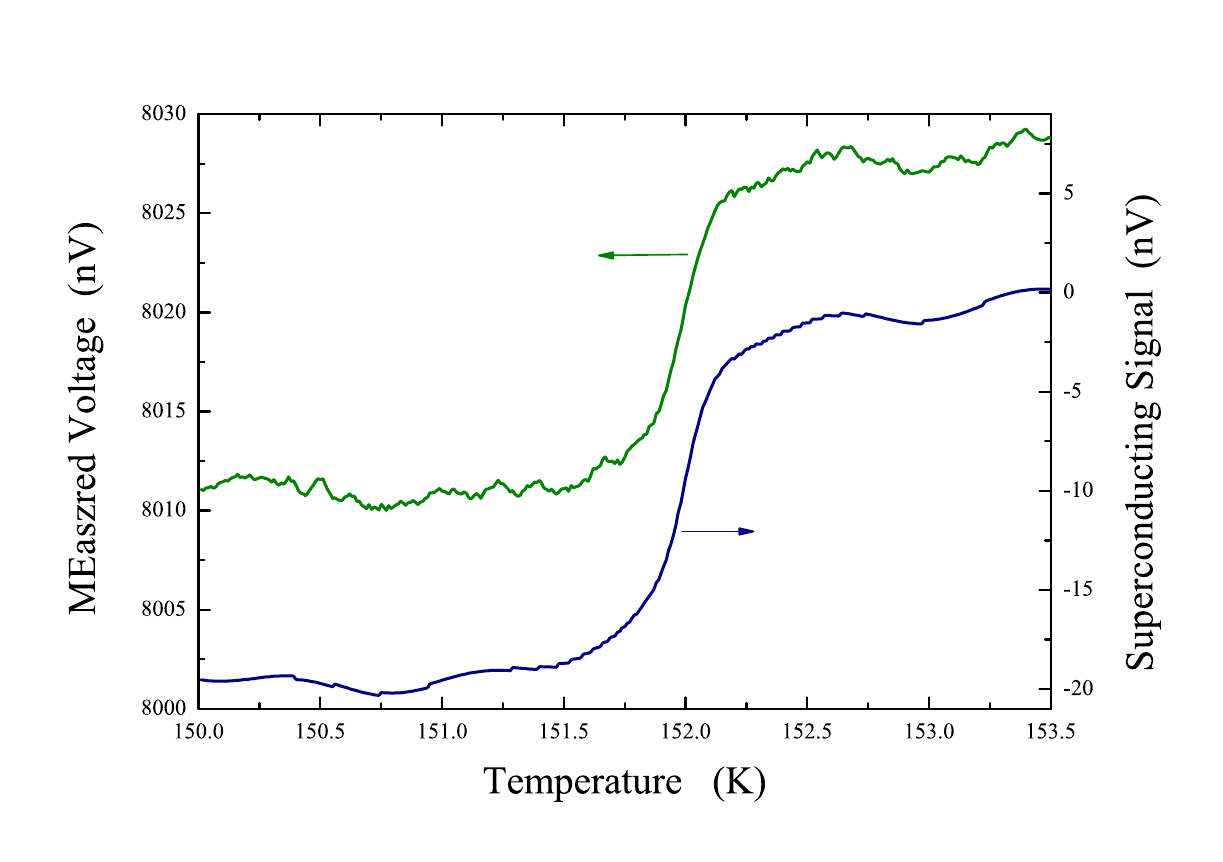}
\end{minipage}
\caption{Simulated ``measured voltage" and ``superconducting" using protocol 1(left panel) and 3 (right panel). 
An excel file with the calculation of the simulation shown in this figure can be downloaded from Ref.~\onlinecite{opendata}}
\label{figure:sim1&2}
\begin{minipage}{2\columnwidth}
\includegraphics[width=0.35\columnwidth]{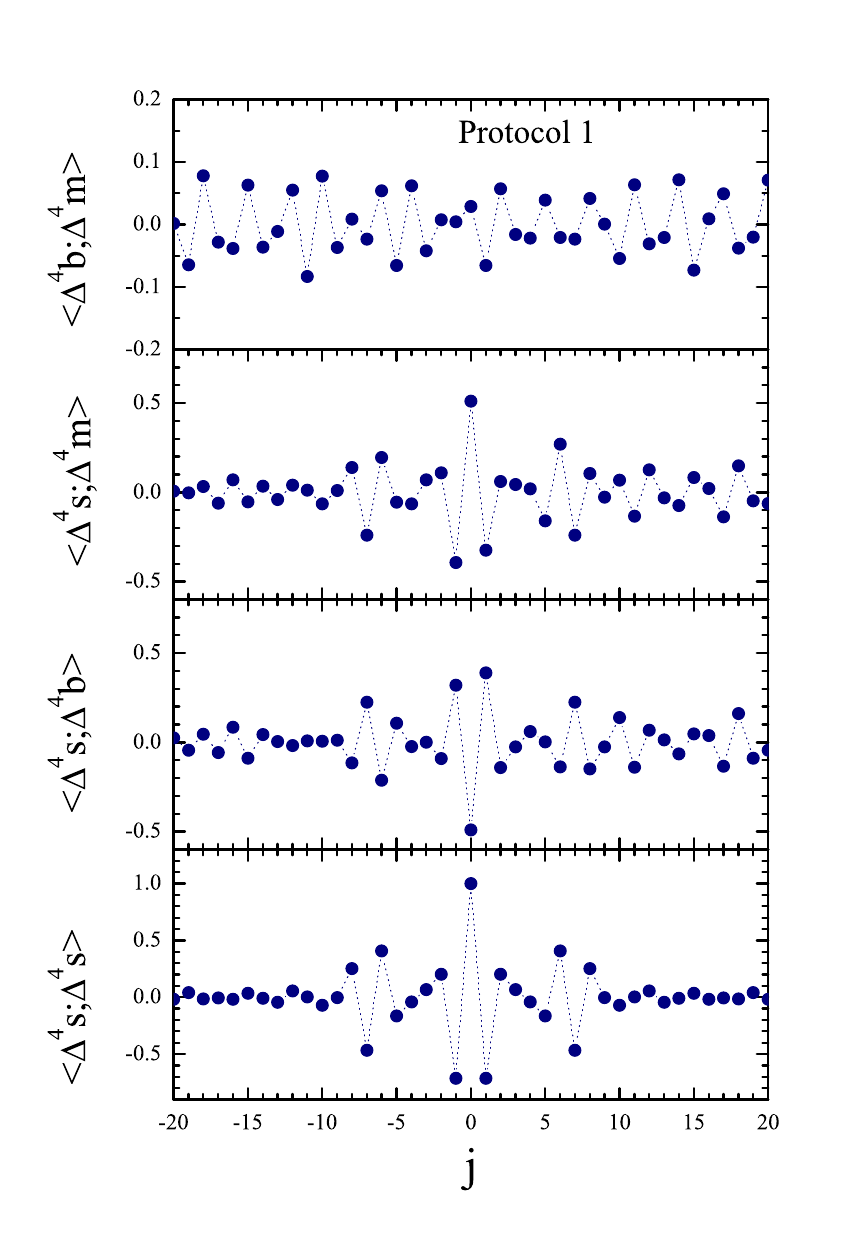}
\includegraphics[width=0.35\columnwidth]{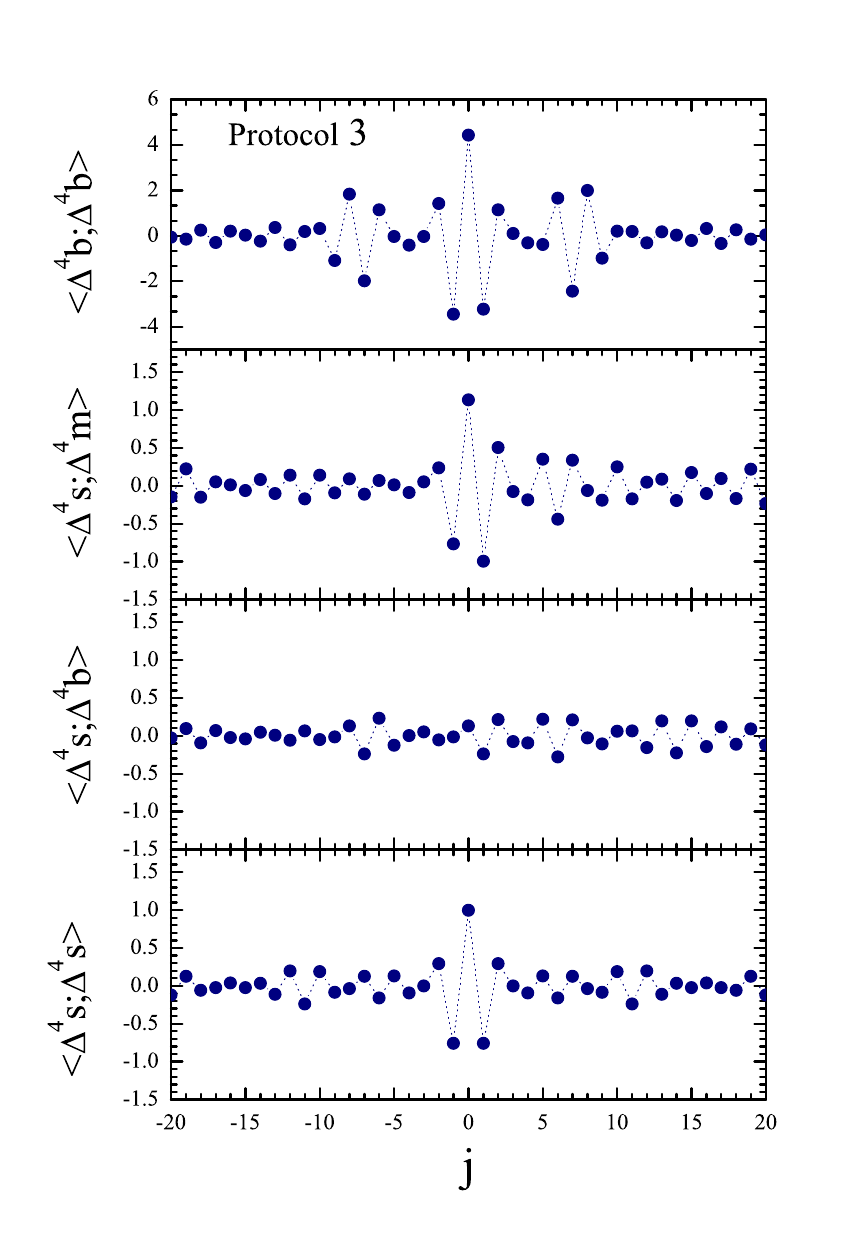}
\end{minipage}
\caption{Correlation functions for $n=4$ of simulated data. Left: see left panel of Fig.~\ref{figure:sim1&2}, Right: see right panel of Fig.~\ref{figure:sim1&2}. From top to bottom : ``Measured voltage" with ``background".  ``Measured voltage" with ``superconducting signal". ``Background" with ``superconducting signal". Autocorrelation of ``superconducting signal".   
All values have been normalized by $g_{sc;sc}^{(4)}(0)$. 
To facilitate comparison of the lower three panels the same vertical range is used.}
\label{figure:sim1&2_correlation_zoomD4}
\end{figure*}

\noindent
In left panel of Fig.~\ref{figure:sim1&2} we show the measured voltage and superconducting signal according to this protocol. The corresponding correlation functions, shown in the left panel of Fig.~\ref{figure:sim1&2_correlation_zoomD4}, have the following features:
\begin{enumerate}
\item
The measured voltage has zero correlation with the background signal. 
\item
The superconducting signal correlates with the background signal. 
\item
The superconducting signal correlates with the measured voltage. 
\end{enumerate}
Comparing this to the above-mentioned data of Ref.~\onlinecite{snider2020} we see that points 1 and 2 are in disagreement. Consequently the data of Ref.~\onlinecite{snider2020} are not obtained with Protocol 1, in contrast to what was stated in this paper.

\vspace{5\baselineskip}
\noindent {\bf Protocol 2, no simulation possible}
\begin{itemize}
\item In Ref.~\onlinecite{dias2022} the authors negate Protocol 1 and say instead that the background signal was not measured but somehow extracted from the measured raw data through a UDB-1 procedure that is left undefined. 
\item Then, the ``superconducting signal" is obtained from 

``superconducting signal" =  ``measured voltage"  - ``background signal"
\end{itemize}
Protocol 2 is invalidated by the fact that there is no correlation between ``superconducting signal" and ``background signal".

\vspace{0.5\baselineskip}
\noindent {\bf Simulation using Protocol 3}
\begin{itemize}
\item
Determine ``superconducting signal" = $P(T)+q(T)$
\item
Determine ``background signal" including random noise~\cite{lock-in},
\item
Calculate ``measured voltage" = ``superconducting signal" + ``background signal"
\end{itemize}
In Fig.~\ref{figure:sim1&2} we show the measured voltage and superconducting signal according to this protocol. The corresponding correlation functions, shown in the right panel of Fig~\ref{figure:sim1&2_correlation_zoomD4}, have the following features:
\begin{enumerate}
\item
The measured voltage has strong correlation with the background signal. The central peak of $g_{bg;mv}(i)$ has satellites in for $i=\pm 7$, which are a consequence of the smoothing of the background data, for which we used 7-point adjacent averaging~\cite{lock-in}.
\item
The superconducting signal has zero correlation with the background signal. 
\item
The superconducting signal correlates with the measured voltage {\it and at $j=0$  the SC-MV correlation has the same amplitude as the SC-SC autocorrelation function}. 
\end{enumerate}
Comparing this to the above-mentioned data of Ref.~\onlinecite{snider2020} we see that there is a match for all three points. Consequently the data of Ref.~\onlinecite{snider2020,dias2021} have been obtained with Protocol 3, in contrast to what was stated in this paper. An immediate consequence is, that for none of the reported pressures the `measured voltage" data provided in Ref.~\onlinecite{dias2021} are raw data.
For the case of 178~GPa, the $g_{sc;sc}^{(4)}(i)$ shows satellites of the central peak for $i=\pm 13$. As we can see from simulation shown in Appendix~\ref{appendix:ED} Fig.~\ref{figure:simulation-BGsmoothing_AA13ScSig} a possible explanation is, that 13-point adjacent average smoothing has been applied to $g_{sc;sc}^{(4)}(i)$ at 178~GPa. 

\section{Summary}\label{section:Summary}
We have analyzed in detail the published ``superconducting signal" and ``measured voltage"  for all pressures reported in Refs.~\cite{snider2020,dias2021}. 
The following {\bf Protocol 3} is to the best of our understanding   the only possible scenario compatible with all reported properties of the data.

\begin{enumerate}
\item The ``superconducting signal" $\chi^{\prime}_{sc}(T)$ is generated as the sum of a quantized component $q(T)$ and a smooth function $P(T)$:  $\chi^{\prime}_{sc}(T)=q(T)+P(T)$.
\item A ``background signal" $\chi^{\prime}_{bg}(T)$ containing noise is determined. 
\item The ``measured  voltage" $\chi^{\prime}_{mv}(T)$ (alternatively labeled ``raw data") is obtained by adding the ``background signal" to the ``superconducting signal": $\chi^{\prime}_{mv}(T)= \chi^{\prime}_{sc}(T)+\chi^{\prime}_{bg}(T)$. 
\end{enumerate}

We have demonstrated  that 
\begin{itemize}
\item For 138, 160, 166, 178, 182 and 189~GPa the ``superconducting signal" exhibits the features expected for point 1 of  the aforementioned protocol. 
\item For 160~GPa the smooth function $P(T)$ is a cubic spline with 15 nodes having ``natural" boundary conditions at the two extremal nodes, and (ii) $q(T) = \Delta_q n(T)$ where $\Delta_q=0.16555$~nV and $n(T)$ is an integer. This component cannot be identified with the background signal, since it departs strongly from the difference ``measured voltage"-``superconducting signal" according to the data reported in Ref.~\onlinecite{dias2021}.
\item The quantized component of the 160~GPa data cannot be identified with the raw data since, other than containing the steep rise at 170~K, it departs strongly from the raw data (``measured voltage") reported in Ref.~\onlinecite{dias2021}.
\item For 138 and 160, 166, 178, 182 and possibly also 189~GPa the ``measured voltage" has been obtained according to point 3 of the protocol. 
\end{itemize}
%
\noindent Recap:

\noindent
{\bf Protocol 1}: In Ref.~\onlinecite{snider2020} the authors say the ``measured voltage" and ``background signal" were independently measured. Then, data is obtained from
``superconducting signal"=``measured voltage" - ``background signal"
\\
This is invalidated by the facts that:
\\
(i)  there is no correlation between ``superconducting signal" and ``background signal"
\\
(ii) there is correlation between ``measured voltage" and ``background signal"
\\
(iii) ``superconducting signal" has much less noise than ``background signal" and ``measured voltage"

\noindent
{\bf Protocol 2}: In Ref.~\onlinecite{dias2022}, the authors negate protocol 1 and say instead that the background signal was not measured but somehow extracted from the measured raw data (``measured voltage") through a ``User Defined Background" (UDB) procedure. Three UDB variants were given in Ref.~\onlinecite{dias2022}.
\\
Then, data is obtained from
``superconducting signal"=``measured voltage" - ``background signal"
\\
The variant UDB-1 is invalidated on the same grounds that there is no correlation between ``superconducting signal" and ``background signal". For UDB-2 and UDB-3 the absence of correlation is trivial since for these variants the background is a straight line.   
These variants are invalidated by the fact that the ``data" have much less noise than the ``raw data": they would have the same noise with these variants.

\noindent
{\bf Protocol 3}: a protocol that is consistent with ``measured voltage" and ``superconducting signal" posted in Ref.~\onlinecite{dias2021} is:
\\
(a) ``superconducting signal" is generated by adding a quantized component and a smooth curve.
\\
(b) ``background signal" containing noise is generated somehow.
\\
(c) ``measured voltage"  is obtained from:
``measured voltage"  = ``superconducting signal" + ``background signal".
 \\ 
This is consistent with all the facts, in particular that:
\\
(i) there is no correlation between ``superconducting signal" and ``background signal".
\\
(ii) there is correlation between ``measured voltage" and ``background signal".
\\
(iii) ``measured voltage" contains the same steps as the ones of ``superconducting signal".

\noindent
Our analysis leads to the conclusion that the only protocol consistent with the facts is protocol 3. In this protocol the ``measured voltage" are not actually measured raw data that reflect the properties of a physical system, contrary to what Refs.~\cite{snider2020},~\cite{dias2021} and~\cite{dias2022} claim.

\vspace{1\baselineskip}
\noindent{\bf Data availability}. The datasets generated and analyzed during the current study are available in Ref.~\onlinecite{opendata}. 

\vspace{0.1\baselineskip}
\noindent{\bf Acknowledgements}. We thank P. Armitage, B. Ramshaw, J. Zaanen and G. Grisonnanche for stimulating discussions and Dukwon for a useful suggestion.

%
\appendix
\section{Effect of smoothing}\label{appendix:smoothing}
If  $\chi^{\prime}_{raw}(T)$ ($\chi^{\prime}_{sm}(T)$) are the susceptibility without (with) smoothing, and the smoothing profile is $f(T)$, the relation between unsmoothed and smoothed function is
\begin{eqnarray}
\chi^{\prime}_{sm}(T_i) = \sum_{-\infty}^{\infty} \chi^{\prime}_{raw}(T_j) f_{i,j}
\end{eqnarray}
The first possibility that comes to mind is the smoothing that results from having the integration time of the lock-in amplifier longer than the sampling time per temperature during the temperature scan. Assuming that the data are measured during a cool-down ramp, the corresponding smoothing profile is
\begin{eqnarray}
f_{i,j} = C H(T_j-T_i) e^{(i-j)/n}
\end{eqnarray}
where $H(x)$ is the Heaviside step function and $C$ a normalization factor. However, from simulations (see Appendix~\ref{appendix:ED} Fig.~\ref{figure:simulation-BGsmoothing}) we can see that  $g_{bg;mv}^{(1)}(j)$ does not have the ``wavy" profile present in the 160 GPa data. If instead we assume gaussian profile
\begin{eqnarray}
f_{i,j} = C e^{-[(i-j)/n]^2}
\end{eqnarray}
the wavy profile is exactly reproduced (see Appendix~\ref{appendix:ED} Fig.~\ref{figure:simulation-BGsmoothing}).
\onecolumngrid
%
%
%
%

\begin{thebibliography}{15}
\bibitem [{loc()}]{lock-in}%
  \BibitemOpen
  \href@noop {} {\bibinfo  {journal} {In the simulations shown in
  Fig.~\ref{figure:sim1&2} the adjacent average over 7 neighboring temperature
  points was taken, as a crude approximation of the effect of the finite
  integration time of a lock-in amplifier}\ }\BibitemShut {NoStop}%
\end{thebibliography}
\clearpage
\section{Extended Data}\label{appendix:ED}
\begin{figure}[!!hb!!]
\centering
\includegraphics[width=\columnwidth]{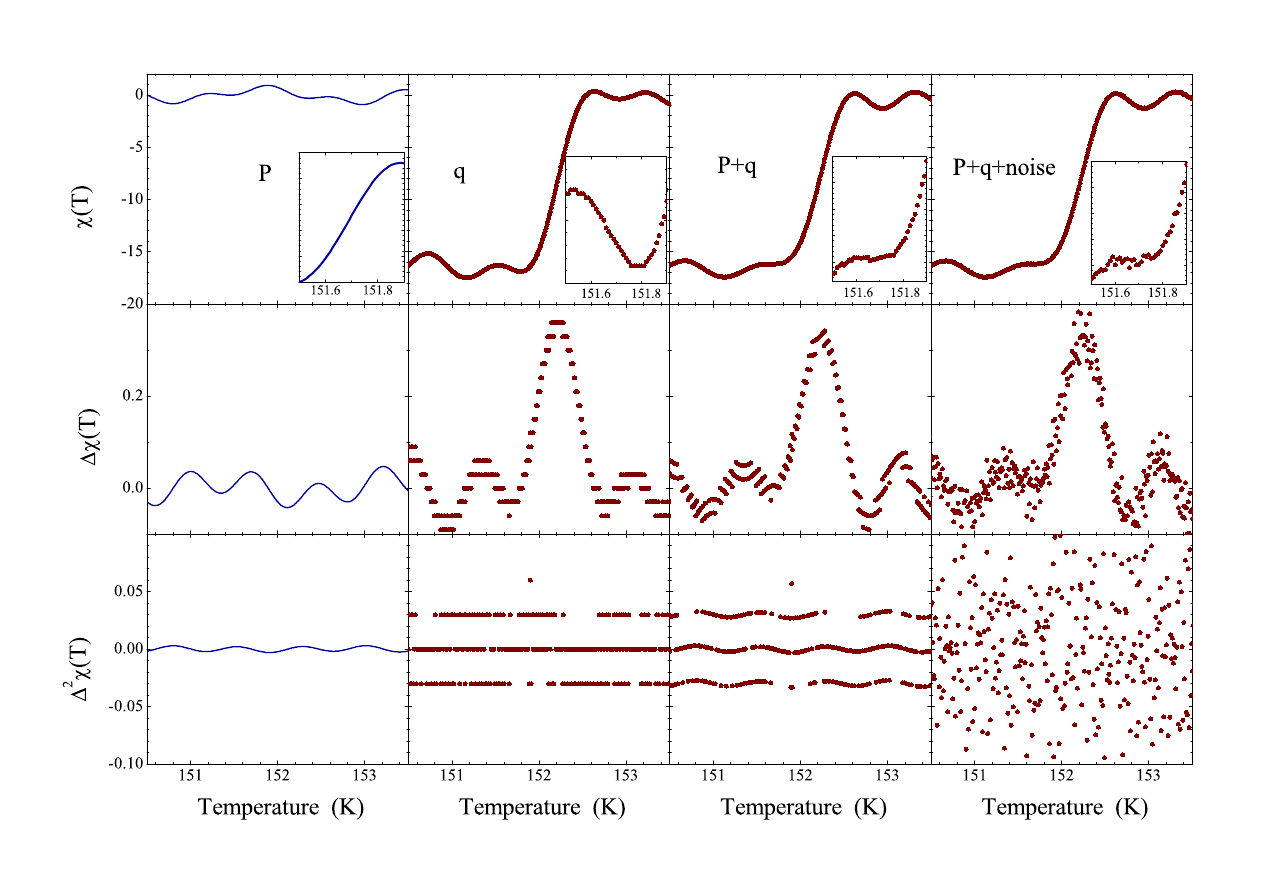}
\caption{Simulated data. 
Top line, from left to right: Smooth function $P(T_j)$, quantized component $q(T_j)=n_j\Delta_q$ with $\Delta_q = 0.03$ nV, 
superconducting signal $\chi^{\prime}_{sc}(j) = P(T_j) + q(T_j)$ and superconducting signal with random noise  $\chi^{\prime}_{sc}(j) + r(j)$ with $r(j) \in \{-0.03;0.03\}$ nV. Second and third line: first discrete differential and second discrete differential.}
\label{figure:simulation_Deltazp03}
\end{figure} 
%
%
\begin{figure}[]
\centering
\includegraphics[width=0.8\columnwidth]{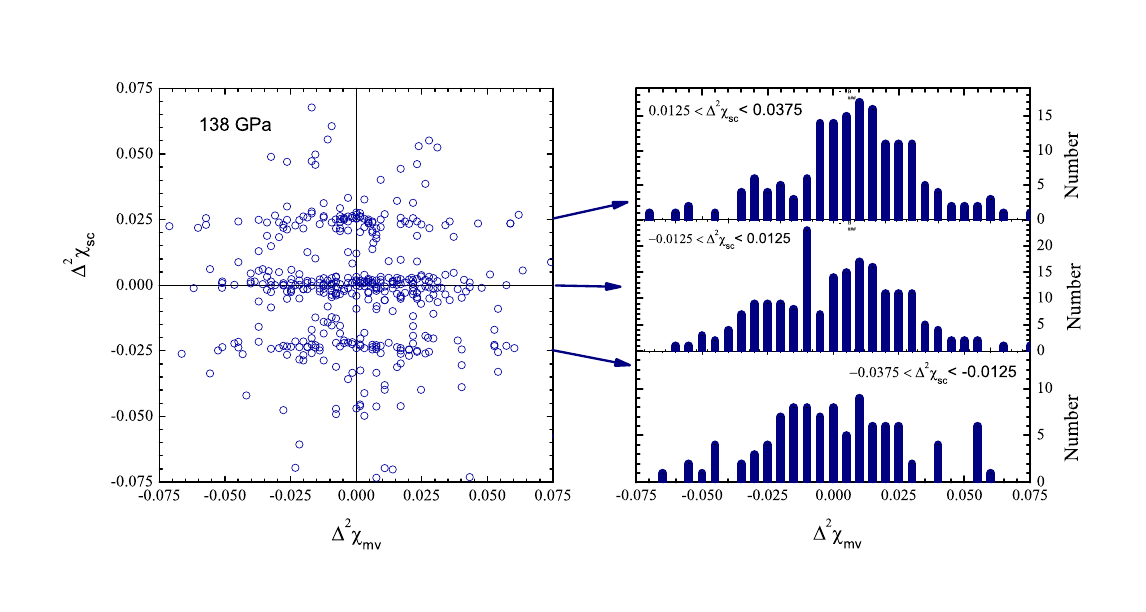}
\caption{Left: $\Delta^2\chi^{\prime}_{sc}$ versus $\Delta^2\chi^{\prime}_{bg}$ at 138~GPa.
Right: Histograms of $\Delta^2\chi^{\prime}_{bg}$ in the slots corresponding to the horizontal stripes.}
\label{figure:correlation_BG_SC_138}
\end{figure} 
\begin{figure}[]
\centering
\includegraphics[width=0.8\columnwidth]{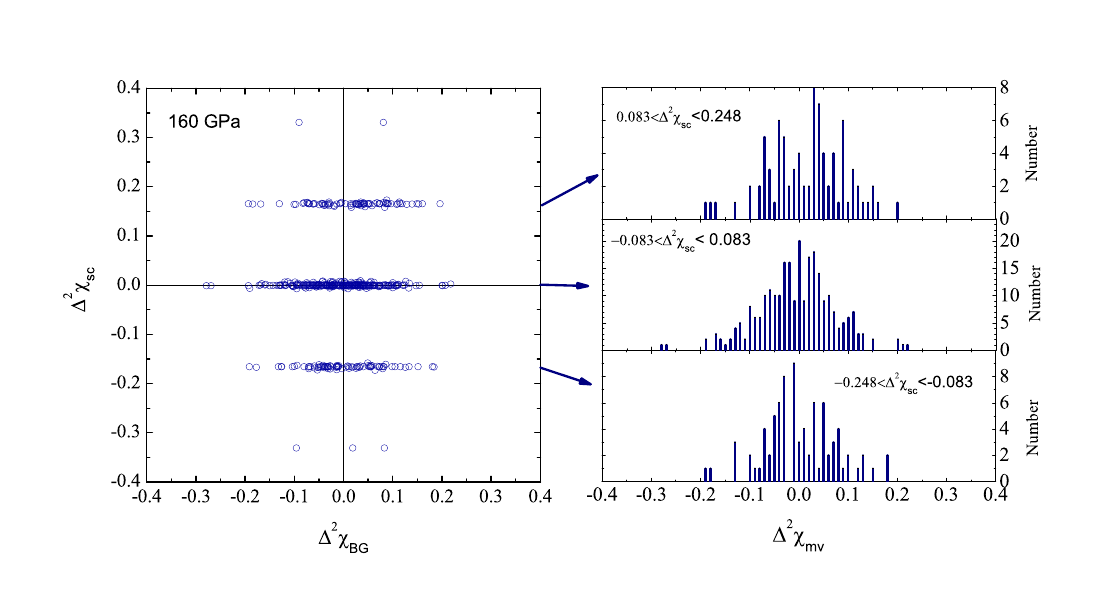}
\caption{Left: $\Delta^2\chi^{\prime}_{sc}$ versus $\Delta^2\chi^{\prime}_{bg}$ at 160~GPa.
Right: Histograms of $\Delta^2\chi^{\prime}_{bg}$ in the slots corresponding to the horizontal stripes.}
\label{figure:correlation_BG_SC_160}
\end{figure} 
\begin{figure}[]
\centering
\includegraphics[width=0.4\columnwidth]{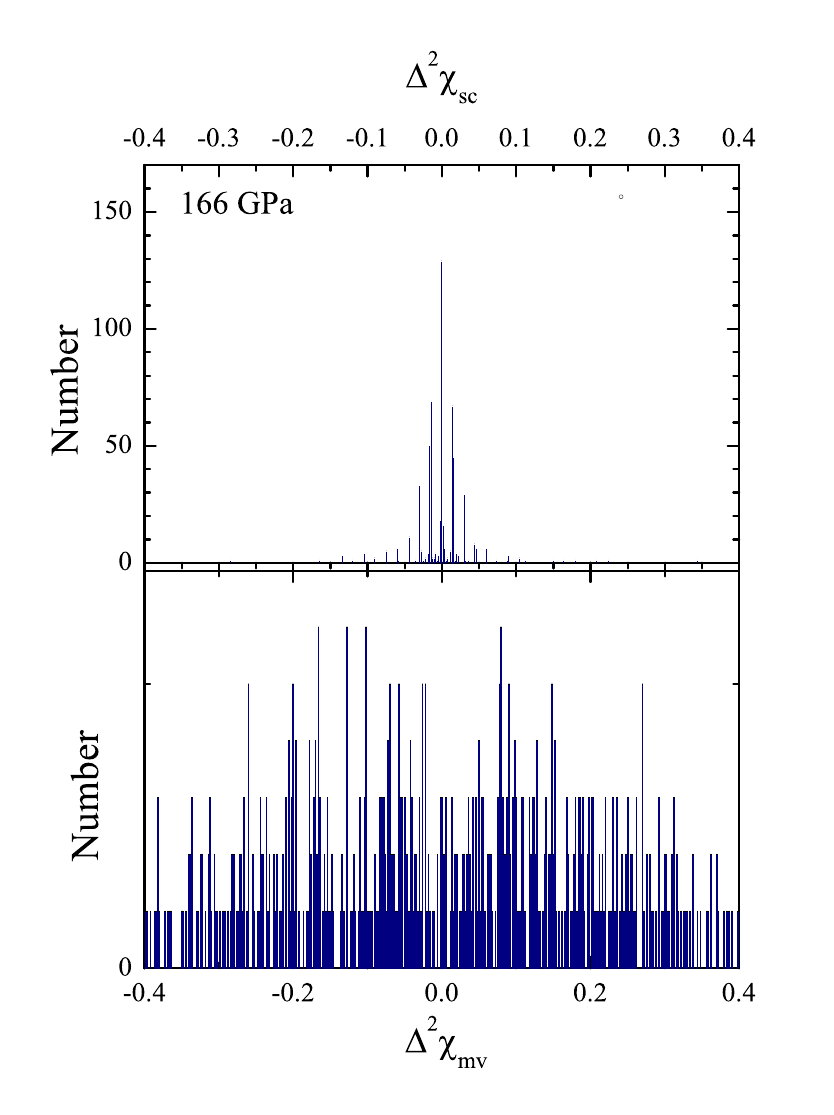}
\includegraphics[width=1\columnwidth]{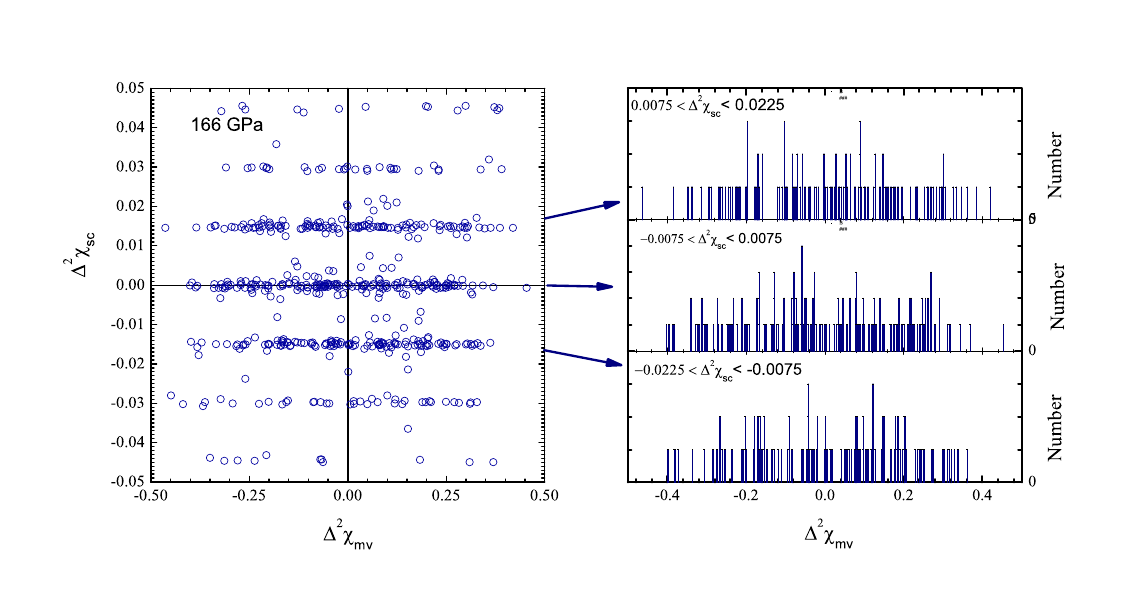}
\caption{Top:
Histograms of $\Delta^2\chi^{\prime}_{sc}$ and $\Delta^2\chi^{\prime}_{mv}$ of the ``measured voltage" at 166~GPa. 
Bottom:
Left: $\Delta^2\chi^{\prime}_{sc}$ versus $\Delta^2\chi^{\prime}_{mv}$.
Right: Histograms of $\Delta^2\chi^{\prime}_{mv}$ in the slots corresponding to the horizontal stripes.}
\label{figure:correlation_MV_SC_166}
\end{figure} 
\begin{figure}[]
\centering
\caption*{\bf Simulation with exponential smoothing of BG: ac susceptibility and correlation functions}
\begin{minipage}{\columnwidth}
\includegraphics[width=0.32\columnwidth]{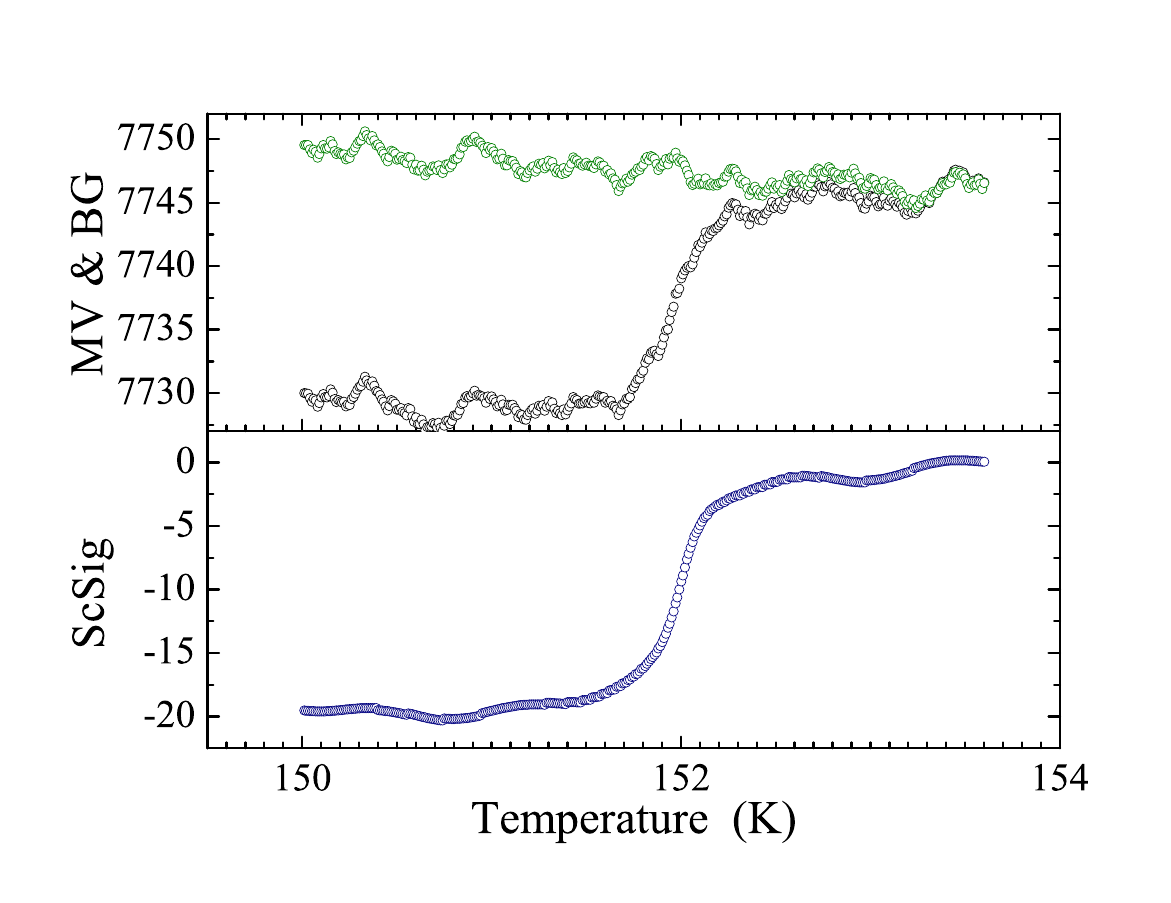}
\includegraphics[width=0.33\columnwidth]{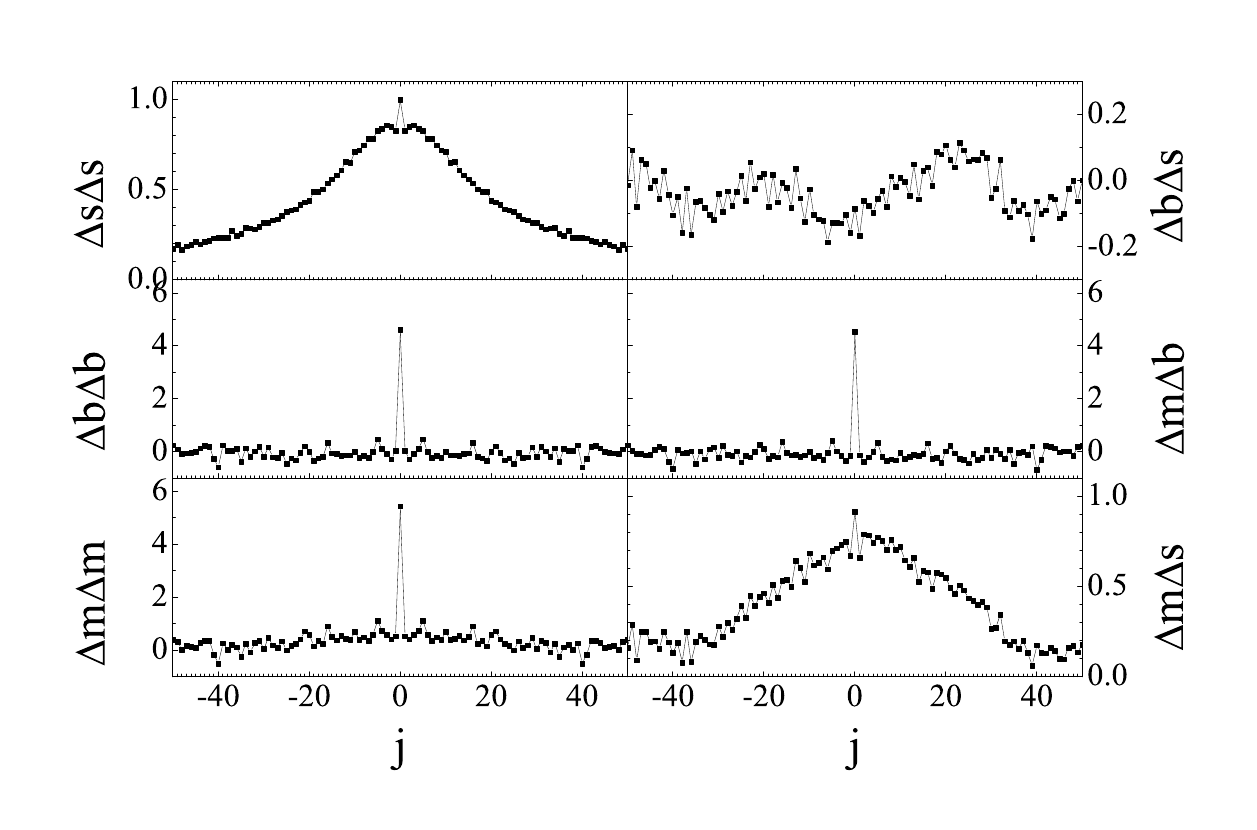}
\includegraphics[width=0.33\columnwidth]{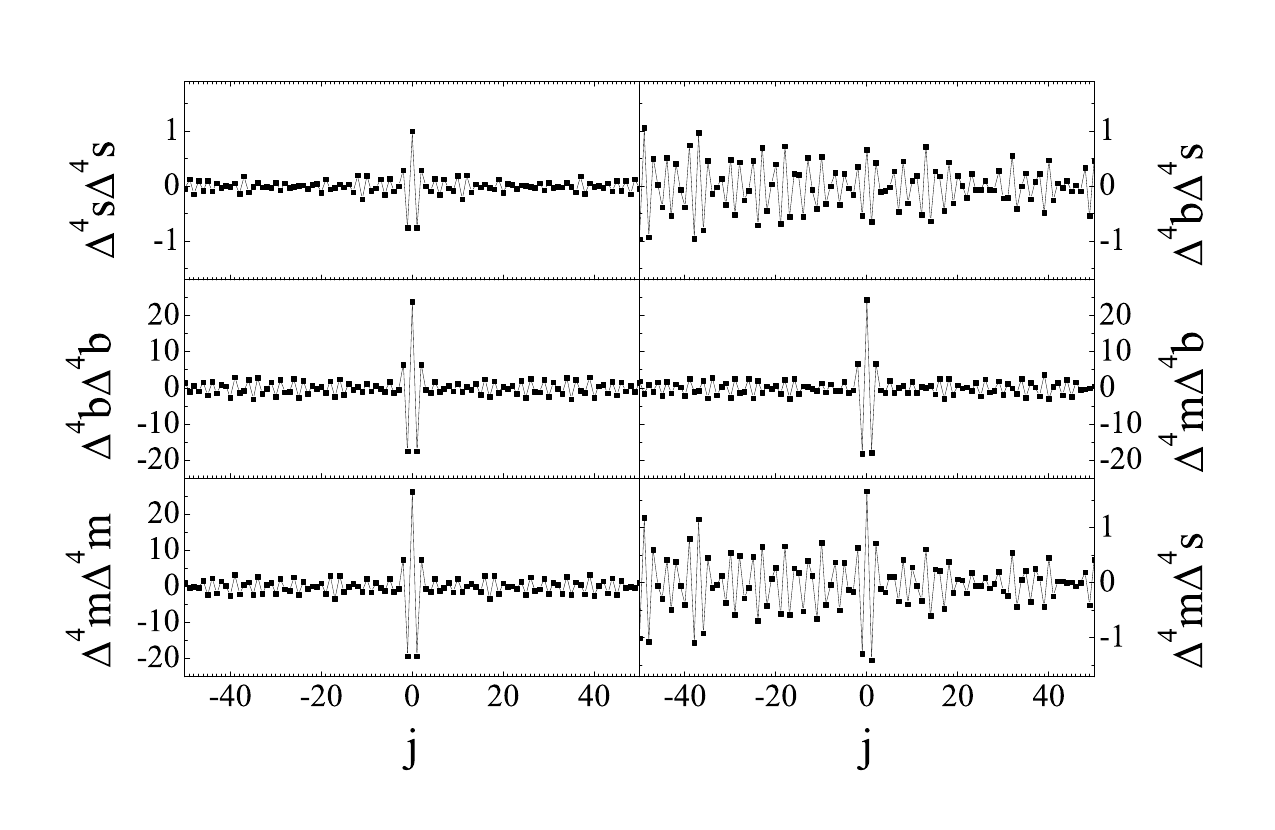}
\end{minipage}
\caption*{\bf 160~GPa: ac susceptibility and correlation functions}
\begin{minipage}{\columnwidth}
\includegraphics[width=0.32\columnwidth]{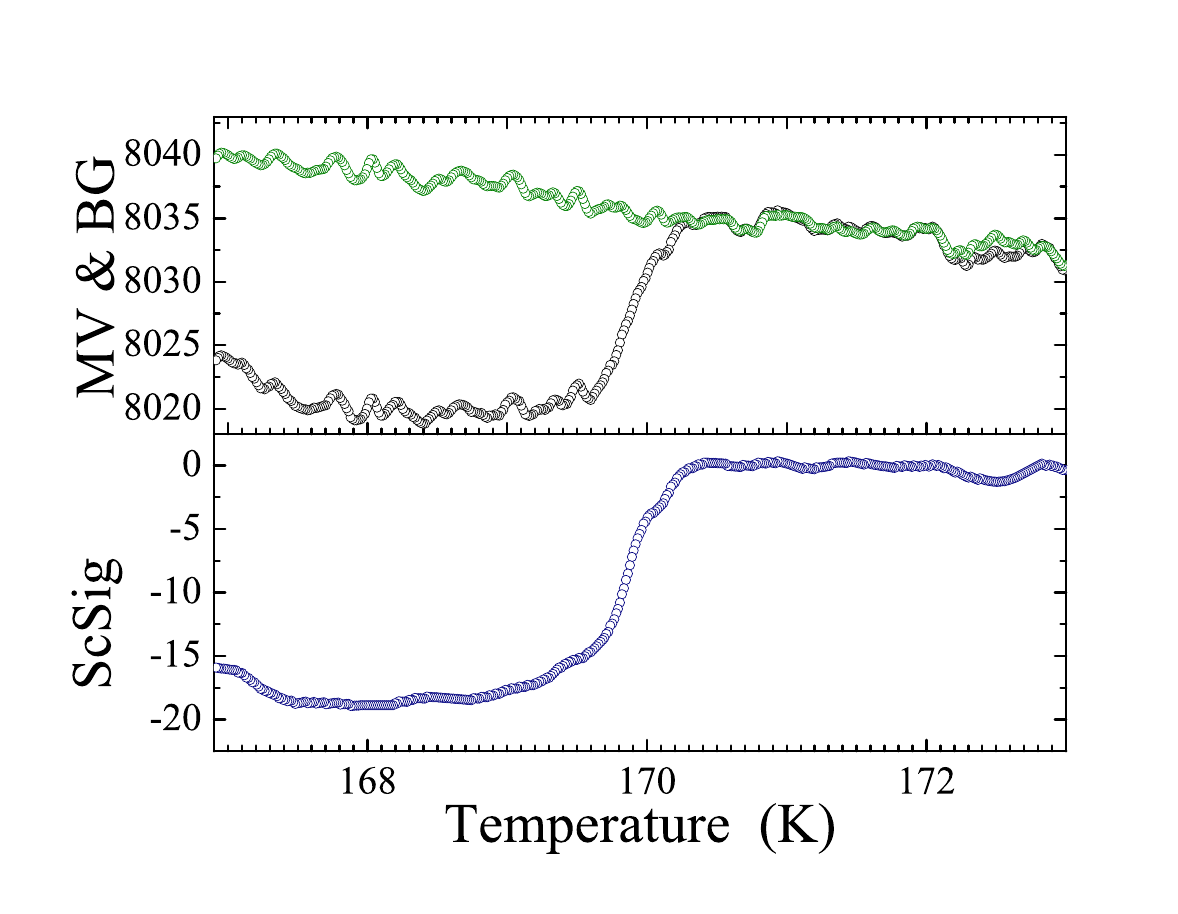}
\includegraphics[width=0.33\columnwidth]{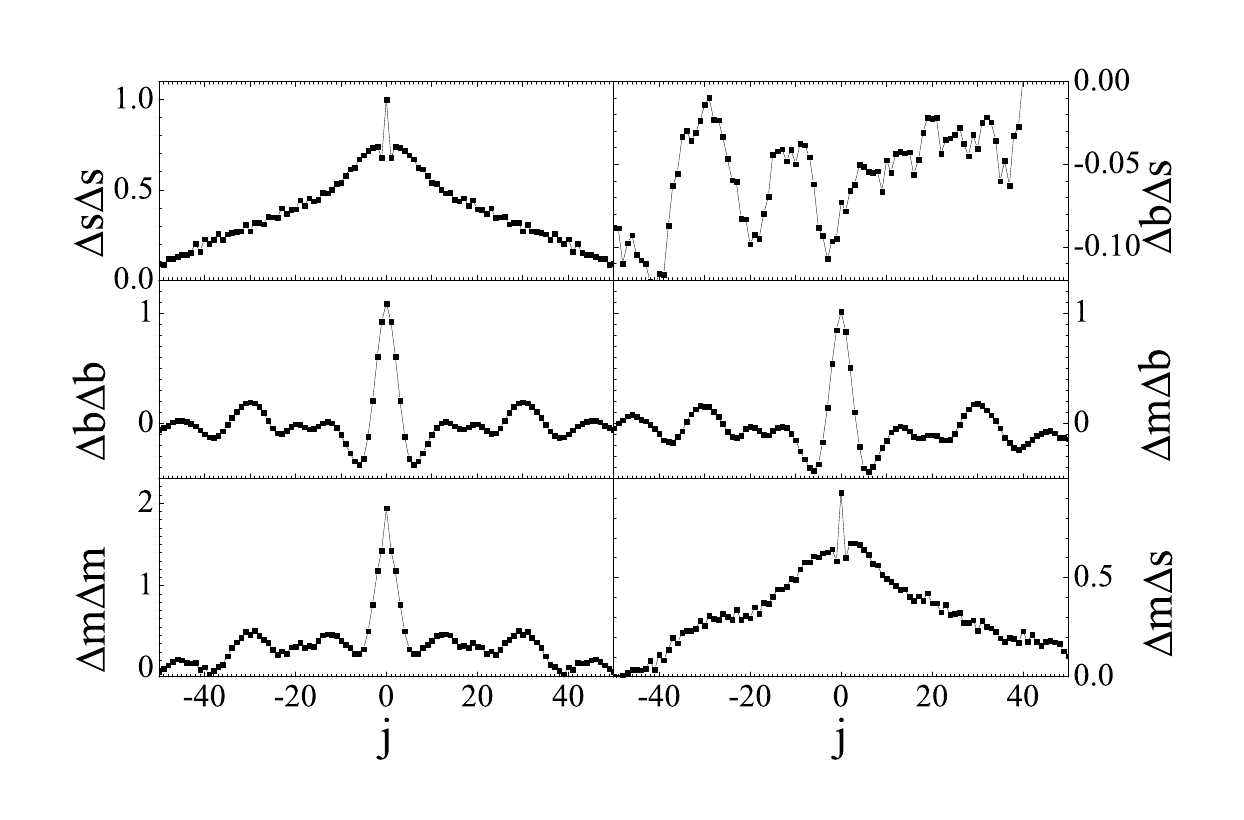}
\includegraphics[width=0.33\columnwidth]{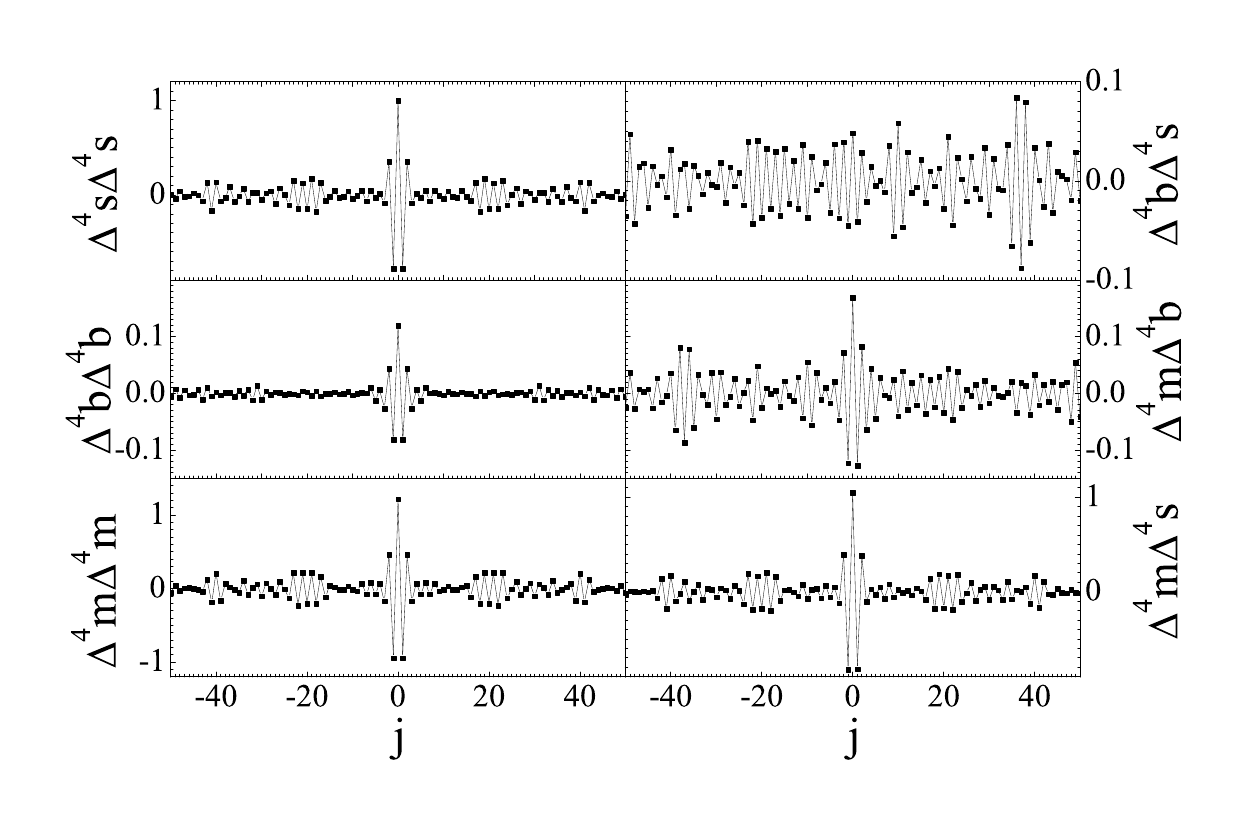}
\end{minipage}
\caption*{\bf Simulation with gaussian smoothing of BG: ac susceptibility and correlation functions}
\begin{minipage}{\columnwidth}
\includegraphics[width=0.32\columnwidth]{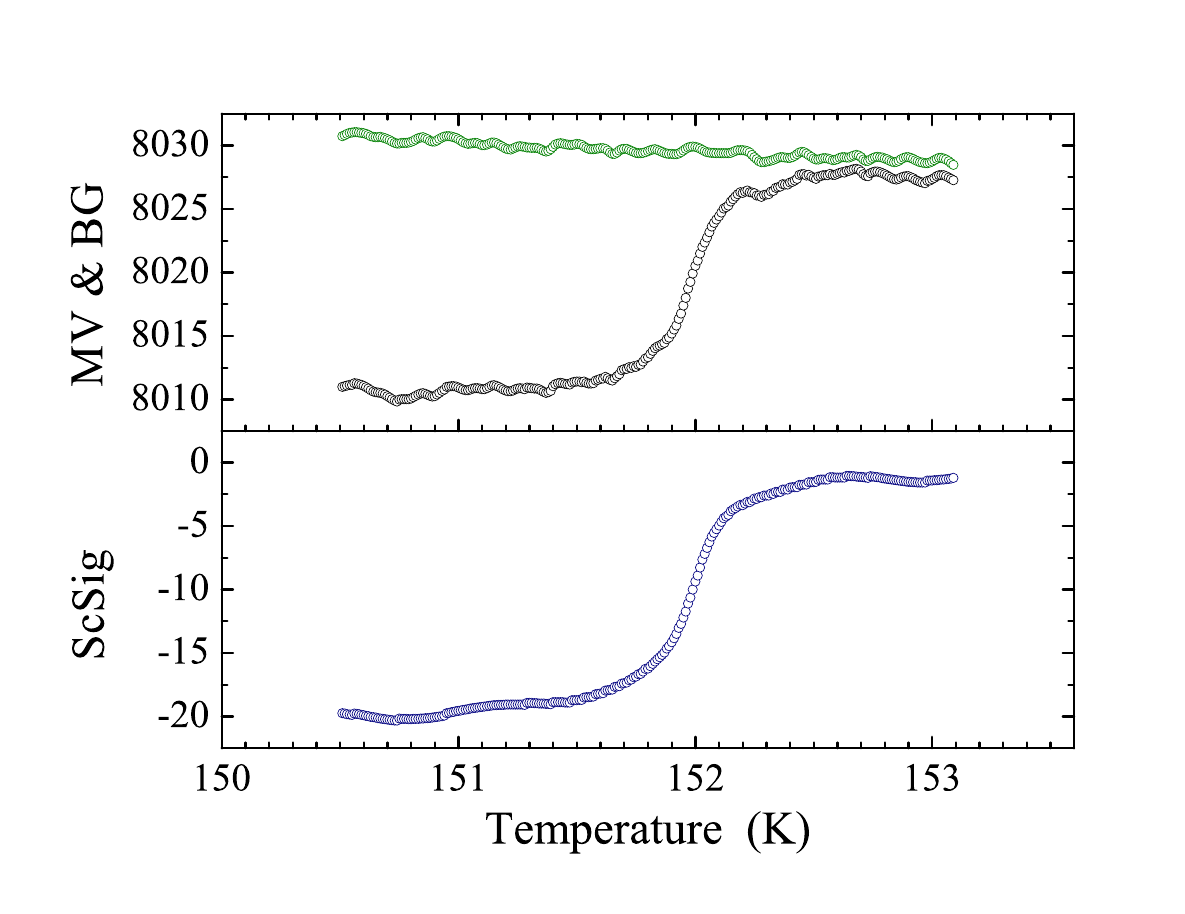}
\includegraphics[width=0.33\columnwidth]{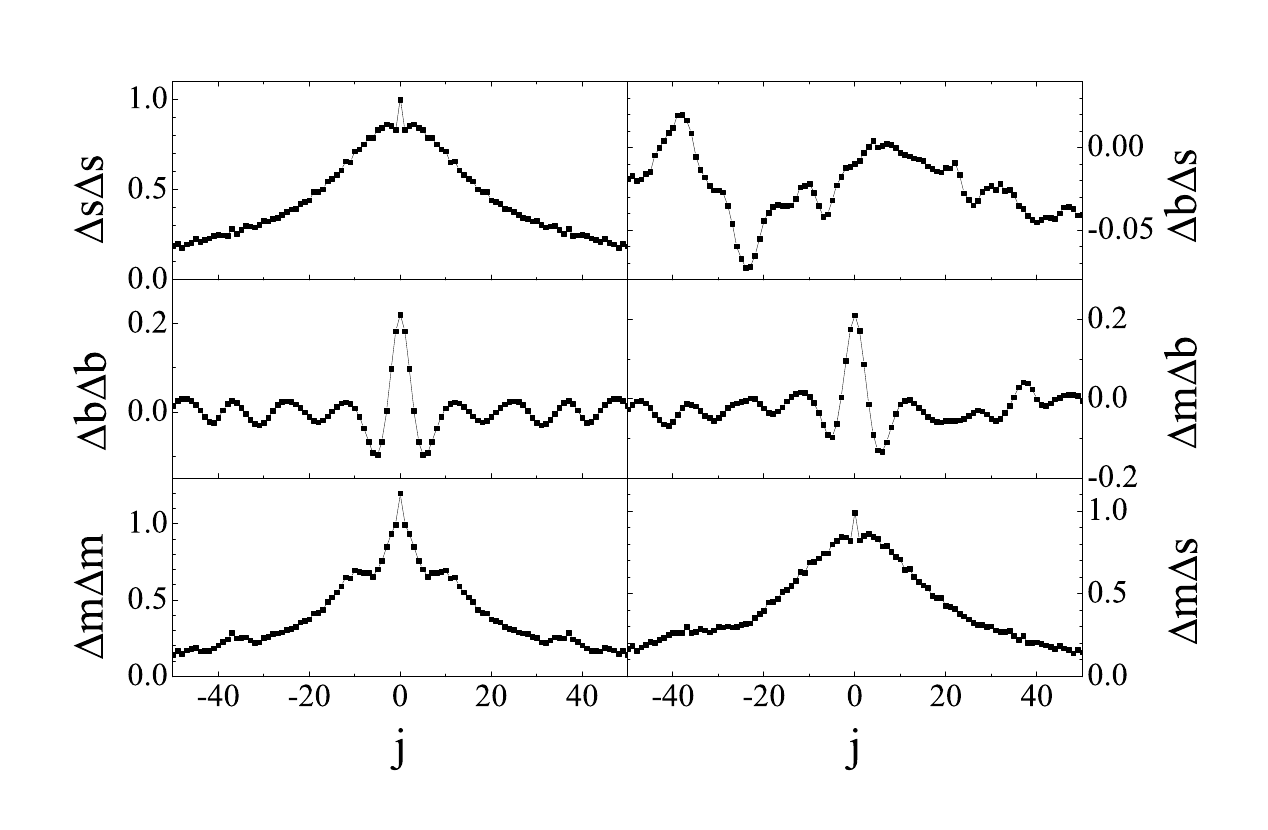}
\includegraphics[width=0.33\columnwidth]{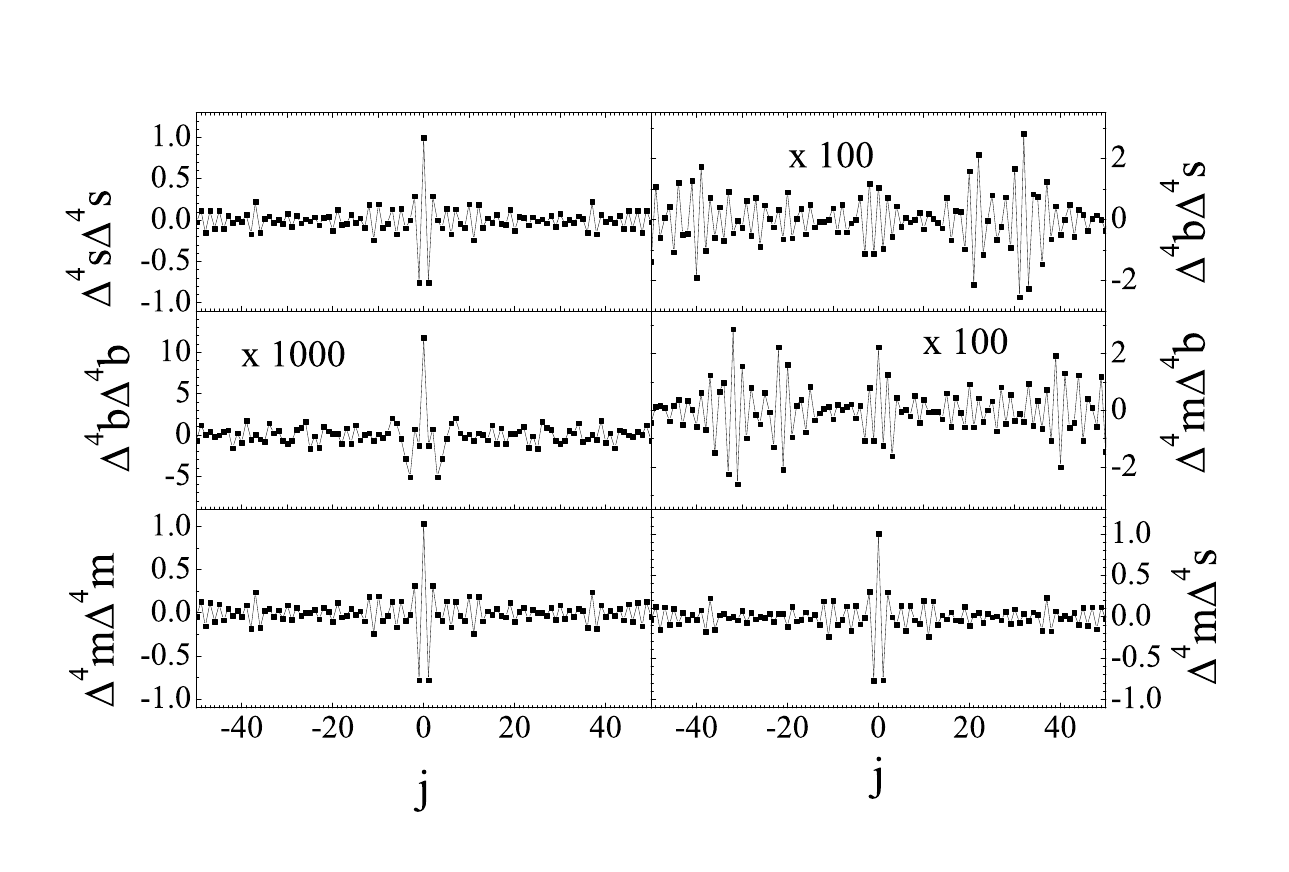}
\end{minipage}
\caption{Simulation of the effect of an exponential smoothing of the background of data (BG) obtained using protocol 3. BG is artificially generated as a smooth function + random noise. A smoothing with $1/e$ distance of 20 data points is applied to BG. The superconducting signal (ScSig) is generated as a smooth function plus quantized function. The measured voltage (MV) is generated using MV = BG + ScSig.
Left: $\chi^{\prime}_{mv}$, $\chi^{\prime}_{bg}$, $\chi^{\prime}_{sc}$.
Middle: correlation functions of the 6 combinations of $\Delta\chi^{\prime}_{mv}$, $\Delta\chi^{\prime}_{bg}$, $\Delta\chi^{\prime}_{sc}$.
Right:  correlation functions of the 6 combinations of $\Delta^4\chi^{\prime}_{mv}$, $\Delta^4\chi^{\prime}_{bg}$, $\Delta^4\chi^{\prime}_{sc}$.
Middle row of panels: Same correlation functions as above for the 160 GPa data reported in Refs.~\cite{snider2020,dias2021}.
Bottom row of panels: Same correlation functions as above for a simulation of the effect of gaussian smoothing of BG of data obtained using protocol 3. BG is artificially generated as a smooth function + random noise. A smoothing with $1/e$ distance of 3 data points is applied to BG. ScSig is generated as a smooth function plus quantized function. MV is generated using MV = BG + ScSig.}
\label{figure:simulation-BGsmoothing}
\end{figure} 
\begin{figure}[]
\centering
\caption*{\bf Simulation with smoothing of BG and ScSig: ac susceptibility and correlation functions}
\begin{minipage}{\columnwidth}
\includegraphics[width=0.32\columnwidth]{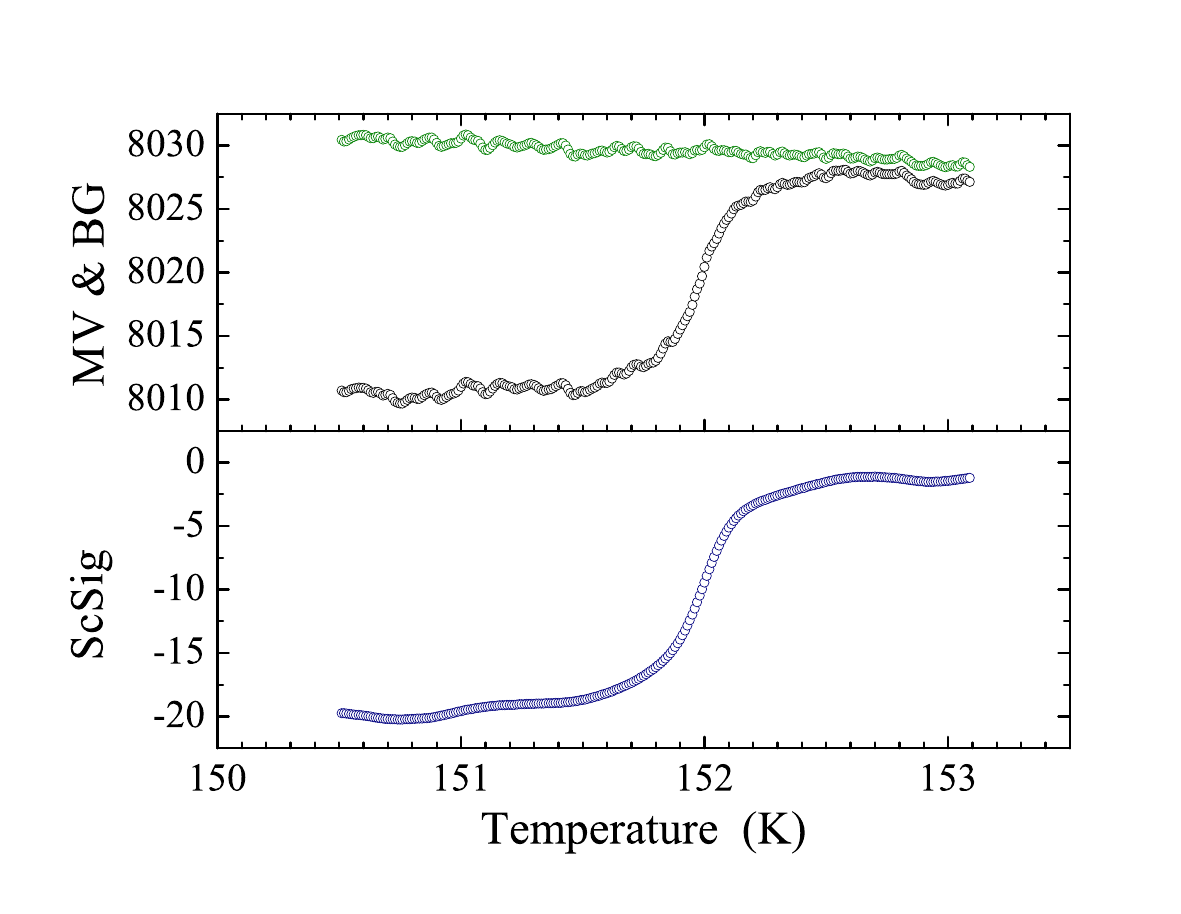}
\includegraphics[width=0.33\columnwidth]{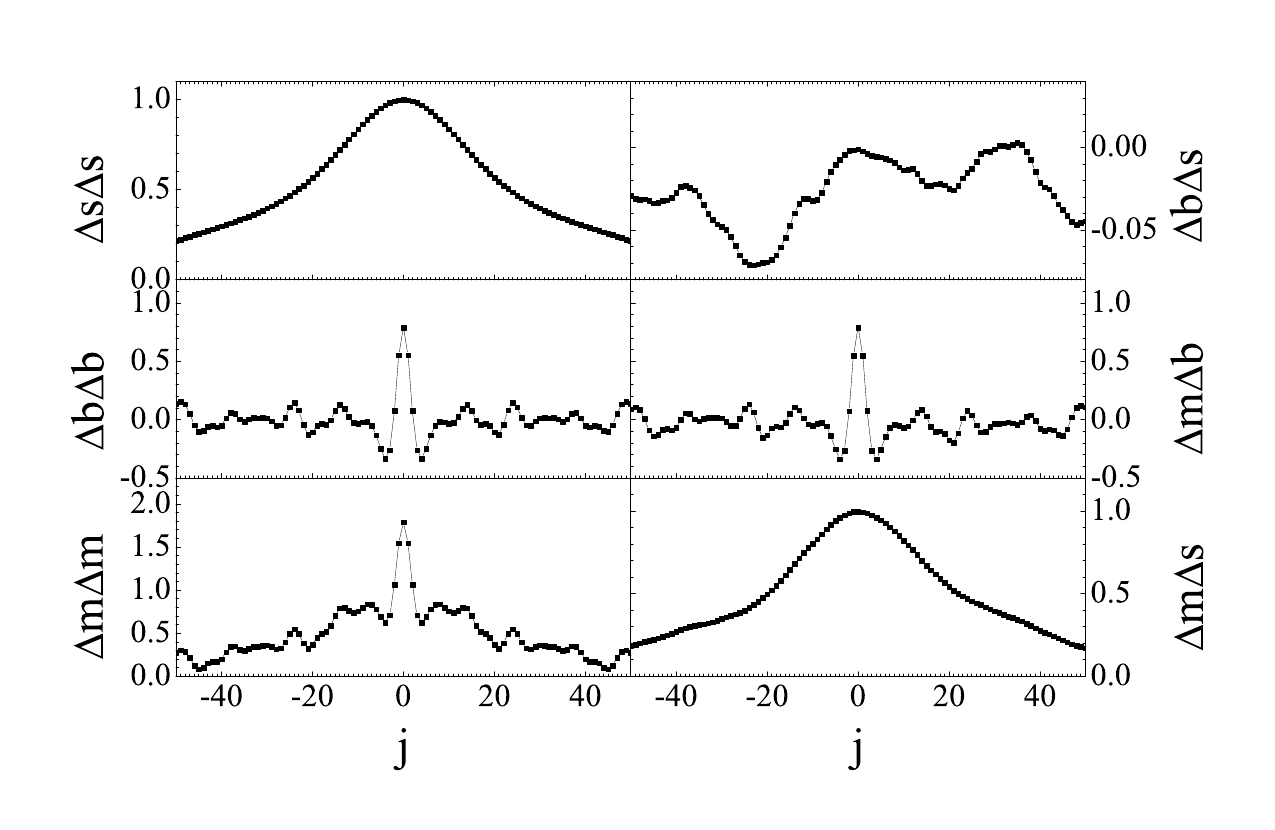}
\includegraphics[width=0.33\columnwidth]{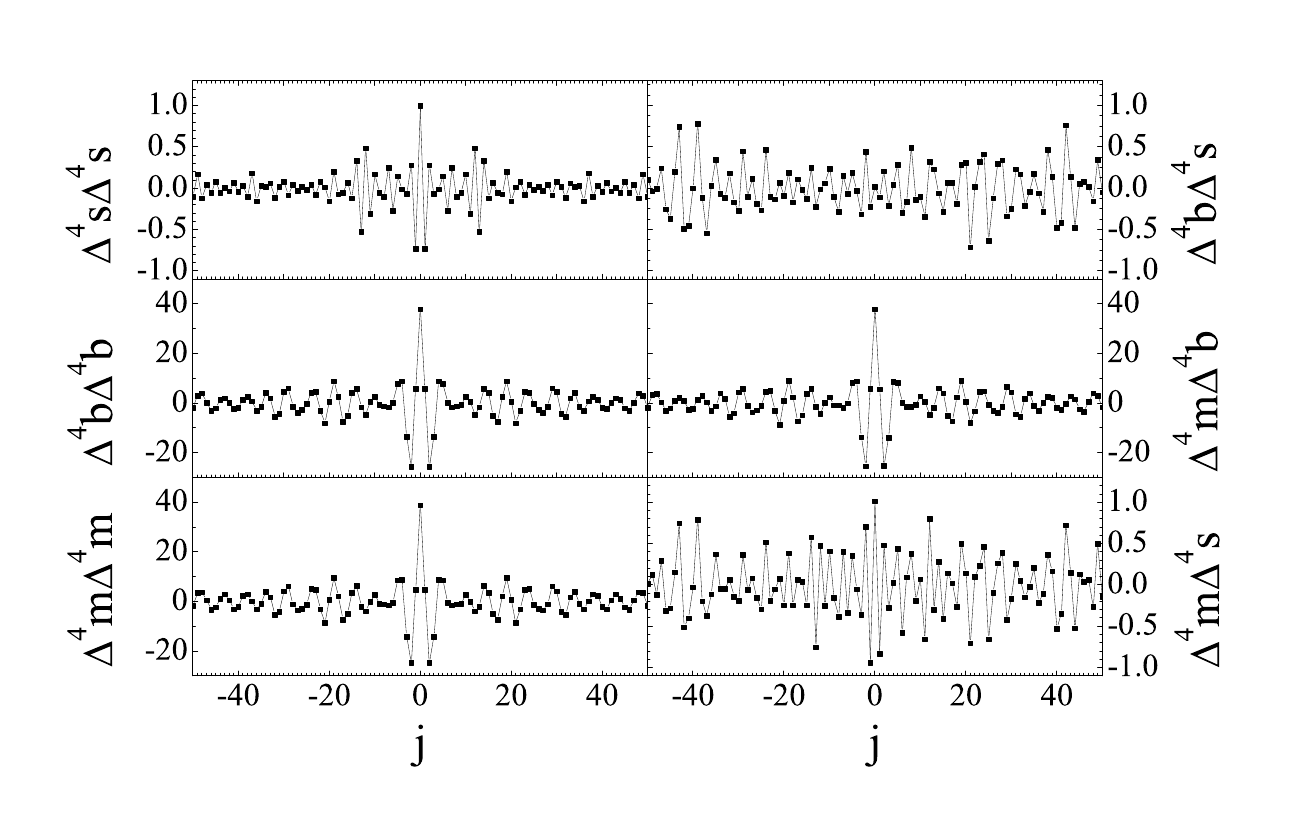}
\end{minipage}
\caption*{\bf 178~GPa: ac susceptibility and correlation functions}
\begin{minipage}{\columnwidth}
\includegraphics[width=0.32\columnwidth]{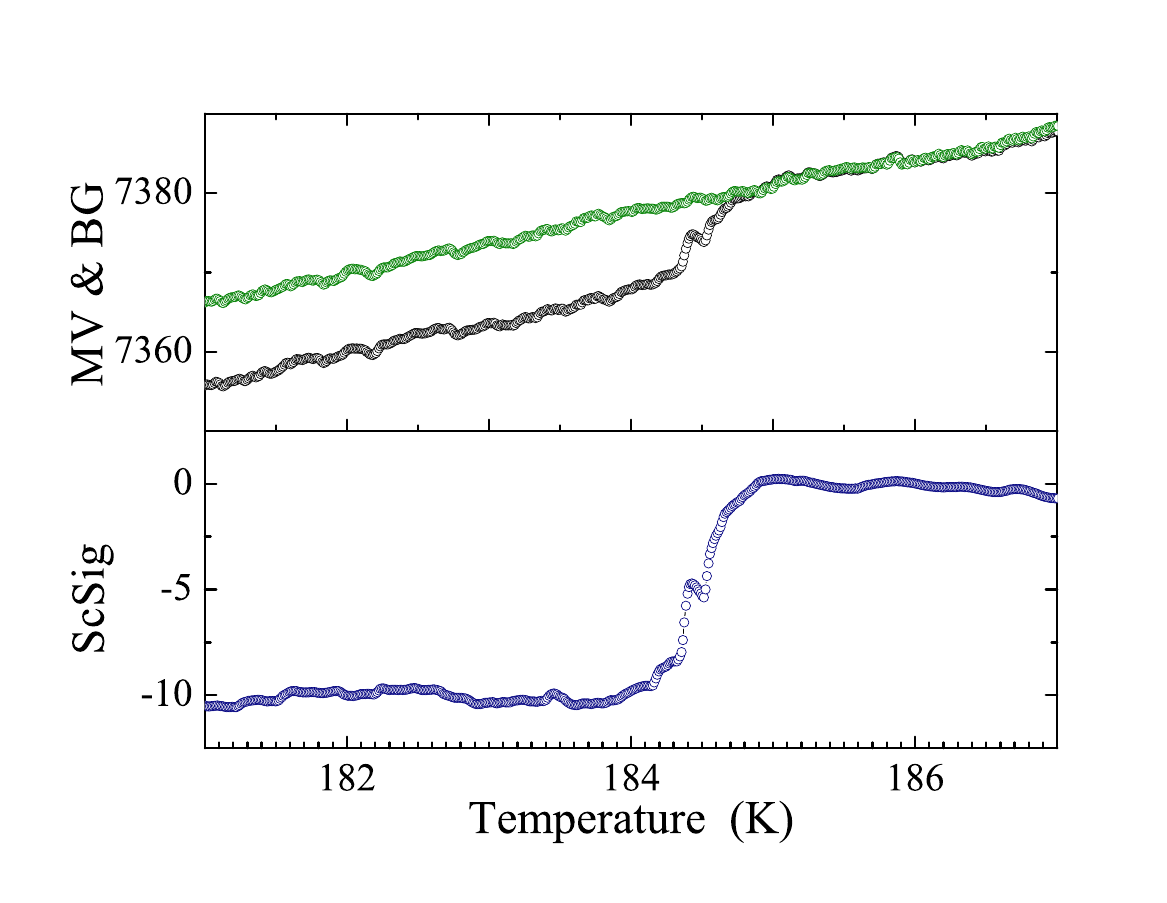}
\includegraphics[width=0.33\columnwidth]{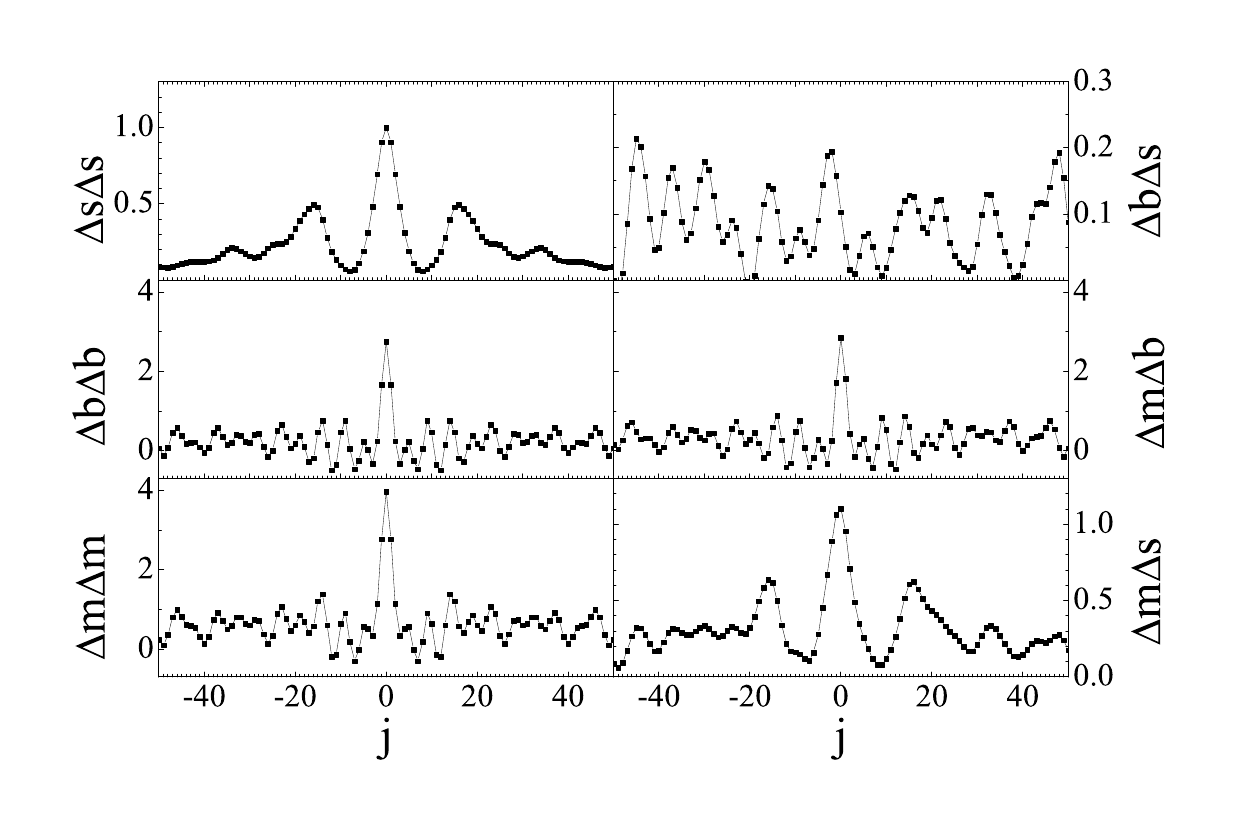}
\includegraphics[width=0.33\columnwidth]{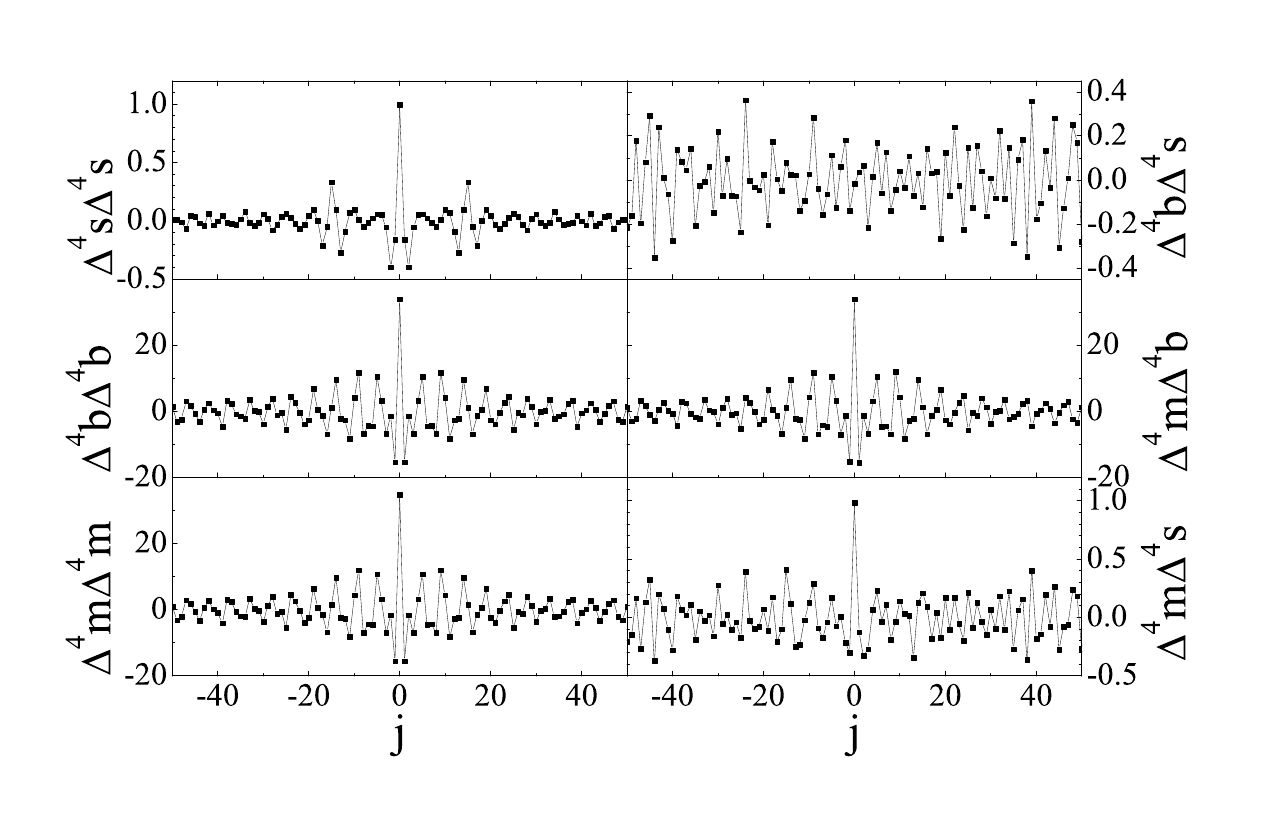}
\end{minipage}
\caption{Top panels: Simulation of the effect of a gaussian smoothing of the background of data (BG) obtained using protocol 3 and Adjacent Averaging of ScSig. BG is artificially generated as a smooth function + random noise. A smoothing with $1/e$ distance of 3 data points is applied to BG. The superconducting signal (ScSig) is generated as a smooth function plus quantized function, and then Adjacent Averaged over 13 points. The measured voltage (MV) is generated using MV = BG + ScSig.
Left: $\chi^{\prime}_{mv}$, $\chi^{\prime}_{bg}$, $\chi^{\prime}_{sc}$.
Middle: correlation functions of the 6 combinations of $\Delta\chi^{\prime}_{mv}$, $\Delta\chi^{\prime}_{bg}$, $\Delta\chi^{\prime}_{sc}$.
Right:  correlation functions of the 6 combinations of $\Delta^4\chi^{\prime}_{mv}$, $\Delta^4\chi^{\prime}_{bg}$, $\Delta^4\chi^{\prime}_{sc}$.
Bottom row of panels: Same correlation functions as above for the 178 GPa data reported in Refs.~\cite{snider2020,dias2021}.}
\label{figure:simulation-BGsmoothing_AA13ScSig}
\end{figure} 
%
%
\begin{figure}[]
\centering
\caption*{\bf 138~GPa: ac susceptibility and correlation functions}
\begin{minipage}{\columnwidth}
\includegraphics[width=0.32\columnwidth]{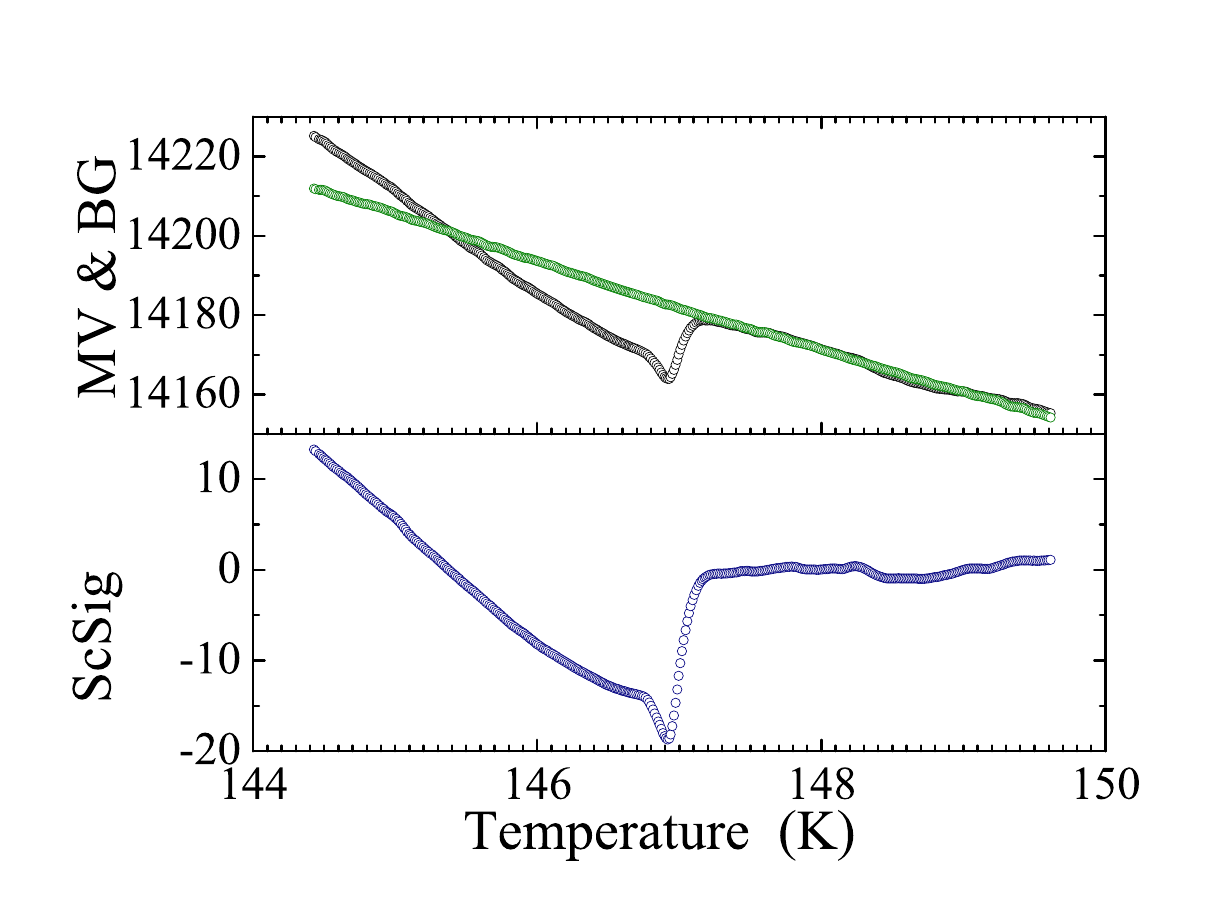}
\includegraphics[width=0.33\columnwidth]{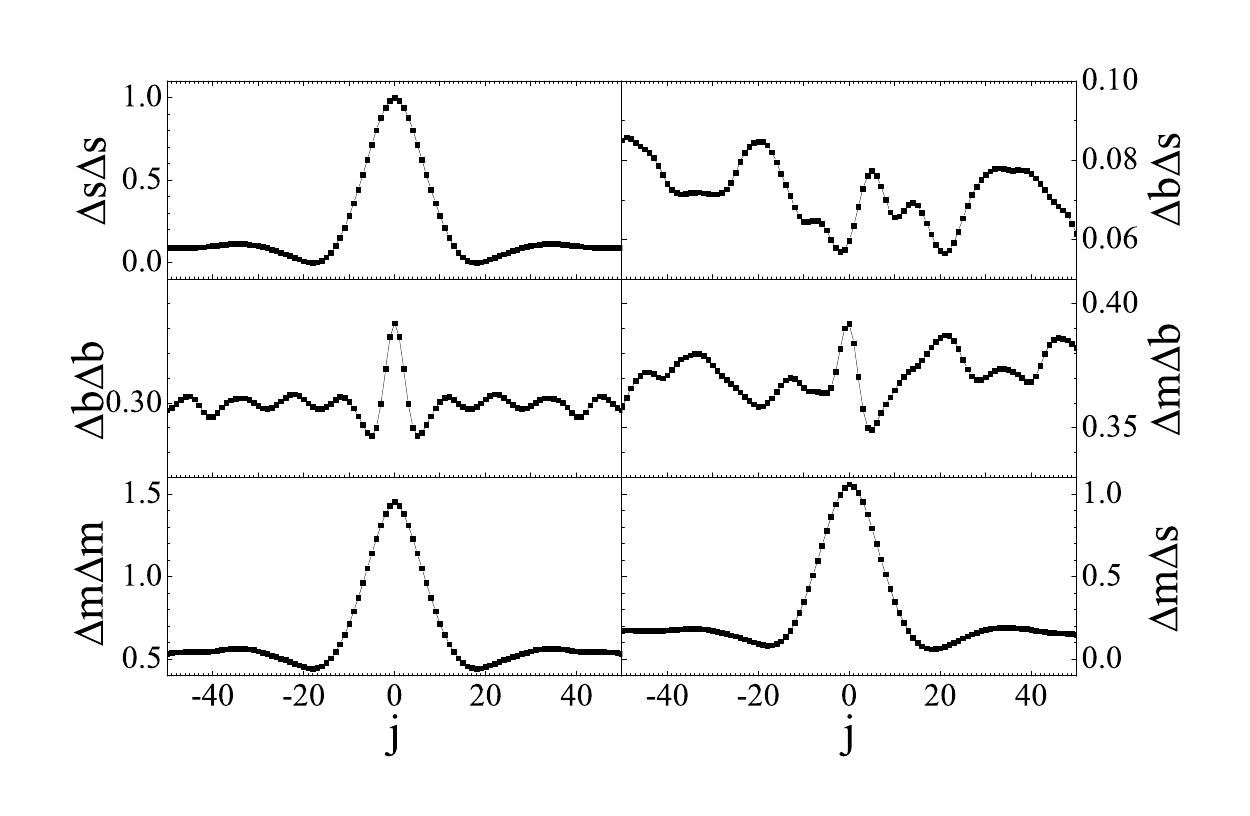}
\includegraphics[width=0.33\columnwidth]{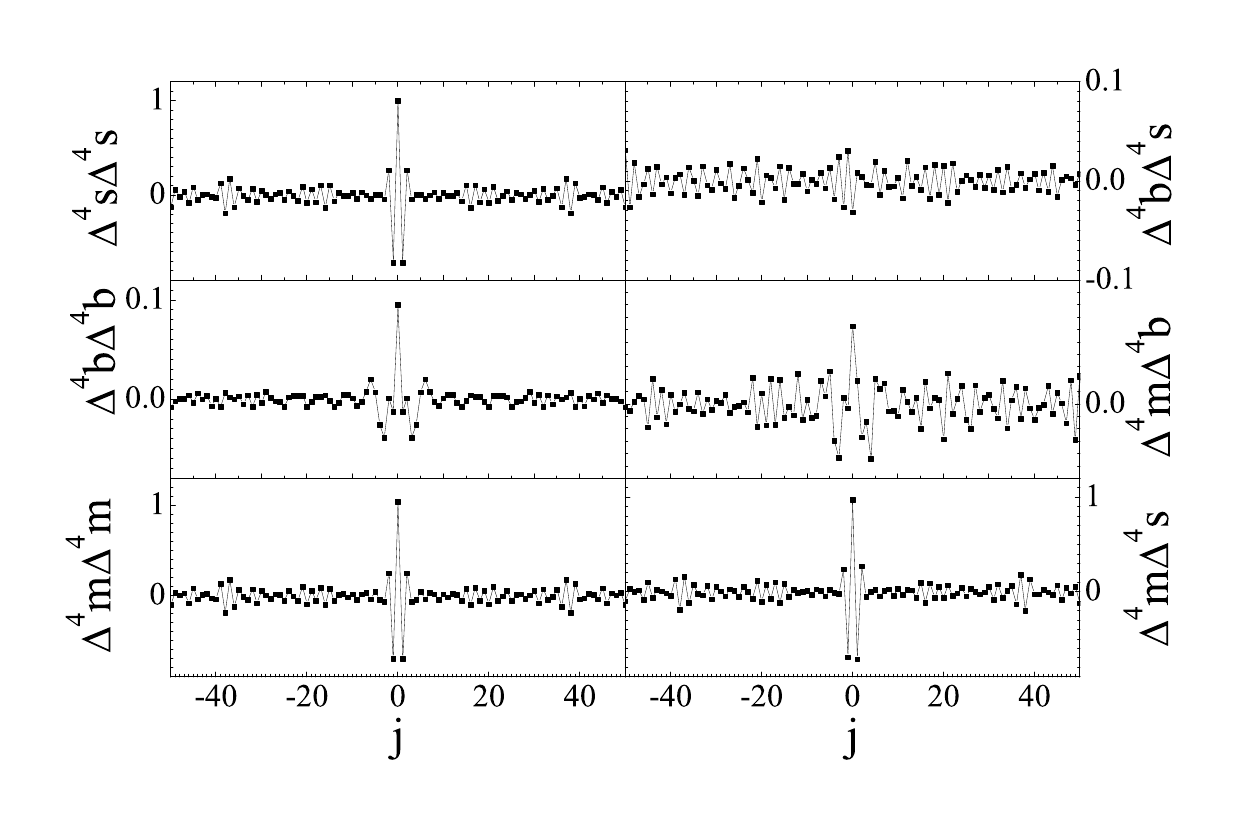}
\end{minipage}
\caption*{\bf 160~GPa: ac susceptibility and correlation functions}
\begin{minipage}{\columnwidth}
\includegraphics[width=0.32\columnwidth]{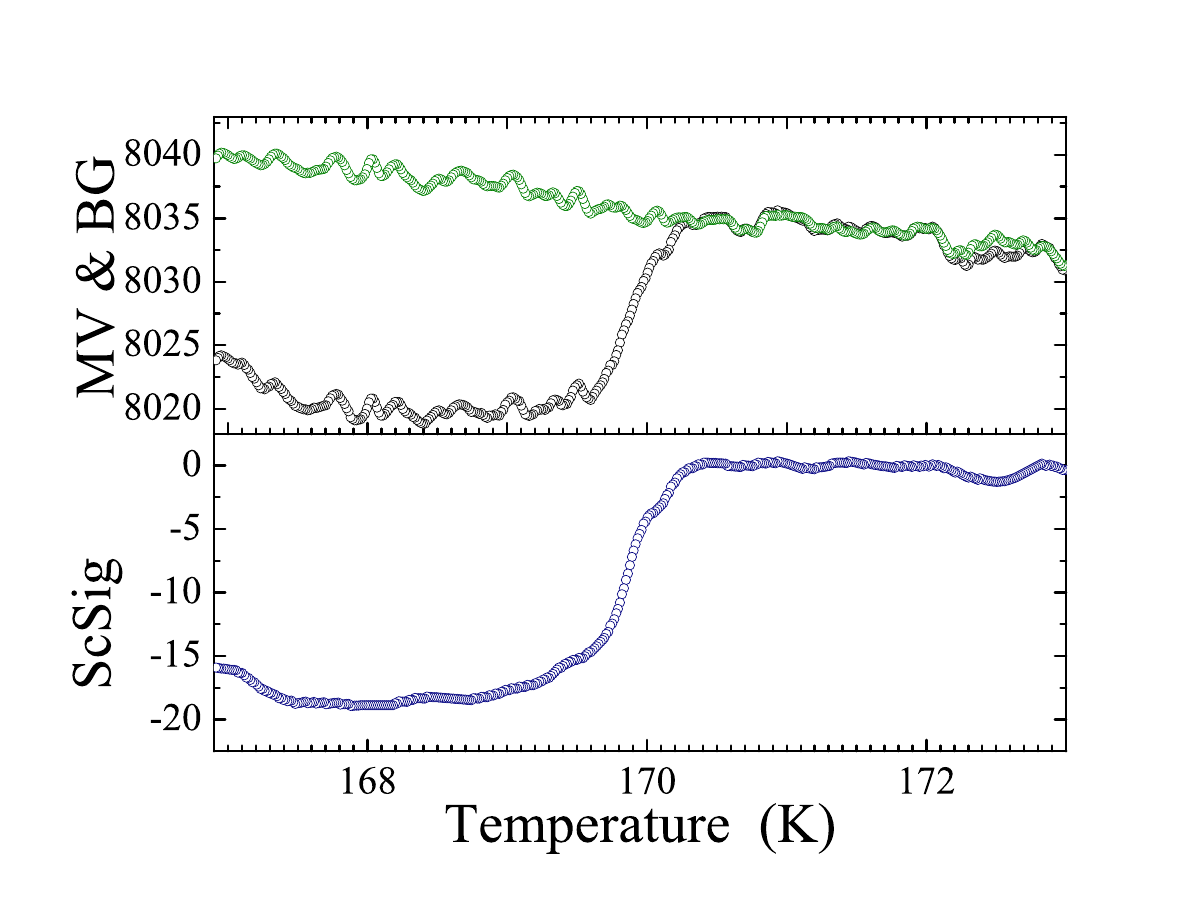}
\includegraphics[width=0.33\columnwidth]{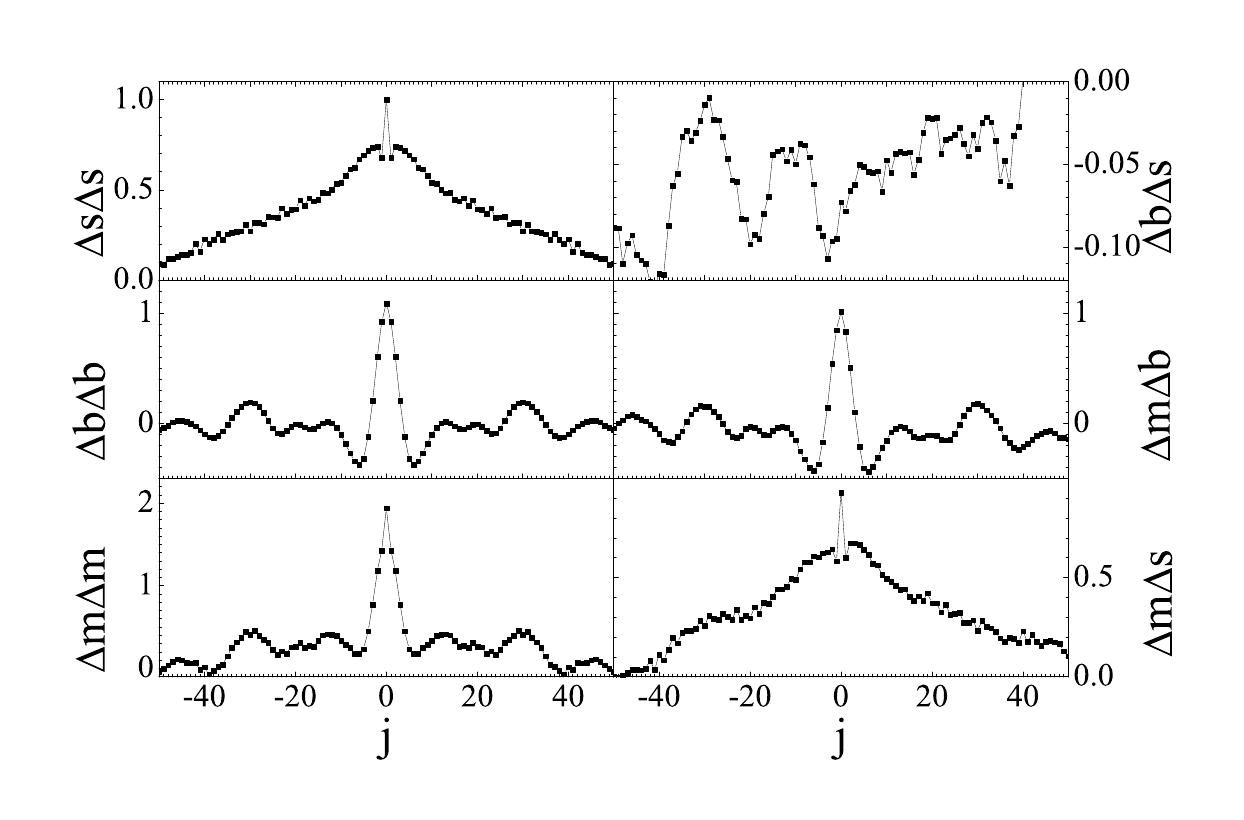}
\includegraphics[width=0.33\columnwidth]{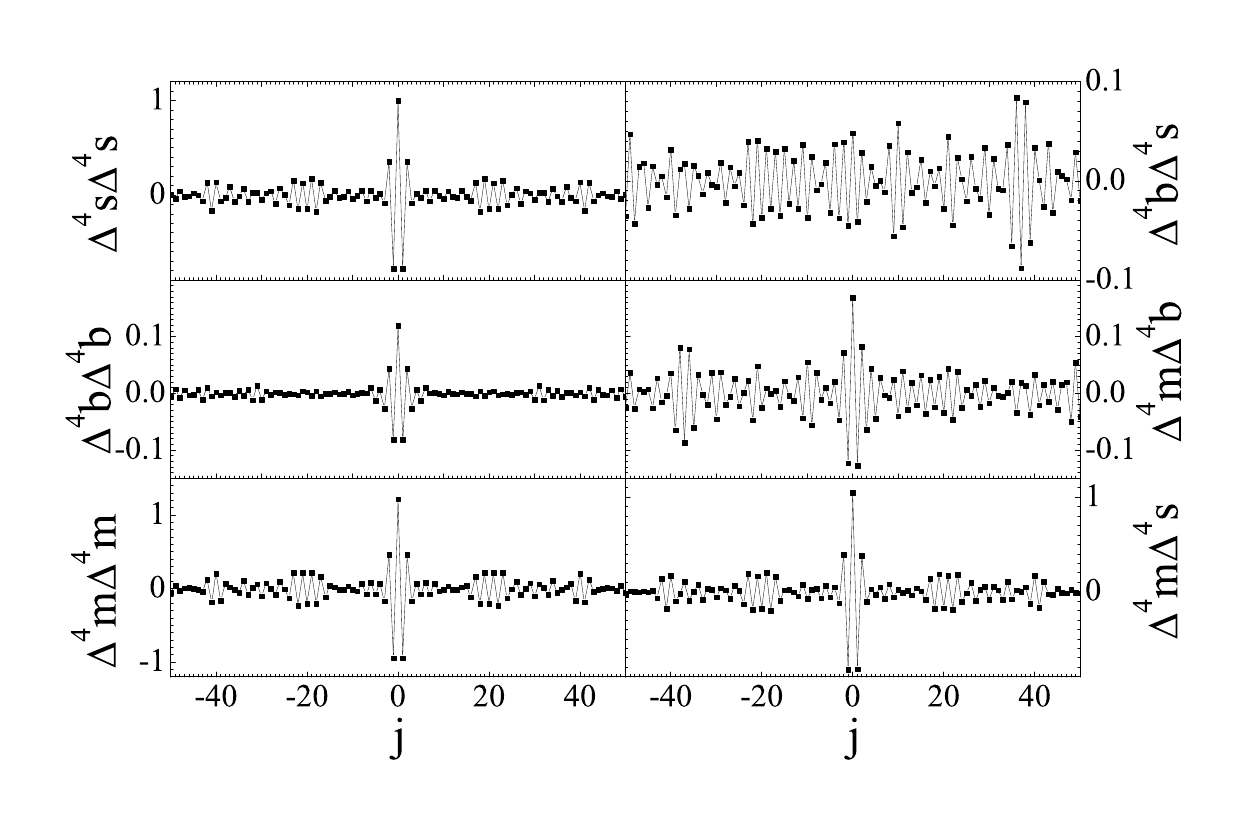}
\end{minipage}
\caption*{\bf 166~GPa: ac susceptibility and correlation functions}
\begin{minipage}{\columnwidth}
\includegraphics[width=0.32\columnwidth]{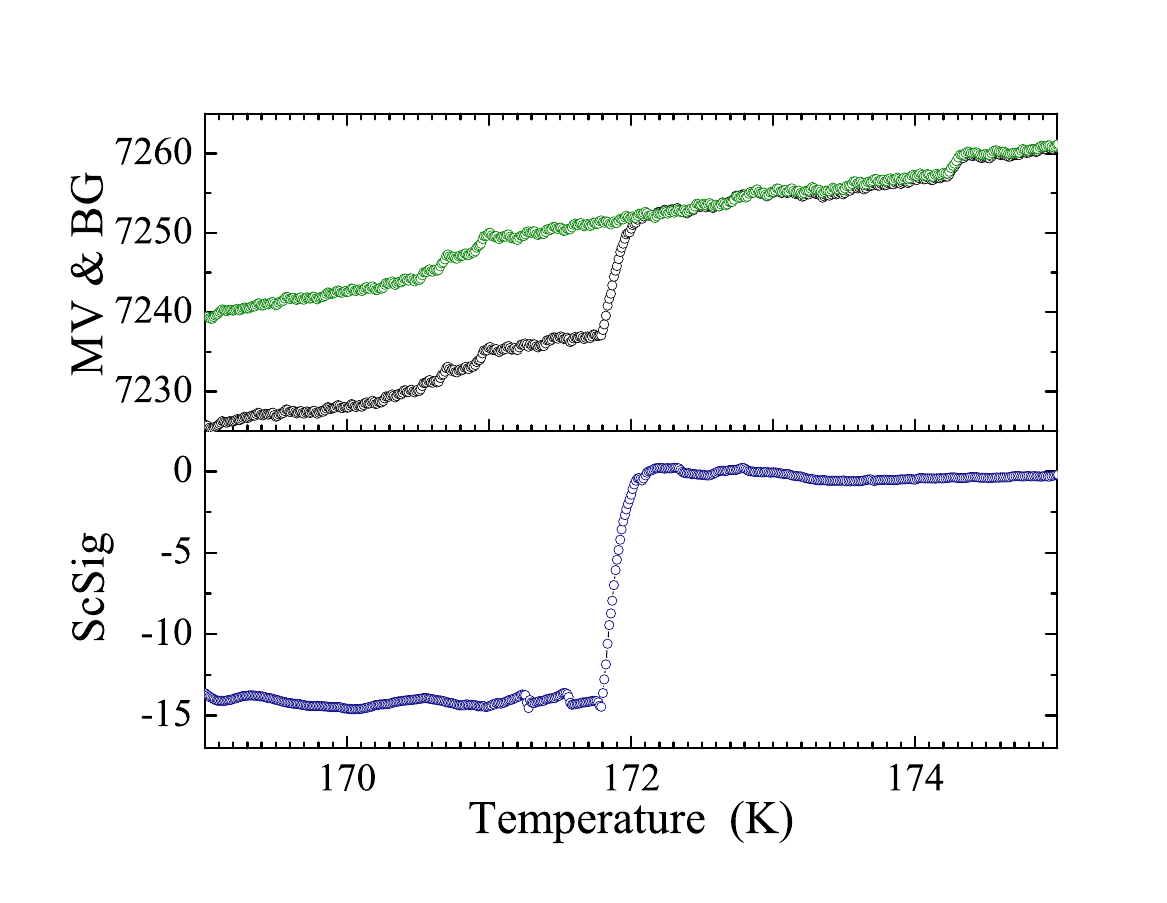}
\includegraphics[width=0.33\columnwidth]{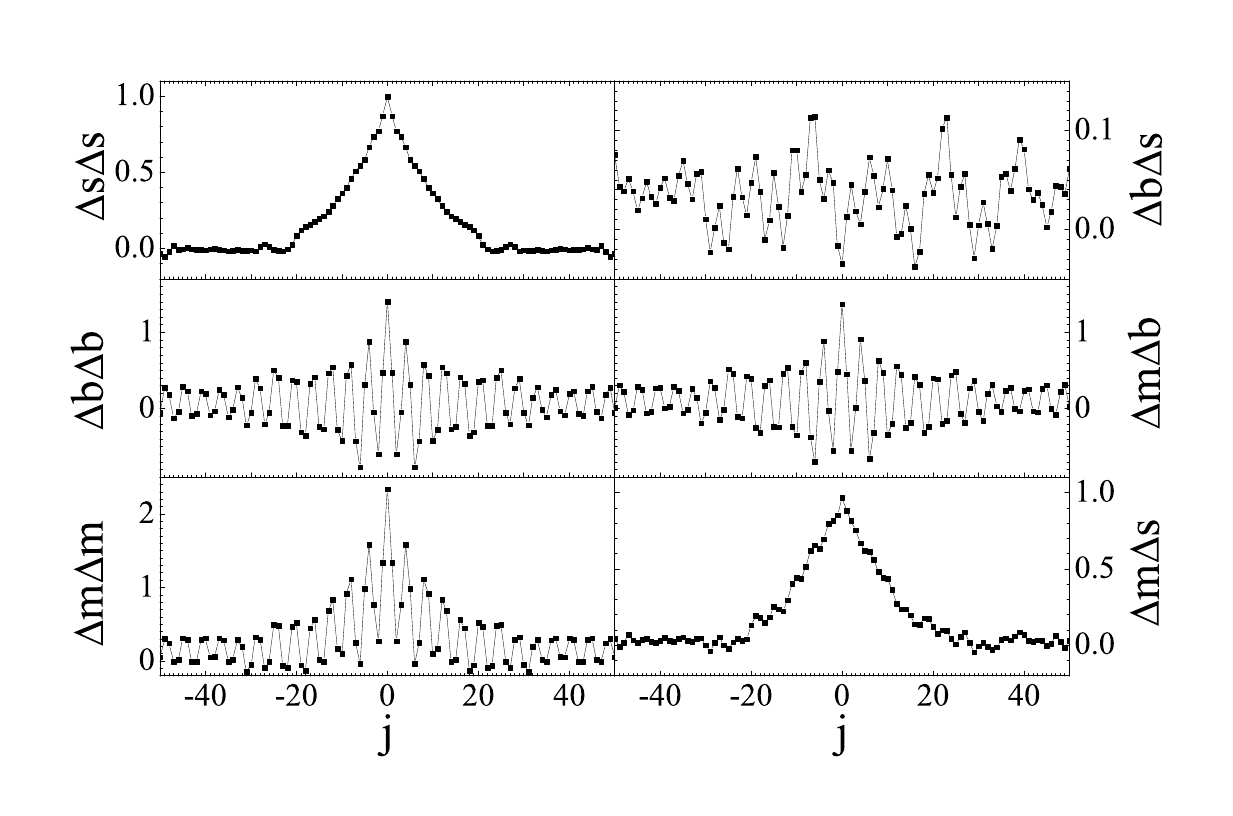}
\includegraphics[width=0.33\columnwidth]{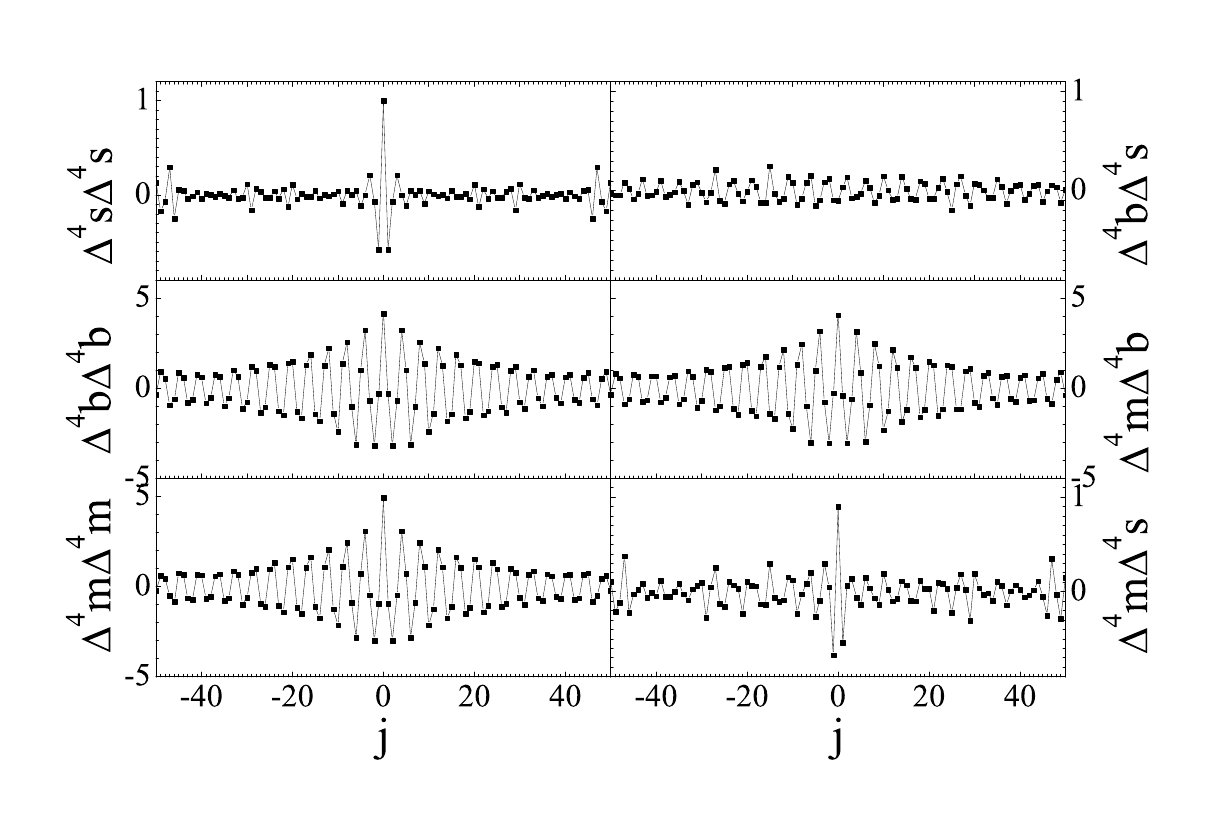}
\end{minipage}
\caption{Left: $\chi^{\prime}_{mv}$, $\chi^{\prime}_{bg}$, $\chi^{\prime}_{sc}$.
Middle: correlation functions of the 6 combinations of $\Delta\chi^{\prime}_{mv}$, $\Delta\chi^{\prime}_{bg}$, $\Delta\chi^{\prime}_{sc}$.
Right:  correlation functions of the 6 combinations of $\Delta^4\chi^{\prime}_{mv}$, $\Delta^4\chi^{\prime}_{bg}$, $\Delta^4\chi^{\prime}_{sc}$.
From top to bottom:138 GPa (top), 160 GPa (middle), 166 GPa (bottom) data reported in Refs.~\cite{snider2020,dias2021}.}
\label{figure:correlations138-160-166}
\end{figure} 
\begin{figure}[]
\centering
\caption*{\bf 178~GPa: ac susceptibility and correlation functions}
\begin{minipage}{\columnwidth}
\includegraphics[width=0.32\columnwidth]{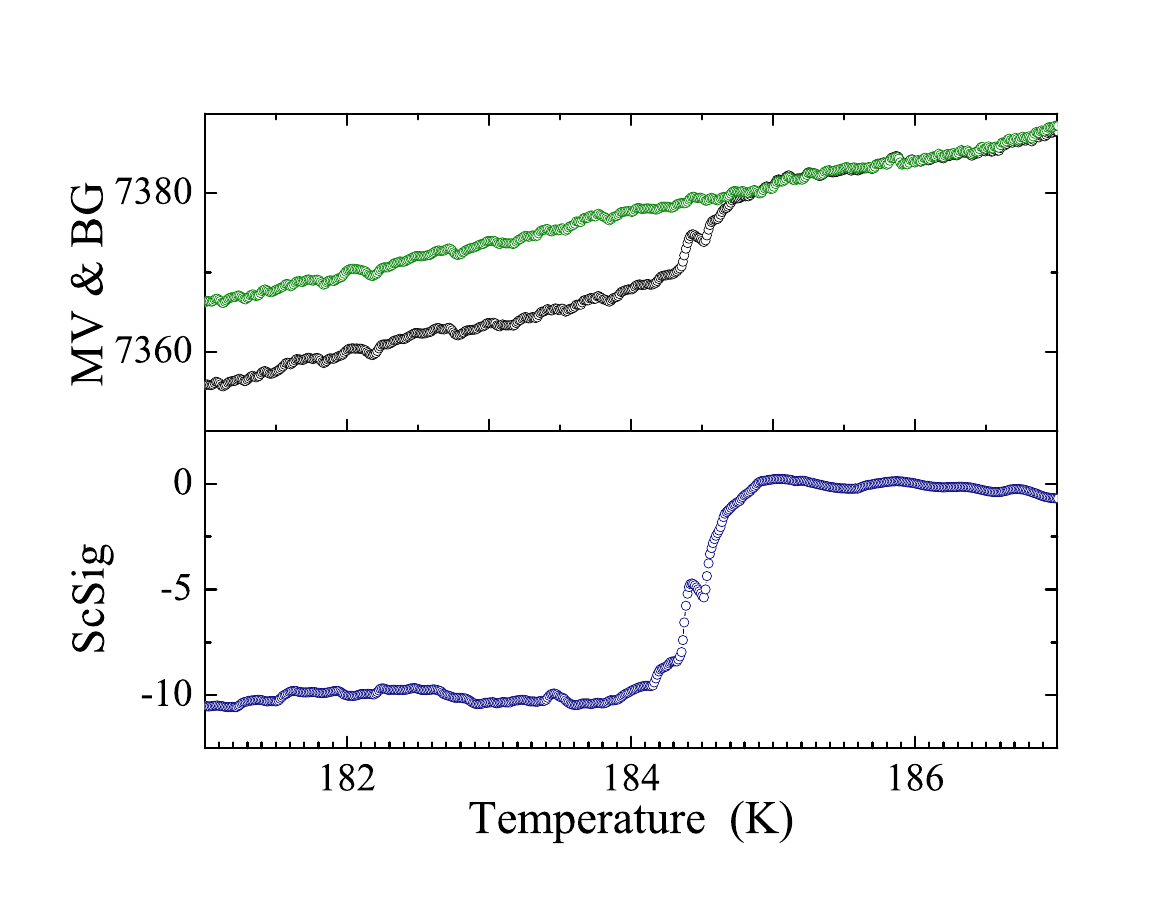}
\includegraphics[width=0.33\columnwidth]{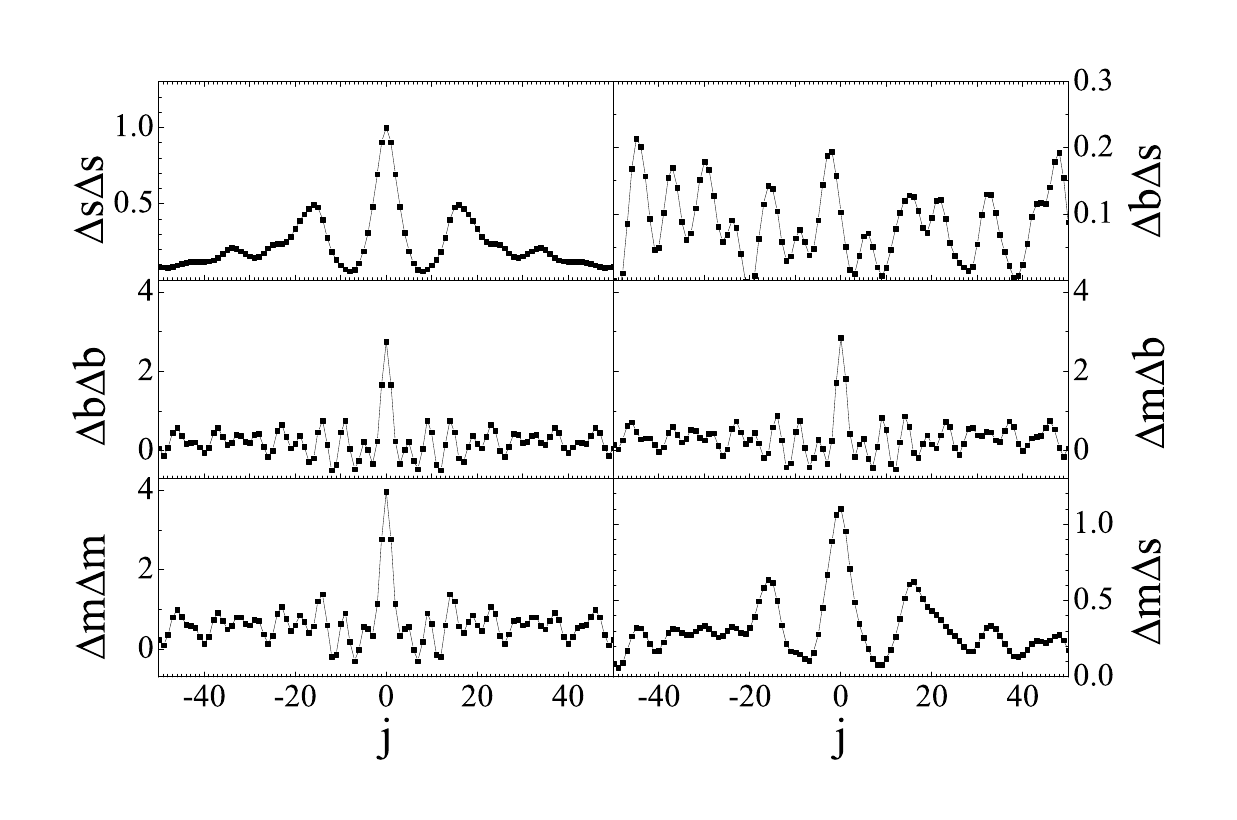}
\includegraphics[width=0.33\columnwidth]{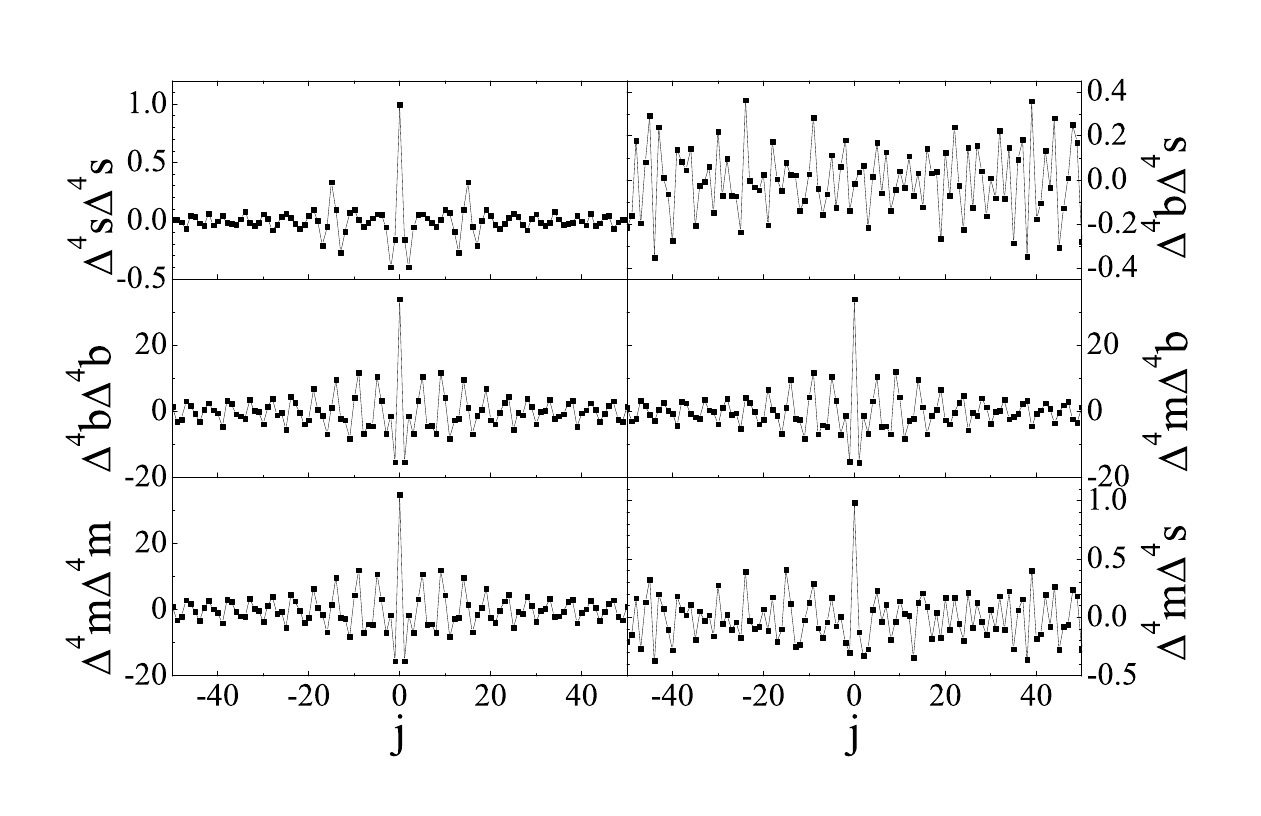}
\end{minipage}
\caption*{\bf 182~GPa: ac susceptibility and correlation functions}
\begin{minipage}{\columnwidth}
\includegraphics[width=0.32\columnwidth]{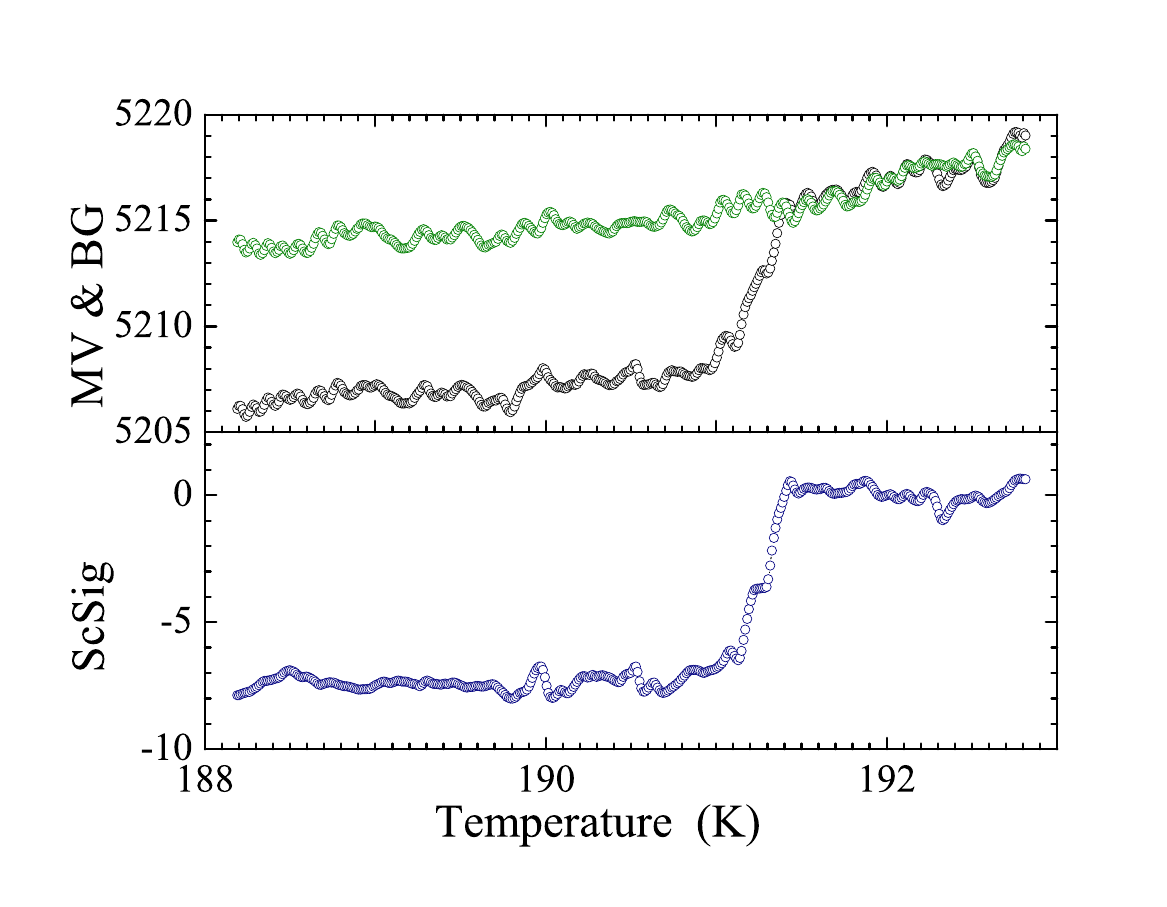}
\includegraphics[width=0.33\columnwidth]{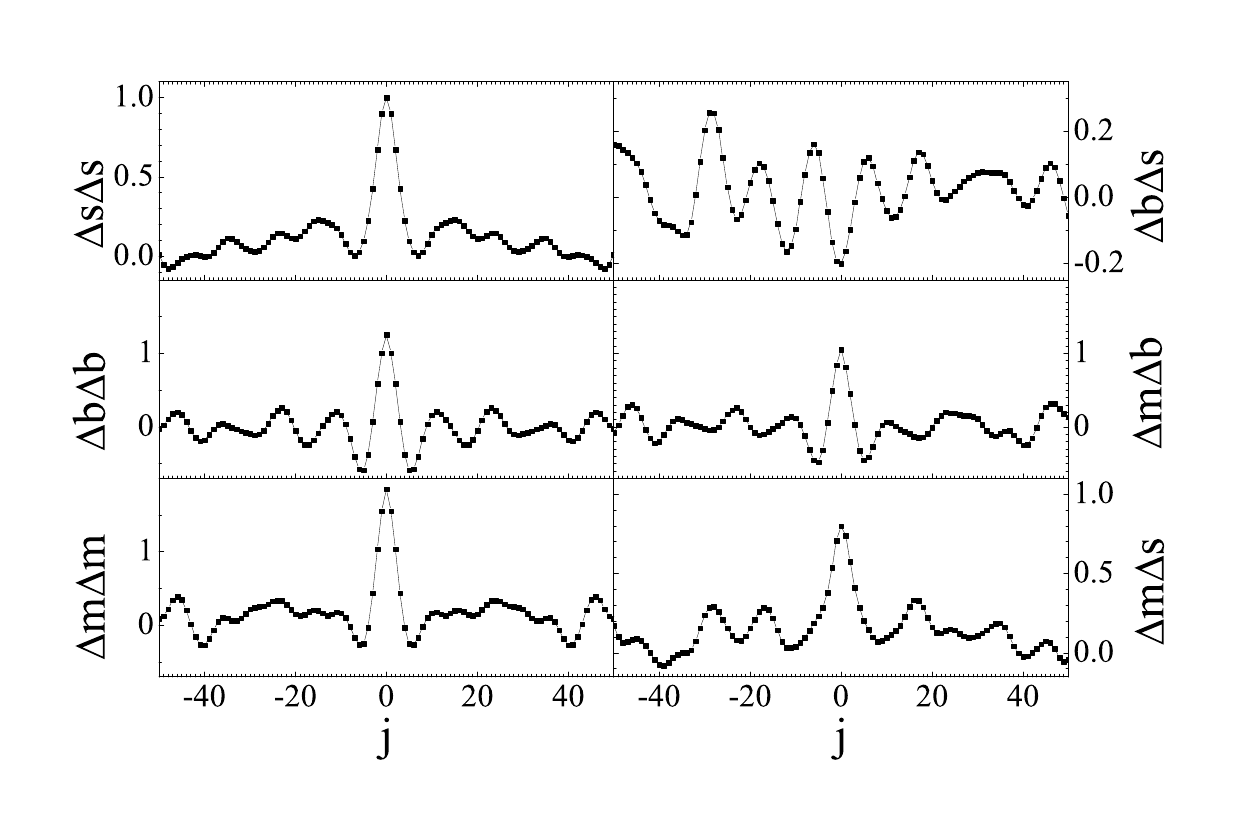}
\includegraphics[width=0.33\columnwidth]{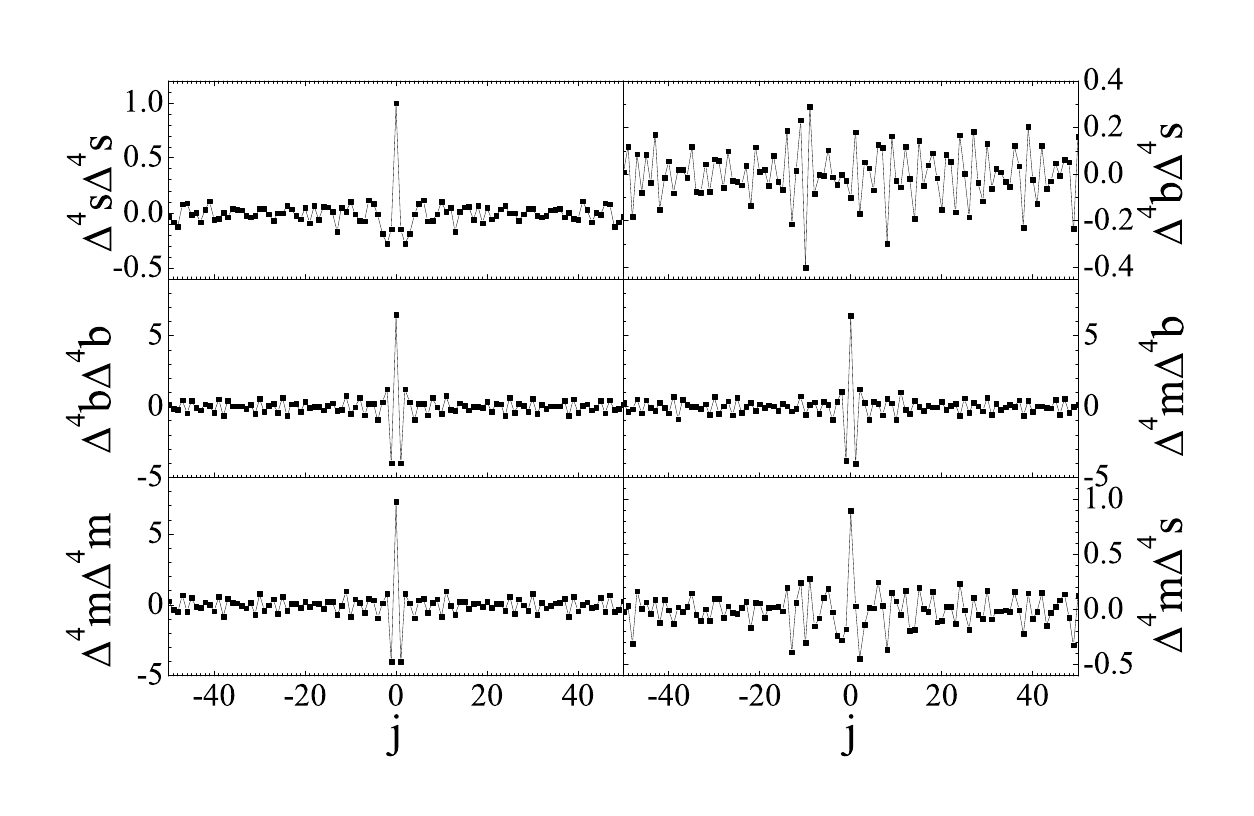}
\end{minipage}
\caption*{\bf 189~GPa: ac susceptibility and correlation functions}
\begin{minipage}{\columnwidth}
\includegraphics[width=0.32\columnwidth]{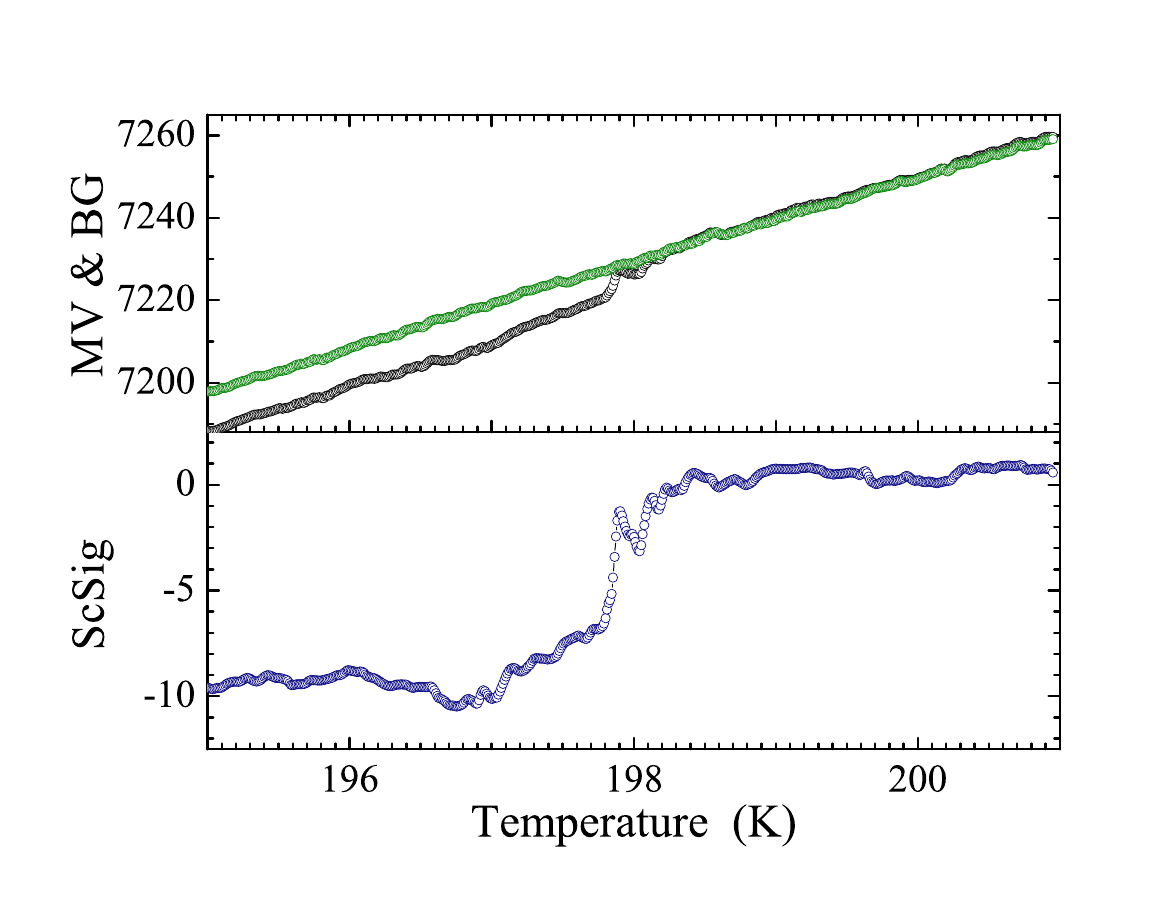}
\includegraphics[width=0.33\columnwidth]{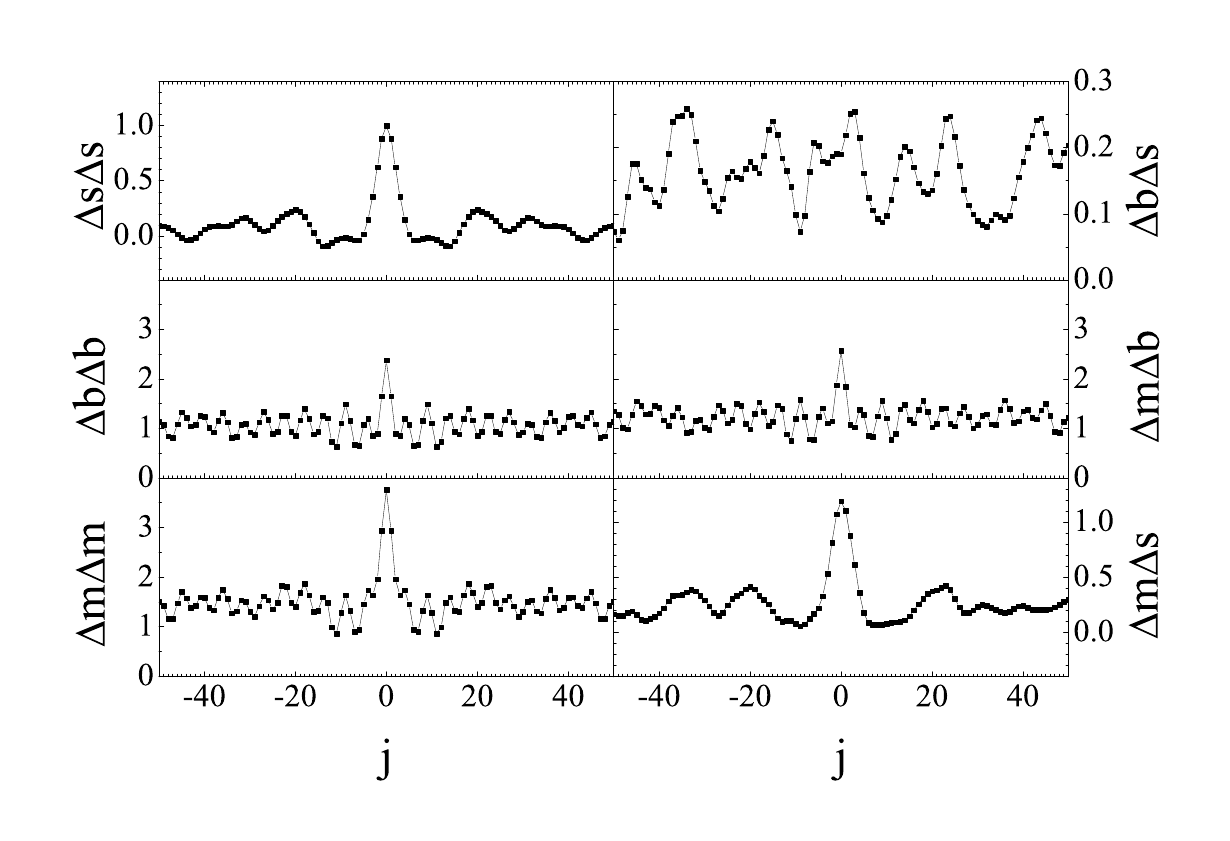}
\includegraphics[width=0.33\columnwidth]{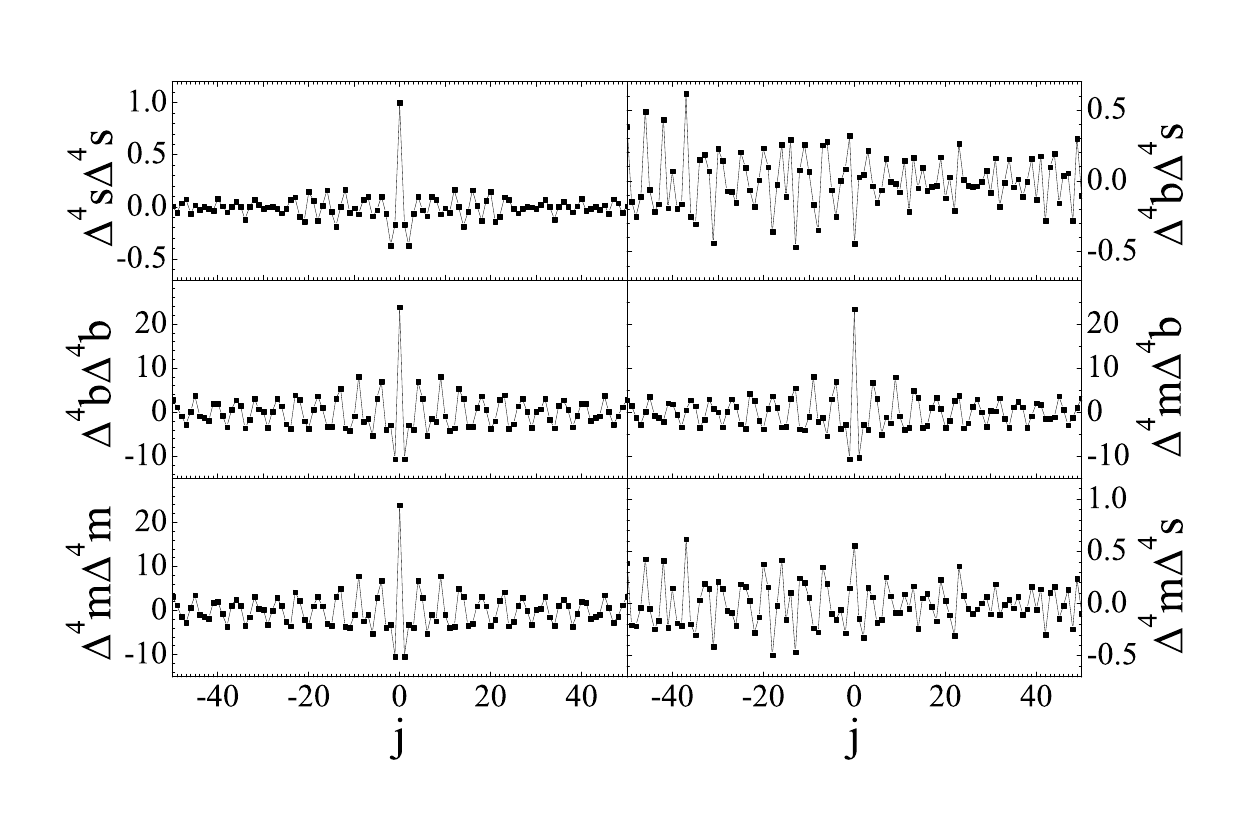}
\end{minipage}
\caption{Left: $\chi^{\prime}_{mv}$, $\chi^{\prime}_{bg}$, $\chi^{\prime}_{sc}$.
Middle: correlation functions of the 6 combinations of $\Delta\chi^{\prime}_{mv}$, $\Delta\chi^{\prime}_{bg}$, $\Delta\chi^{\prime}_{sc}$.
Right:  correlation functions of the 6 combinations of $\Delta^4\chi^{\prime}_{mv}$, $\Delta^4\chi^{\prime}_{bg}$, $\Delta^4\chi^{\prime}_{sc}$.
From top to bottom:178 GPa (top), 182 GPa (middle), 189 GPa (bottom) data reported in Refs.~\cite{snider2020,dias2021}.}
\label{figure:correlations178-182-189}
\end{figure} 
\end{document}